\documentclass[floatfix,aps,prd,twocolumn,letterpaper,lengthcheck,superscriptaddress,showpacs,amssymb,amsmath,amsfonts,nofootinbib,altaffilletter,nopreprintnumbers,showpacs,longbibliography]{revtex4-1}
\usepackage[utf8]{inputenc}
\usepackage{xspace}
\usepackage{acronym}
\usepackage{booktabs}
\usepackage{todonotes}
\usepackage{graphicx}
\usepackage{amssymb}
\usepackage{color}
\usepackage{ulem}
\usepackage{wasysym}
\usepackage{url}
\usepackage[colorlinks, linkcolor=blue, pdfborder={0 0 0}, breaklinks=true]{hyperref}
\definecolor{lightgray}{gray}{0.9} 
\usepackage{tabularx}
\usepackage{multirow}
\usepackage{scrextend}
\usepackage{textgreek}
\usepackage{graphicx}
\usepackage[caption=false]{subfig}
\usepackage{mathtools}
\usepackage{float}
\usepackage{soul}
\usepackage{booktabs}
\usepackage{enumitem}
\usepackage{dcolumn}
\usepackage{bm}

\usepackage{lineno} 

\DeclarePairedDelimiterX{\norm}[1]{\lVert}{\rVert}{#1}
\begin{document}

\preprint{APS/123-QED}
\title{Robustness of Deep Learning Models to Precession in Gravitational-Wave Searches for Intermediate-Mass Black Hole Binaries} 

\author{Quirijn Meijer}\email[Corresponding author: ]{r.h.a.j.meijer@uu.nl}
\affiliation{Institute for Gravitational and Subatomic Physics (GRASP), Department of Physics, Utrecht University,
Princetonplein 1, 3584CC Utrecht, The Netherlands}
\affiliation{Nikhef, Science Park 105, 1098XG Amsterdam, The Netherlands}

\author{Marc van der Sluys}
\affiliation{Institute for Gravitational and Subatomic Physics (GRASP), Department of Physics, Utrecht University,
Princetonplein 1, 3584CC Utrecht, The Netherlands}
\affiliation{Nikhef, Science Park 105, 1098XG Amsterdam, The Netherlands}

\author{Sarah Caudill}
\affiliation{Department of Physics, University of Massachusetts, Dartmouth, Massachusetts 02747, USA}
\affiliation{Center for Scientific Computing and Visualization Research, University of Massachusetts, Dartmouth, Massachusetts 02747, USA}

\date{\today}

\begin{abstract}
\noindent Gravitational-wave searches for signals of intermediate-mass black hole binaries are hindered by detector glitches, as the increased masses from stellar-mass systems hinder current generation detectors from observing the inspiral phase of the binary evolution. This causes the waveforms to strongly resemble glitches, which are of similar duration within a similar frequency band. Additionally, precession of the orbital plane of a binary black hole may further warp signal waveforms. In this work three neural network-based classifiers for the task of distinguishing between signals and glitches are introduced, with each following different training regimes to study the impact of precession on the classifiers. Although all classifiers show highly accurate performance, the classifier found to perform best was trained following the principle of curriculum learning, where new examples are introduced only after the mastery of easier preceding examples. This classifier obtains an accuracy of approximately $95\%$ on a synthetic test set consisting of signals and glitches injected into coloured noise from the O3 LIGO Hanford power spectral density. The model is compared to matched filtering, the state-of-the-art in modelled gravitational-wave searches, and analysed in search of particular sensitivities to black hole binary parameters. It was found that while the classifier is affected by the total mass of a system, the prediction of a misclassification is most strongly determined by visibility through the signal-to-noise ratio. The analysis of the three classifiers demonstrates that precession is handled differently depending on the training regime, meaning the architecture is not fully robust to precession and advancements can still be made through the development of training routines.
\end{abstract}

\maketitle

\section{Introduction}\label{sec:intro}

Black holes, since the prediction of their existence as solutions of the Einstein field equations in general relativity \cite{wald2010general}, have been indirectly observed in astronomy through electromagnetic radiation or gravitational interaction with nearby objects \cite{Gillessen_2009, Ghez_1998, Grumiller:2022qhx, doi:10.1126/science.147.3656.394, 1996ASPC..102..369G, 1994RPPh...57..417G}. The first of such observations came when Cygnus~X-1 was discovered as an X-ray source in 1964, which was later recognised as being a black hole \cite{doi:10.1126/science.147.3656.394, 1975ApL....16....9S}. In 2015, the first black hole binary was observed through the messenger of gravitational waves in the event GW150914 \cite{PhysRevLett.116.061102}. In the years following this first observation, the LIGO-Virgo-KAGRA collaboration has confirmed over $90$ detections of gravitational waves over three observational runs \cite{gwtc21, PhysRevX.9.031040, PhysRevX.11.021053, theligoscientificcollaboration2021gwtc3}, with the fourth observational run currently on-going. 

Among these detections was the event GW190521 \cite{Abbott_2020}. Before 2020, a mass gap spanning the range between stellar-mass black holes and supermassive black holes \cite{Greene_2020, Abbott_2022} was thought to exist, called the range of intermediate-mass black holes. The detection of GW190521 in 2020, however, has placed a black hole resulting from a merger within the intermediate-mass range. Observing black holes within this mass range is of interest as they may serve as precursors to supermassive black holes \cite{2001ApJ...552..459H, 2003ApJ...582..559V, Gerosa_2021}, the analysis of which in turn may further the scientific understanding of galaxy formation and evolution \cite{Greene_2020}. Furthermore, intermediate-mass black holes are considered as possible ultraluminous X-ray sources \cite{Dewangan_2006, Madhusudhan_2006}

The signals of intermediate-mass black hole binaries as measured by current-generation gravitational-wave detectors will be of short duration, as most of the signal falls outside of the sensitive bands. Because of this, the signals can easily be confused with glitches, bursts of non-Gaussian noise that are not of astrophysical origin \cite{Cabero_2019}. Current searches are therefore antagonised by glitches and rely on several types of statistical tests to mitigate them \cite{Cabero_2019, ghosh2023unmasking, Chandra_2022}. Despite the systems in place the classification problem remains difficult.

In this work the problem of discriminating between gravitational-wave signals from intermediate-mass black hole binaries and glitches is approached using machine learning. The resulting models can potentially be used as additional statistics in searches. In particular, our work uses convolutional neural networks \cite{10.5555/1162264}. Convolutional neural networks have previously succesfully been applied to the study of black hole binaries, and more generally compact binaries, as well as to glitches and their separation from such signals \cite{Wei_2021, PhysRevD.101.104003, george2017deeplearningrealtimegravitational, GEORGE201864, PhysRevD.100.063015, Fan_2019, Verma_2022, PhysRevD.103.062004, Jadhav_2023, QIU2023137850, PhysRevD.109.022006, PhysRevD.104.064046, PhysRevD.107.024007, Koyama:2024zos, PhysRevD.97.101501, PhysRevD.106.023027, lopez2022simulating}. In this work three such models obtained with different training regimes are introduced, and shown to perform with high accuracies. Other studies typically assume non-eccentricity and no precession of the orbital plane to be present. Precession can distort signals through amplitude and phase modulations \cite{PhysRevD.49.6274, PhysRevD.86.104063}, which may cause even higher similarity to glitches. In this paper the robustness of the neural networks to the added effects of precession is tested for the first time. A comparison in performance is made between the neural network models and matched filtering, a signal processing algorithm that relies on the convolution of known signal models with data \cite{helstrom1960statistical}, used in current state-of-the-art detection pipelines. This comparison establishes a baseline for the possible improvements enabled by the use of machine learning for this task. The behaviour of the models is interpreted through several data analysis methods in an attempt to reveal differences between the classes of signals and glitches that other methods are not able to probe. Within the class of signals, the impact of precession on model sensitivity is tested and compared to other system variables to both interpret the models and to test their robustness to precession.

This work is organised into six sections. Sec.~\ref{sec:imbh} introduces intermediate-mass black hole binaries and the state-of-the-art for their searches. In Sec.~\ref{sec:wavenet}, the architecture of the neural networks is discussed. Sec.~\ref{sec:methodology} covers the used methodology, from the construction of the dataset to the training of the models and their evaluation. The results are collected in Sec~\ref{sec:results}, leading up to the conclusions in Sec~\ref{sec:conclusions}.

\section{Intermediate-Mass Black Hole Binary Searches}\label{sec:imbh}

Compact binary coalescence describes the coalescence of two compact objects such as black holes or neutron stars in a binary system \cite{10.1093/acprof:oso/9780198570745.001.0001}. In this work, the study of such processes is limited to binary black holes, as neutron stars add the consideration of tidal deformation, and no neutron stars exist within the intermediate-mass range. Stellar-mass black holes are abundantly represented in the current gravitational-wave transient catalogues \cite{gwtc21, PhysRevX.9.031040, PhysRevX.11.021053, theligoscientificcollaboration2021gwtc3}, and supermassive black holes too are well known, with the supermassive black hole at the center of the Messier~87 galaxy and Sagittarius~$\textup{A}^{*}$ at the center of the Milky Way having been observed by the Event Horizon Telescope \cite{https://doi.org/10.18727/0722-6691/5150, sagvii, sagviii}. These observations bound the mass gap of intermediate-mass black holes between $10^{2}$ and $10^{5}~M_{\odot}$ \cite{imbho3}. Proof of black holes with masses within this gap was found only recently with the detection of GW190521 \cite{Abbott_2020}, which also holds the record for largest recorded progenitor masses in a binary system detected through gravitational waves. During this event two black holes with masses of $85~M_{\odot}$ and $66~M_{\odot}$ merged into one black hole with a mass of $142~M_{\odot}$, meaning $9$ solar masses of the total mass were radiated away in the form of gravitational-wave emission that was subsequently registered by the LIGO and Virgo detectors. Besides establishing their existence, the observation of this event also proves that mergers are a possible formation mechanism for intermediate-mass black holes. It was previously conjectured that black holes in this range could form through mergers, as the formation through gravitational collapse is unlikely \cite{1964ApJS....9..201F, PhysRevLett.18.379, 1984ApJ...280..825B, Woosley_2007, 2021ApJ...912L..31W}. An alternative explanation is that these black holes have formed during cosmological inflation, making them primordial black holes \cite{Greene_2012, 10.1111/j.1365-2966.2008.13523.x}.

Intermediate-mass black holes are interesting as they may serve as precursors to supermassive black holes \cite{2001ApJ...552..459H, 2003ApJ...582..559V, Gerosa_2021}. A stellar-mass black hole would have to accrete mass at the Eddington limit for a billion years to reach the status of a supermassive black hole \cite{2001ApJ...552..459H}. If instead an intermediate-mass black hole were to be taken for the seed, the required time could greatly decrease. The study of intermediate-mass black holes is therefore strongly coupled to that of both stellar-mass black holes and supermassive black holes.

Inspiraling binary black holes may exhibit precession of the orbital plane \cite{PhysRevD.49.6274}, occuring if the total spin vector of the masses admits anything other than a $z$-component in the frame where the orbital angular momentum is aligned with the $z$-axis. Because of this, binary black holes that do not exhibit precession can be referred to as (anti-)aligned spin systems. A binary black hole system is then defined by $15$ parameters. The two black holes each have a mass, with an additional set of six parameters for the spins. There are then two sky location parameters, the luminosity distance, the phase of coalescence, the inclination, polarisation angle, and finally, the time of arrival \cite{PhysRevD.49.6274}. With these parameters given, traditionally, the waveform evolution of the system is modelled best by numerical methods \cite{baumgarte2010numerical, Healy_2017}. In many applications however, waveform approximants are used. Approximants are models that approximate numerical waveforms to high accuracy at much lower cost \cite{Buonanno_1999, PhysRevD.80.084043, PhysRevD.102.064001, Garcia-Quiros:2020qpx, PhysRevD.103.104056, PhysRevD.99.124051, PhysRevD.106.024020}. In this work the waveform model used is the IMRPhenomTP approximant \cite{Estell_s_2021}, a time-domain model that supports precession. Precession manifests itself in the waveform as modulations in the amplitude and phase \cite{PhysRevD.49.6274, PhysRevD.86.104063}.

Due to the increased masses when compared to stellar-mass binary black holes, the signals of intermediate-mass black hole binaries sweep through the sensitive band of current generation detectors with a shorter duration, allowing for the measurement of only a limited number of wave cycles that start from a lower frequency \cite{imbho3}. This is the result of the waveform duration being inversely proportional to the chirp mass, which increases with increases of the component masses \cite{10.1093/acprof:oso/9780198570745.001.0001}. As a consequence the signal is dominated by the merger and ringdown phases \cite{imbho3}, as the inspiral will fall below the low-frequency cutoff \cite{Aasi_2014}. The short duration causes signals to be confused for glitches \cite{Chandra_2022, ghosh2023unmasking, imbho3}, with blip glitches being a particular example \cite{Cabero_2019}. Blip glitches are defined as transient bursts of non-Gaussian noise that are not of astrophyisical origin, with a duration close to $25$~ms and frequencies concentrated between $30$ and $250$ Hz \cite{Abbott_2016}. This frequency range largely overlaps with the band for intermediate-mass black hole binaries \cite{Abbott_2016}. As an example, in the search for intermediate-mass black holes in data from the third observational run \cite{imbho3}, $200214\_224526$ was mistakenly identified as a signal candidate by the coherent WaveBurst pipeline \cite{drago2021coherent}. In truth this turned out to be a blip glitch. Current systems in place to filter out glitches from searches rely on vetoes, gating, and statistical coherence and coincidence tests between detectors \cite{Chandra_2022, ghosh2023unmasking, imbho3}. There is however reason to believe these safeguards are not sufficient, and precision is required in the analysis of intermediate-mass black hole signals. Their rarity in observation warrants care so that signals are not mistaken for glitches or vice versa, as glitches may impact intermediate-mass black hole population studies.

Aside from unmodelled searches \cite{imbho3, Abbott_2020}, the current state-of-the-art for modelled intermediate-mass black hole searches relies on matched filtering \cite{imbho3, helstrom1960statistical}, in particular on a quantity called the signal-to-noise-ratio (SNR). The SNR is computed as an integral over two frequency series $h$ and $s$ as

\begin{equation}
    \int_{0}^{\infty} \frac{h(f) s(f)^{*}}{\textup{PSD}(f)} \mathrm{d}f,
    \label{eq:snr}
\end{equation}

\noindent where $h$ is typically a pre-computed waveform, and $s$ is detector data. The asterisk denotes the complex conjugate, and the PSD is the power spectral density that characterises the sensitivity of the detector at different frequencies \cite{10.1093/acprof:oso/9780198570745.001.0001}. In modelled searches a large set of template waveforms is generated for different parameters, and the set of all such templates is called a template bank. Given a frame of detector data, the SNR can then be computed for each of the templates, and a template is registered as a trigger if the SNR exceeds a set threshold \cite{imbho3}. Gravitational-wave detection pipelines then analyse these triggers on multiple statistics in order to identify candidates and possible detections \cite{cannon2020gstlal, Usman_2016, Cabero_2019}.

Efforts have been made to improve on matched filtering in gravitational-wave searches using convolutional neural networks. The process of matched filtering has been approximated using convolutional neural networks \cite{PhysRevLett.120.141103, PhysRevD.109.043009, 2020FrPhy..1514601L}, and work has been done to apply neural networks to searches for compact binary coalescences \cite{Wei_2021, PhysRevD.101.104003, george2017deeplearningrealtimegravitational, GEORGE201864, PhysRevD.100.063015, Fan_2019, Verma_2022, PhysRevD.103.062004, Jadhav_2023, QIU2023137850}. In addition, research that includes general transient gravitational-wave signals \cite{skliris2024realtimedetectionunmodelledgravitationalwave, PhysRevD.109.022006, PhysRevD.107.024007} and glitches \cite{PhysRevD.109.022006, PhysRevD.104.064046, PhysRevD.107.024007, Koyama:2024zos, PhysRevD.97.101501, PhysRevD.106.023027, lopez2022simulating} has been done.

\section{WaveNet}\label{sec:wavenet}

Originally designed for the generation of audio speech waveforms, \texttt{WaveNet} is a convolutional neural network \cite{10.5555/1162264} architecture designed to learn sequential data characteristics \cite{WaveNet}. The architecture achieves this by using causal convolutions, which impose a canonical time direction on the data, and allow for better predictions to be made. Such concepts are widely used in speech recognition, where recurrent neural networks have been shown to be highly effective \cite{graves2013speech} by directing the flow of information between layers. The \texttt{WaveNet} architecture uses dilation in its causal convolutions to capture this effect. Dilations allow the network to have a wide receptive net reaching back through its layers at a limited cost, as information is dilated by removing neurons from the feed-forward loop. The name of the dilations is derived from the increasing number of neurons that are disregarded in every subsequent layer.

The \texttt{WaveNet} architecture has previously seen success both in the generation of triggers for intermediate-mass black holes \cite{Wei_2021} and the classification problem of distinguishing between cosmic string cusp signals and blip glitches \cite{PhysRevD.109.022006}. In the current work the architecture from \cite{PhysRevD.109.022006} is used, with only the dimensionality of the dense layer being changed, as the throughput data will be of a lower sampling rate. The architecture is therefore adapted from waveform generation to binary classification on datasets generated by current-generation gravitational-wave observatories. The neural networks are implemented in \texttt{PyTorch} \cite{PyTorch} and run on the LIGO Data Grid, specifically on an NVIDIA A30 GPU.

\section{Methodology}\label{sec:methodology}

The first objective is the training of three classifiers that are able to distinguish the signals of intermediate-mass black holes from blip glitches when both are injected into detector noise. Of the three classifiers, one was trained exclusively on an aligned-spin dataset, one was trained on a precessing-spin dataset, and the third classifier was trained on both datasets through the principles of curriculum learning \cite{curriculum} and transfer learning \cite{transfer}. Curriculum learning assumes that a neural network, like humans in education curricula, will be more receptive to the learning of new concepts if they are gradually introduced, starting from easier examples. In the current context, the aligned-spin dataset is considered the easier of the two, meaning the third classifier was trained on the aligned-spin dataset before being transferred to the precessing-spin dataset to resume training. Having three classifiers enables a survey of the architecture robustness to precession.

The second objective is then to compare and interpret the three classifiers, thus testing the robustness of the architecture to the effects of precession, comparing to the state-of-the-art of matched filtering, and dissecting the behaviour of the models.

From this section onwards, the aligned-spin examples, precessing-spin examples and glitch examples will be referred to as the $A$, $P$ and $G$-sets respectively. The resulting classifiers trained on these examples will be labeled $\mathcal{A}$ for the aligned-spin classifier, $\mathcal{P}$ for the precessing-spin classifier, and $\mathcal{C}$ for the curriculum learning classifier. This leads to six different pairings of classifiers and datasets, as the glitches are always assumed to be included, and every classifier will be interpreted as a function from the datasets into the unit interval $[0, 1]$. The pairings are shown ahead in Table \ref{tab:confusion}.

In this section this methodology will be covered, split over three subsections. First, the generation of the datasets is treated. Second, the training on these datasets is explained. Lastly, the methods for the evaluation of the results are described.

\subsection{Data}

Three types of data were generated. The first types are the waveforms of the intermediate-mass black hole binaries, consisting of signals with either aligned spins, which form the first type, or precessing spins that form the second type. The injection tables containing the parameters for the binaries were generated using the \texttt{inspinj} function from \texttt{lalapps} \cite{lalsuite}, resulting in $50,000$ parameter samples of both types. The parameter distributions were set to uniform save for the distance, which was set to be logarithmically distributed on $[200, 220]$~Mpc. This range is chosen to target a specific distribution in SNR. The specified total mass range was set to $[150, 600]~M_{\odot}$ with mass ratio within $[1, 10]$ and spin magnitudes within $[0, 0.99]$. The inclusion or exclusion of precession is regulated through a flag that forces aligned spins when set. The waveforms for the rows in the injection tables were generated in the time domain using the IMRPhenomTP approximant \cite{Estell_s_2021}.

For the third type a total of $50,000$ blip glitches were generated using the \texttt{gengli} package \cite{gengliweb, lopez2022simulating}. The \texttt{gengli} package strives to characterise the O2 blip glitch population of LIGO Hanford and Livingston, allowing the generation of unique new blip glitches through a generative adversarial network.

\begin{figure}[!]
    \includegraphics[width=1\columnwidth]{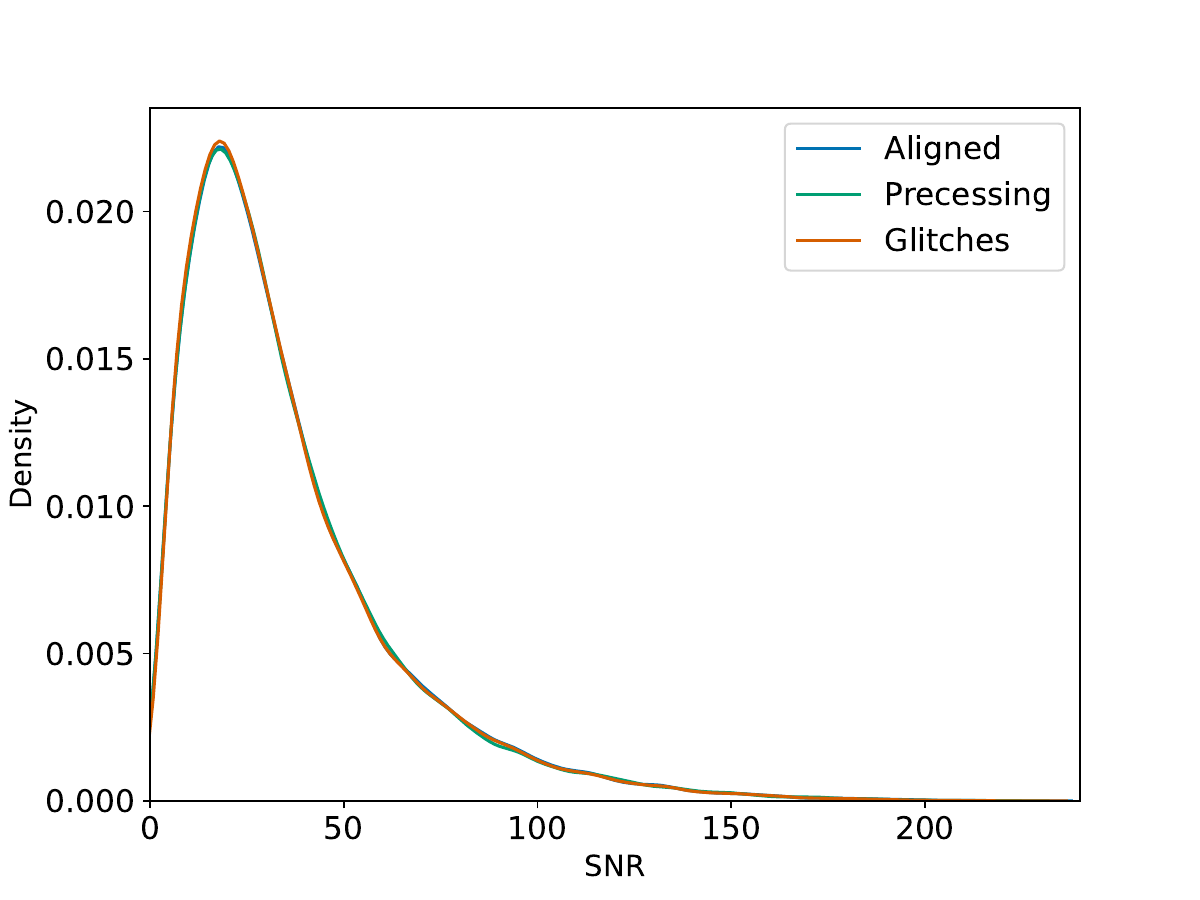}%
    \caption{The SNR distributions of the aligned spin waveforms, precessing spin waveforms and glitch waveforms. The datasets were generated in such a way that each purposefully follows the same distribution.}
    \label{fig:snr_dist}
\end{figure}

The two signal types were injected into noise using \texttt{PyCBC} \cite{pyCBC}. For every signal waveform, four seconds of Gaussian noise was generated at a sampling rate of $4096$~Hz. The positive class $A$ of strains including aligned spins was then constructed by projecting the waveforms onto the LIGO Hanford detector and injecting them into the noise at randomly drawn times, and the same was done for the precessing-spin waveforms to form the second positive class of strains $P$. Colouring was done using the realised O3 Hanford PSD \cite{O3ASD}, based on the first three months of O3 data \cite{Abbott_2020b}. At the end of the injection procedures the strains were whitened. 

Glitches were injected following a similar procedure to form the negative class $G$ of glitch strains, with the difference that the glitches output by \texttt{gengli} are output in whitened form. 

During the injection procedures, the optimum SNR was computed for all examples. The resulting distributions are shown in Fig.~\ref{fig:snr_dist}.

Additionally, \texttt{gengli} allows for the computation of similarity metrics between two glitches. These metrics are the Wasserstein distance, the match, and the cross-covariance \cite{lopez2022simulating}. In order to assign every individual glitch a value on these bivariate metrics, $400$ typical blip glitches were identified and selected as reference points. In this way the distance metrics can be reduced to a univariate function. For each glitch and each metric, the metric was computed for the glitch and every one of the references, and the average was then taken. The result is one average value per glitch and metric. Small values of these metrics imply expected glitch morphologies, with higher values corresponding to glitches that deviate from the norm.

From these strains two balanced datasets $A \cup G$ and $P \cup G$ with a total size of $100,000$ examples each were created, sharing the same glitches to allow for better comparisons, in particular in the learning phase. These datasets were split into balanced training, validation and test sets of sizes $70,000/10,000/20,000$. Effectively, every classifier is therefore called on three equally sized sets of examples: the aligned-spin examples in $A$, the precessing-spin examples in $P$, and the glitch examples in $G$.

\subsection{Training}

As in preceding work \cite{PhysRevD.109.022006}, the networks were trained using stochastic gradient descent using the \texttt{AdamW} optimiser \cite{AdamW} with a learning rate of $10^{-4}$ and weight decay of $10^{-3}$. Tuning of these parameters was shown not to improve results. The batch size was set to $30$, and each classifier is trained for $20$ epochs before interpreting the training and validation cross-entropy losses. As the classifiers are functions from the data into the unit interval due to the final softmax layer, a threshold for the classification needs to be chosen.

In the case of the curriculum classifier, the training phase is started on the aligned-spin set $A \cup G$, before transferring the classifier to continue the training phase on the precessing-spin set $P \cup G$ without overfitting visually setting in. Through this process, it can be measured how much is left to learn from the precessing-spin examples after having learned the aligned-spin examples.

\subsection{Evaluation}

In order to probe the behaviour of the resulting classifiers, three data analysis methods were employed. These methods will be described in this subsection, and were used to compare the classifiers both to matched filtering and to each other to infer how the different training recipes have affected the classifiers.

The first method is that of principal component analysis (PCA) \cite{hastie2009elements}. PCA is a linear dimensionality reduction method that linearly projects data onto a subspace spanned by an ordered basis of vectors called the principal components, where the ordering in the basis is determined by the explanatory power of the basis elements as measured by variance. As such, lower-dimensional approximations can be constructed by ommitting basis elements at the cost of the variance being reconstructed. In this work, the PCA implementation in \texttt{scikit-learn} \cite{pedregosa2011scikit} is used to investigate the sensitivity of the classifier to waveform parameters.

The second method is an alternative dimensionality reduction called t-distributed stochastic neighbourhood embedding, or t-SNE \cite{vanDerMaaten2008}. Whereas PCA is linear, t-SNE relaxes the assumption of linearity, and is therefore part of a class of algorithms called non-linear dimensionality reduction methods, or alternatively, manifold learning \cite{nldrbook, 2024AnRSA..1140522M}. In contrast to PCA that captures global relations to minimise variance, t-SNE attempts to balance local and global relations, guided through a hyperparameter called the perplexity. The perplexity can be interpreted as an estimate of the number of neighbours each datapoint should have. The algorithm then operates by converting the Euclidean distances in the high-dimensional space to probability distributions that represent pairwise similarities, and it is assumed that similar distributions exist in the lower-dimensional space of latent variables. Typically this algorithm is used for visualisation and the latter space is set to be two-dimensional. If these distributions can be similarly realised, then the dimensionality reduction must be possible and successful. Therefore, t-SNE attempts to maximise this similarity by minimising the Kullback-Leibler divergence \cite{Kullback51klDivergence} between the distributions using gradient descent. This algorithm is also implemented in \texttt{scikit-learn}, and is used in this work to search for clusters of misclassifications in the waveform parameter space.

The third and last method is that of bivariate correlation, in particular through the Pearson and Spearman correlation coefficients \cite{corrcoefficients}. The Pearson coefficient is a normalised measure of the covariance that can detect linear relations. As a covariance-based statistic it is parametric in the statistical sense, as it requires assumptions to be made on the distribution. Alternatively, the Spearman coefficient is computed in a similar manner, but using ranking statistics. It is therefore non-parametric and moreover only assumes monotonicity, meaning this coefficient can capture non-linear relations. As there is no a priori reason to assume either coefficient will perform better for this work, both are used to investigate the correlation of classifier outputs with different waveform parameters and derived quantities.

\section{Results}\label{sec:results}

Relying on the methods described in the methodology, this section presents the results, starting with the proceedings from the training phases and model selection. Once the models have been established in the first subsection, a comparison is made to matched filtering in the second subsection, and in the last two subsections the input data is related to the classifier outputs. This is done first by investigating the extreme examples in the false classes of the confusion matrix, and secondly by testing the sensitivity of the classifiers to the different waveform parameters. This relates the misclassifications to the parameters in both directions. By comparing the classifiers during these analyses the robustness against precession is evaluated.

\subsection{Model selection}

\begin{figure*}[!]
    \includegraphics[width=1\textwidth]{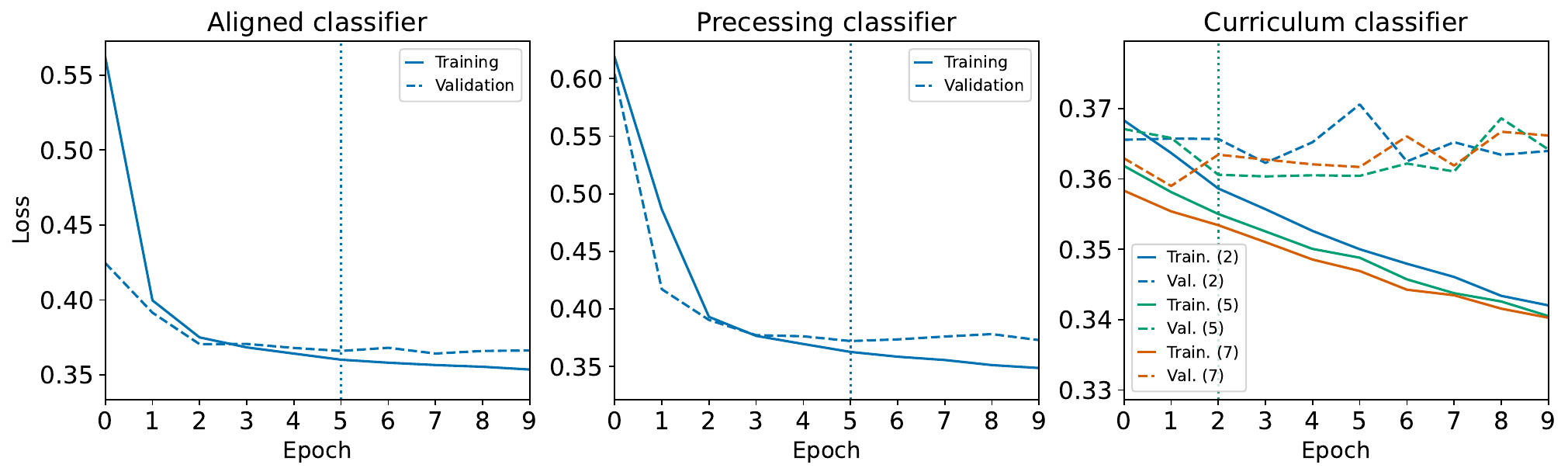}%
    \caption{The training and validation losses for the three classifiers, with the stopping times shown as a vertical dotted line. For the curriculum classifier, the losses for three different stopping times of the aligned-spin classifier (which were then used as the starting point for the training on the precessing-spin dataset) are shown. Since the classifier ultimately chosen stopped its training on the aligned-spin dataset at the fifth epoch, the stopping time is shown in green accordingly.}
    \label{fig:losses}
\end{figure*}

The training and validation losses for the three different classifiers are shown in Fig.~\ref{fig:losses}. Even though each classifier was trained for $20$ epochs, overfitting had set in before training had reached the last epoch, and therefore only the results of the first $10$ epochs are shown for each. Although both the true positive rate and false positive rate are optimised, if in the following discussion two classifiers show equal overall performance, the classifier with the lowest true positive rate will be favoured. The reason for this is that a pipeline potentially depending on this model for triggers would likely rather veto false positives than miss true positives.

In the case of the aligned-spin classifier, the classifiers corresponding to epochs $2$, $5$ and $7$ appear to be the most promising candidates based on the training and validation losses. Comparing statistics computed from the confusion matrix, it turns out that the true positive rate is identical for the three, but the false positive rate is lowest at epoch $5$. Although the absolute difference is not much, the accuracy is highest in epochs $5$ and $7$. Therefore, epoch $5$ was chosen as the stopping time.

Following identical reasoning for the precessing-spin classifier, based on the losses, the best stopping times appear to be either epoch $3$ or $5$. Comparing on the same metrics as before, although the false positive rate is comparable for both classifiers, the true positive rate and accuracy are highest for epoch $5$. Therefore here too, epoch $5$ was selected as the stopping time.

Three training phases were started for the curriculum classifier, corresponding to the stopping time candidates for the aligned-spin classifier, which were used as the starting times. The reason for starting three such phases is that it is unclear to what degree the knowledge of precession should supersede the knowledge that has already been extracted from the aligned-spin dataset, as the former can serve either to overwrite or as complement. The different phases lead to three candidates, identified by ordered pairs of the stopping time for the aligned-spin classifier (or starting time for the curriculum classifier) and the stopping time of the curriculum classifier: $[2, 3]$, $[5, 2]$ and $[7, 1]$. For all three, the corresponding classifiers obtain approximately equal local optima in their respective accuracies. The true positive rates and false positive rates are correlated and of comparable values on an absolute scale, meaning a variation in epoch that leaves the local optimum in accuracy is not rewarded with a better trade-off between these rates. Based on the classification problem at hand, the classifier with the highest true positive rate is chosen, which is $[5, 2]$.

Tuning of the treshold has lead to no improvements, as the outputs of all classifiers are heavily concentrated on the outer edges of the unit interval. The treshold is therefore set at $0.5$, with the output being rounded to a positive classification label of $1$ if and only if the output strictly exceeds $0.5$. Per complement, the output is rounded to a negative classification label of $0$ otherwise.

\begin{table}[!]
	\begin{tabular}{|c c c c c|}
		\hline
		\quad \textbf{Set} \quad & \quad \textbf{TP} \quad & \quad \textbf{FN} \quad & \quad \textbf{TN} \quad & \quad \textbf{FP} \quad \\
		\hline
		$\mathcal{A}[A \cup G]$ & $9144$ & $856$ & $9851$ & $149$ \\
		\hline
		$\mathcal{A}[P \cup G]$ & $9097$ & $903$ & $9851$ & $149$ \\
		\hline
		$\mathcal{P}[A \cup G]$ & $9037$ & $963$ & $9783$ & $217$ \\
		\hline
		$\mathcal{P}[P \cup G]$ & $8993$ & $1007$ & $9783$ & $217$ \\
		\hline
		$\mathcal{C}[A \cup G]$ & $9332$ & $668$ & $9762$ & $238$ \\
		\hline
            $\mathcal{C}[P \cup G]$ & $9254$ & $746$ & $9763$ & $237$ \\
		\hline
	\end{tabular}
	\caption{The confusion matrix of the three classifiers over the two test sets, using the notation introduced in Sec. \ref{sec:methodology}. The table shows the true positives (TP), false negatives (FN), true negatives (TN) and false positives (FP).}
	\label{tab:confusion}
\end{table}

\begin{table}[!]
	\begin{tabular}{|c c c c|}
		\hline
		\quad \textbf{Set} \quad & \quad \textbf{Accuracy} \quad & \quad \textbf{TPR} \quad & \quad \textbf{FPR} \quad \\
		\hline
		$\mathcal{A}[A \cup G]$ & $0.9498$ & $0.9144$ & $0.0149$ \\
		\hline
		$\mathcal{A}[P \cup G]$ & $0.9474$ & $0.9097$ & $0.0149$ \\
		\hline
		$\mathcal{P}[A \cup G]$ & $0.9410$ & $0.9037$ & $0.0217$ \\
		\hline
		$\mathcal{P}[P \cup G]$ & $0.9388$ & $0.8993$ & $0.0217$ \\
		\hline
		$\mathcal{C}[A \cup G]$ & $0.9547$ & $0.9332$ & $0.0238$ \\
		\hline
            $\mathcal{C}[P \cup G]$ & $0.9509$ & $0.9254$ & $0.0237$ \\
		\hline
	\end{tabular}
	\caption{The metrics of the three classifiers over the two test sets, using the notation introduced in Sec. \ref{sec:methodology}. These metrics were computed from the confusion matrices as $(\textup{TP}+\textup{TN)}(\textup{TP}+\textup{FN}+\textup{FP}+\textup{TN})^{-1}$ for the accuracy, $\textup{TP}(\textup{TP}+\textup{FN})^{-1}$ for the true positive rate (TPR), and $\textup{FP}(\textup{FP}+\textup{TN})^{-1}$ for the false positive rate (FPR).}
	\label{tab:metrics}
\end{table}

The confusion matrices for the three classifiers on the test sets are shown in Table~\ref{tab:confusion}, and the statistics computed from these matrices are are shown in Table~\ref{tab:metrics}. It can be seen from the latter table that the true positive rate for the sets $A \cup G$ and $P \cup G$ is highest for the curriculum classifier, although the false positive rate is also highest for this classifier. Regardless, the accuracy is highest for the curriculum classifier on both the aligned-spin and precessing-spin datasets. This is also supported by the area under the curve (AUC) values of the receiver operator characteristic (ROC) curves, evaluating to $0.9486$ for the aligned-spin classifier, $0.9399$ for the precessing-spin classifier, and $0.9527$ for the curriculum classifier. The curriculum classifier can therefore be appointed as the classifier that performs best, beating the classifiers that were trained on exclusively aligned-spin or precessing-spin examples.

Between the aligned-spin and precessing-spin classifiers, the first achieves a higher accuracy on the set $P \cup G$ than the latter does. This is remarkable. Closer inspection learns that this stems from the FPR. Although the TPR are close together, the FPR for the precessing-spin classifier is higher on $P \cup G$, affecting its accuracy. This means the precessing-spin classifier is more confused by glitches, and that this is the main driver for its weaker performance in comparison.

\begin{figure}[!]
    \includegraphics[width=1\columnwidth]{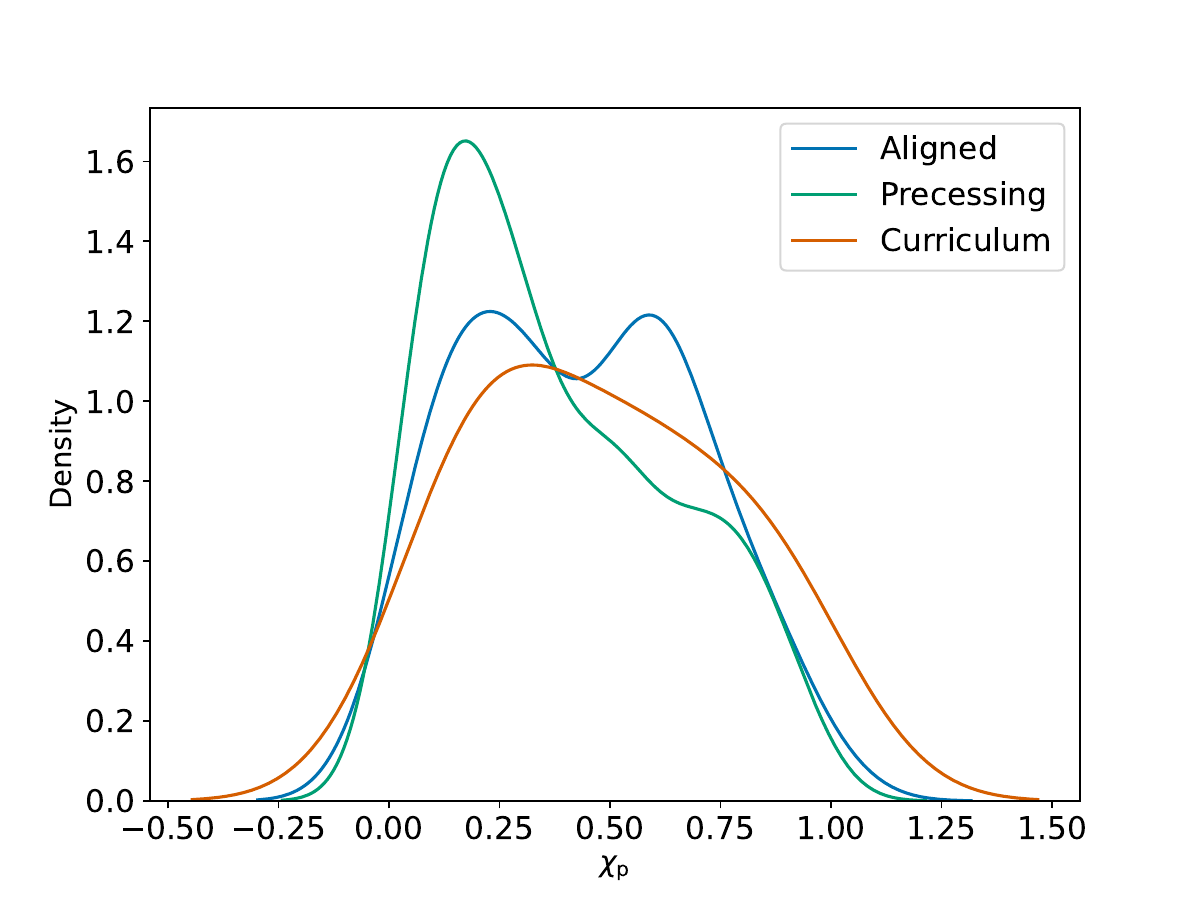}%
    \caption{The distribution of the false negatives unique to each of the three classifiers over the values of the effective precessing spin parameter $\chi_{\textup{p}}$. This plot shows that the false negatives are affected by precession in ways unique to each of the classifiers.}
    \label{fig:fn_chip}
\end{figure}

Returning to the curriculum classifier, it is difficult to discern what it is exactly that the curriculum classifier has learned that makes it outperform the other classifiers on the positive examples, in particular why it performs better on the aligned-spin dataset than the aligned-spin classifier does. However, these classifiers share many false negatives. This observation will be revisited in the later analysis in this section. When taking the false negatives unique to each classifier and considering their distribution over the effective precessing spin parameter $\chi_{\textup{p}}$ \cite{Schmidt_2015} computed for the injected waveforms, a clear difference can be seen, as shown in Fig. \ref{fig:fn_chip}. For the precessing-spin classifier, the false negatives are concentrated at a lower value of $\chi_{\textup{p}}$, indicating that this classifier might have accurately learned the effects of high precession to the waveforms. For the aligned-spin classifier, the distribution overtakes that of the precessing-spin classifier, meaning this classifier has comparatively more difficulty at higher values of $\chi_{\textup{p}}$. This is a natural consequence of precessing-spin data having been withheld from the aligned-spin classifier. The distribution for the curriculum classifier looks more evenly distributed, which could suggest that this classifier has succesfully absorbed the information it was provided by the curriculum.

\subsection{Comparison to matched filtering}

In order to compare the classifiers to matched filtering, the performance of both methods was evaluated on the test set $A \cup P \cup G$. It should be noted that this comparison is only informative for this comparison, and does not allow conclusions to be drawn for matched filtering based searches in general. In a pipeline there would be mitigation of false alarms, for instance through coincidence in the network of detectors. 

For both the aligned-spin and precessing-spin examples, the choice was made to use the injected signal as the template. This means the previously computed optimum SNR can be used, and also means there is no possible disadvantage for matched filtering in this process. For the glitch examples no theoretical optimum filter exists, as these examples do not contain a signal in the first place. Instead, a small bank of $20$ templates was constructed, consisting of $10$ templates uniformly sampled from the aligned-spin waveforms, and $10$ templates uniformly sampled from the precessing-spin waveforms. This bank was then ran against the glitches in the test set. The use of a larger template bank would be unfair to matched filtering, as the template bank puts a lower bound on the SNR that matched filtering will find. The inclusion of other templates could therefore only increase the SNR, leading to more false positives, as any trigger produced is per definition erroneous.

\begin{figure}[!]
    \includegraphics[width=1\columnwidth]{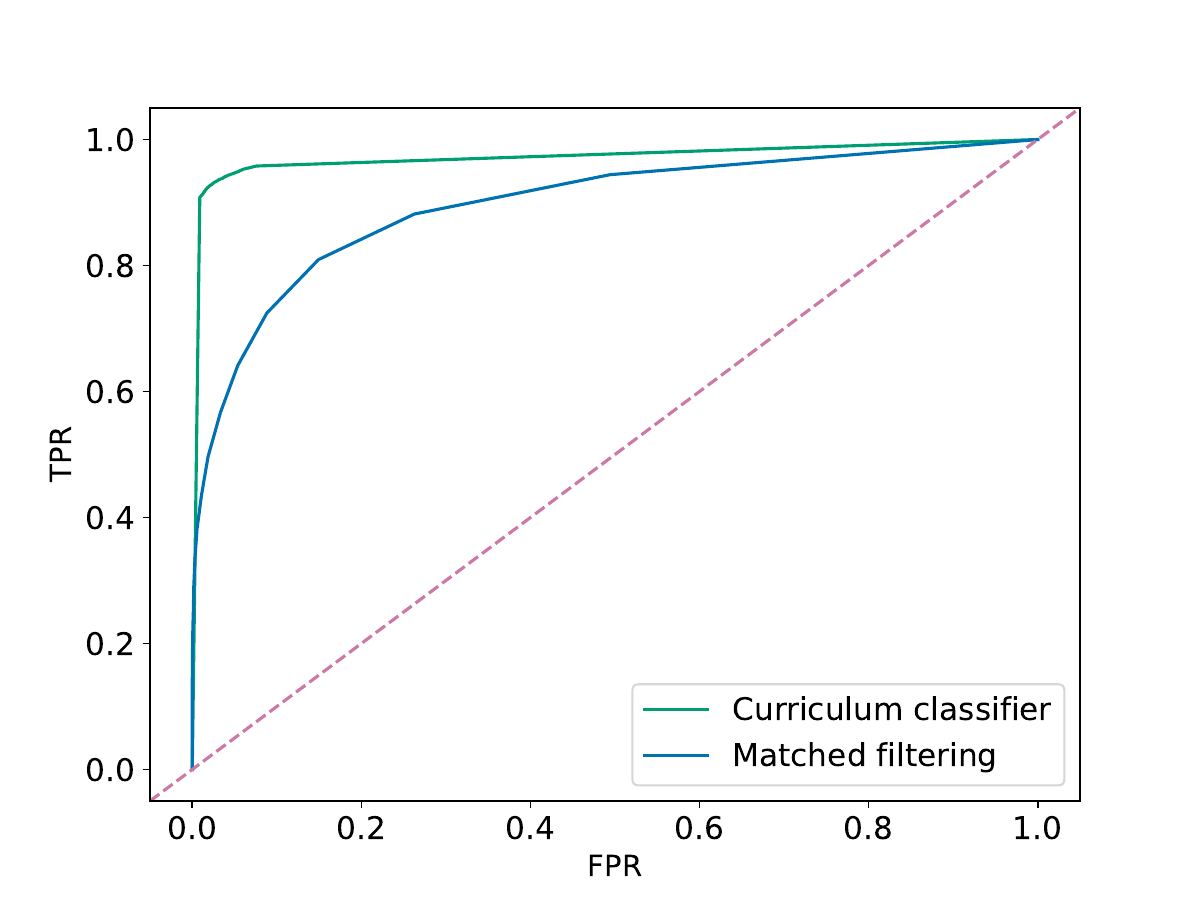}%
    \caption{The Receiver Operating Characteristic curves of the curriculum classifier and matched filtering evaluated on the held-out test set. The diagonal represents a random classifier.}
    \label{fig:mf_curriculum_roc}
\end{figure}

The ROC-curves for both the curriculum classifier and matched filtering obtained on the test set are shown in Fig.~\ref{fig:mf_curriculum_roc}. This figure shows that the performance of matched filtering is strictly dominated by that of the curriculum classifier. In line with the statistics from the previous subsection, this classifier performs extremely well on the test set.

For the aligned-spin and precessing-spin classifiers, the ROC-curves are highly similar, and are therefore not shown. This is supported by the similarity in AUC-values given for the three classifiers in the previous subsection. In terms of robustness, little can be deduced from a comparison to matched filtering. Due to the very similar performance, and the objections to an increase in the size of the template bank used, no statistically sound conclusions can be drawn for the signals.

For the glitches an interesting observation is revealed. Computing the $90$-th percentile of the SNR values within the glitches gives an SNR threshold of $\rho^{*} \approx 16.43$, which can be used to set a bound for the most confident false positives of matched filtering. Comparing the average value of the glitch metrics within these false positives to the average values over all glitches, the cross-covariance and match do not differ significantly. The value of the Wasserstein metric however, which on all glitches ranges between $1$ and $120$ with an average of $17.66$, is on average $48.76$ for the false positives of matched filtering with $\rho^{*}$ as the threshold. As large values of the Wasserstein metric imply anomalous glitches \cite{PhysRevD.106.023027}, matched filtering seems particularly sensitive to glitches deemed anomalous on this metric. This is supported by the lack in difference of the average match scores, as the match is computed based on the same inner product (Eq.~\ref{eq:snr}) that is used by matched filtering in the computation of the SNR.

The comparisons in this subsection show that on the test set, matched filtering is strongly outperformed by the three classifiers, and that this can possibly be predicted by the values on the Wasserstein metric.

\subsection{Extreme examples}

In this subsection extreme examples of misclassifications are considered in order to deepen the understanding of the classifier behaviours through these pathologies. A pathological example is defined for this purpose as either a signal or a glitch example that contains a high SNR injection that should be clear to the classifier but is not, as evidenced by a classifier output far-removed from the true label of the strain. An example would be a signal with high SNR for which the strain with true label $1$ is assigned a classifier output near $0$. This definition will be used in this subsection. The pathological examples should in theory capture the morphologies that are most confusing to the classifiers. Studying these examples should therefore reveal information on parameters and statistics that lead to extreme misclassifications made by the models, and can be used to direct further investigation of specific parameters, as will be done in the next subsection on parameter sensitivity. It is also a vehicle for discovering differences between the three classifiers, allowing to survey the robustness.

A number of interesting pathological examples surface. The aligned-spin classifier and curriculum classifier share one false negative in the aligned-spin set that has a particularly low classifier output, even when compared to the other extreme false negatives. The precessing-spin classifier, however, assigns this same example an output near $1$. When comparing the injection values for this example to the average values in the false negatives, it becomes apparent that the masses and inclinations are much higher at a total mass of approximately $592 M_{\odot}$, and inclination of $2.99$ respectively. This places this particular example at the outer edges of the sample ranges, likely causing the classification differences. 

Furthermore, there is one false negative in the aligned-spin set that is shared by all three classifiers. This example has a very high primary mass of $497 M_{\odot}$ but a relatively low secondary mass of $52 M_{\odot}$. This leads to a high mass ratio and chirp mass. The signal is therefore of short duration, which is a probable cause for the confusion of the three classifiers. 

In general the aligned-spin classifier and curriculum classifier share many examples. Further investigating, most false negatives show a high primary mass, but the secondary mass is mostly higher for the aligned-spin and curriculum classifiers than they are for the precessing-spin classifier. This marks the mass ratio as a property of interest for general misclassifications, as well as for tests of robustness. Cross-comparing, both the aligned-spin classifier and curriculum classifier did fairly well on the false negatives of the precessing-spin classifier. This leads to the observation that the former two classifiers have learned signatures that likely save them from making the same mistakes as the precessing-spin classifier. The curriculum classifier performed well on the false negatives of the aligned-spin classifier as well, suggesting that within the confines of these extreme examples, this classifier has eliminated at least some of the difficulties that plague both the aligned-spin and precessing-spin classifiers.

The inclination is also a noteworthy injection parameter. On the pathological false negatives of the precessing-spin dataset the inclination value is on average lower for the aligned-spin classifier ($1.21$) and curriculum classifier ($0.9$) than it is for the precessing-spin classifier ($1.55$), meaning more of the pathological false negatives of the precessing-spin classifier are closer to being edge-on. This difference is not present to the same extent in the false negatives of the aligned-spin dataset. One possible explanation could be that the inclusion of precession has made the classifier more sensitive to modulation, considering the inclination as a constant modulation.

\begin{figure}[!]
    \includegraphics[width=1\columnwidth]{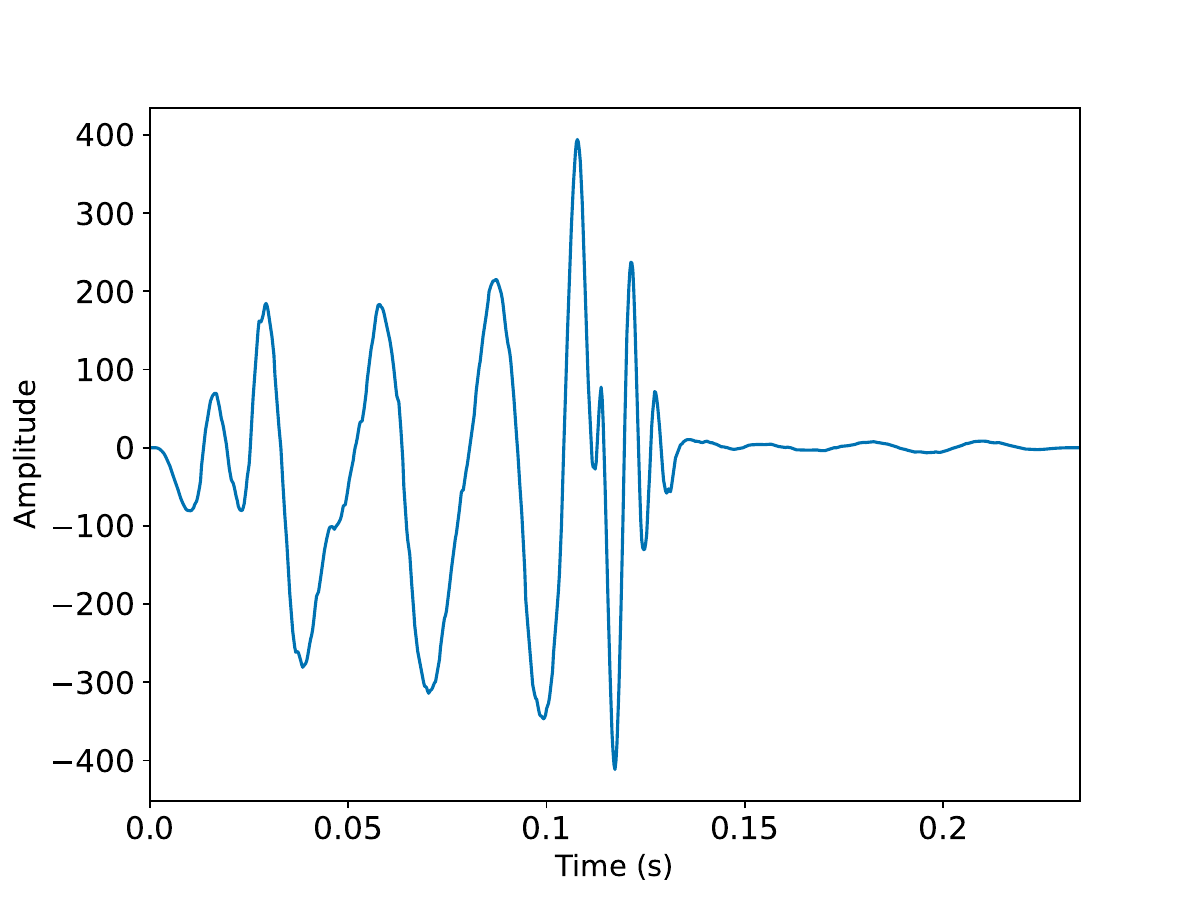}%
    \caption{An example of an extreme false positive that is shared by the three classifiers. This glitch waveform resembles the waveform of an intermediate-mass black hole binary, where the inspiral phases are difficult to detect.}
    \label{fig:glitch_fp}
\end{figure}

In the case of glitches, the false positives have a large intersection, with only the precessing spin classifier making unique mistakes. There is however nothing in terms of the glitch similarity metrics that could distinguish these mistakes. Although the classifiers performed very differently on the signals, this does not seem to be the case for the glitches. As the glitches are shared between the datasets, the classifiers have likely learned similar features for these glitches. All of the found false positives strongly resemble signals, as is shown for one particular glitch example in Fig.~\ref{fig:glitch_fp}.

The results from this investigation of the most extreme examples show that for the signals, the mass ratio and inclination are parameters of interest, as these parameters are higher on average for the false negatives. It is also retrieved that the duration plays a role. This mirrors the duration being a factor in matched filtering searches. In the case of the neural networks, it is possible the short duration signals do not exhibit enough of the features the neural networks have learned to make a positive classification. The misclassified glitches largely intersect, and the three classifiers show near-identical performance on the extreme examples. This likely means that even though the three classifiers do show different behaviours on the signals, the different training regimes has not effected the features learned to identify glitches. Based on the fact that the curriculum classifier performed well on the false negatives of both the aligned-spin and precessing-spin classifiers, one can argue that the curriculum classifier has successfully absorbed information that the aligned-spin classifier did not learn. This can be seen as support for the architecture not being fully robust to precession.

\subsection{Parameter sensitivity}\label{subsec:paramsens}

The sensitivity of the classifiers to injection parameters can be investigated by relating the parameters to the classifier outputs. In this subsection this analysis is done first for the curriculum classifier, as it has the highest discriminating power, seemingly capturing the difference between the injected signals and glitches best. The obtained information can then be used for a statistical comparison for parameter sensitivity between the three classifiers as a test of robustness.

\begin{figure}[!]
    \includegraphics[width=1\columnwidth]{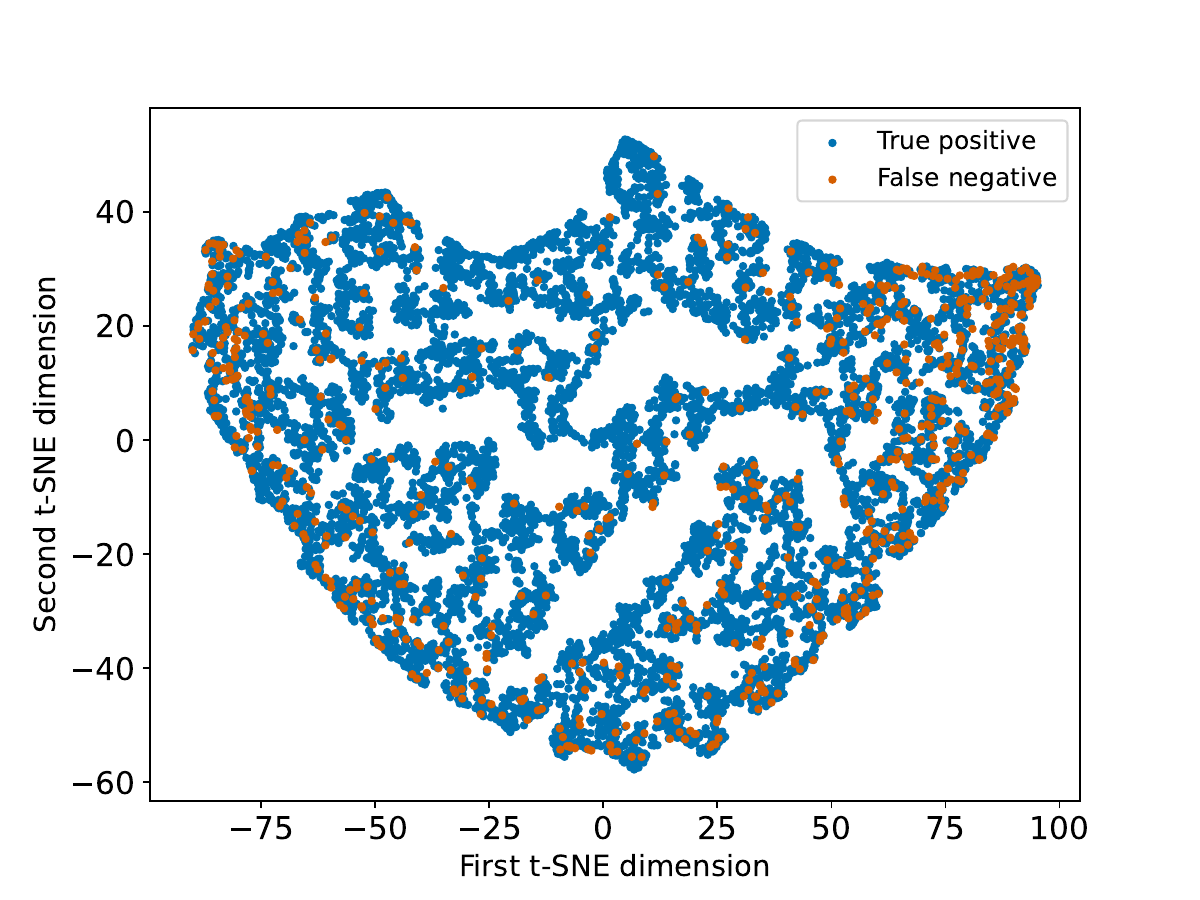}%
    \caption{The true positives and false negatives of the curriculum classifier projected onto a plane using t-SNE with a perplexity value of $50$. Different hyperparameters did not result in noticeable differences in the separation of the classes.}
    \label{fig:tsne}
\end{figure}

First, it is tested if separated clusters can be found in the class of signal examples using PCA and t-SNE. The existence of such clusters would give indication of sensitive regions for either the true positives or false negatives within the parameter space. Since every signal example that is assigned an output by the classifier was generated from a row in the injection table, the binary label prediction can be directly connected to the waveform parameters. The classifications can therefore be treated as data points residing in the parameter space, with each corresponding to either a true positive or false negative. This dataset was input to the t-SNE algorithm which visually maps the data down to a $2$-dimensional space, as shown in Fig.~\ref{fig:tsne}. As can be seen from the figure, the output reveals no separated clusters of either class. 

\begin{figure}[!]
    \includegraphics[width=1\columnwidth]{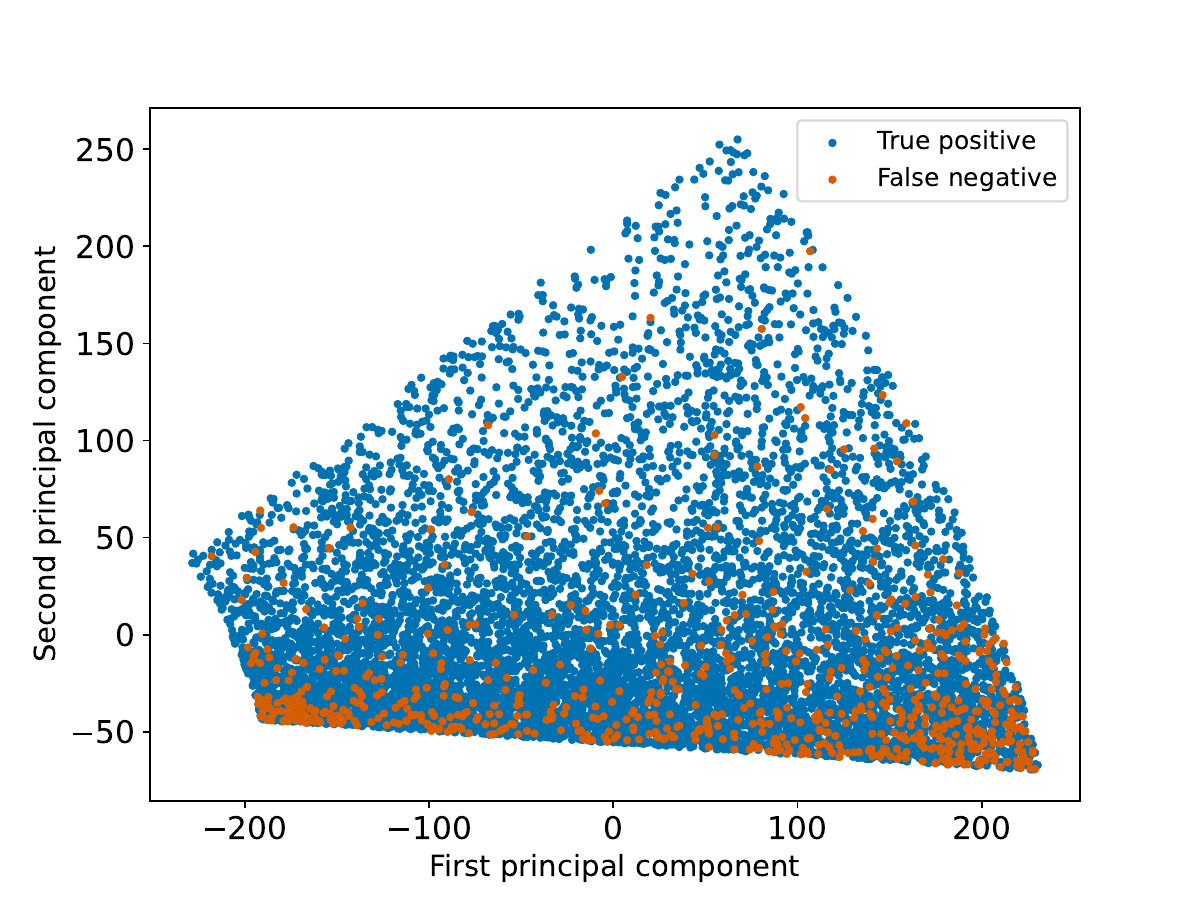}%
    \caption{The true positives and false negatives of the curriculum classifier projected onto a plane using PCA. As for t-SNE, there is no separation between the classes.}
    \label{fig:pca}
\end{figure}

As an additional test, the same data was input to PCA, which captured nearly the full variance ($99.6\%$) with two principal components. Here too however, the classes show no separation, even though the density of the false negatives does seem to correlate with the second principal component. 

The results from both t-SNE and PCA suggest that no separated clusters of classifications exist, or that they can not be captured by these two algorithms. This reduces the current study of sensitivity to outlier analysis using individual statistics, rather than studying specific regions of the parameter space.

\begin{figure}[!]
    \includegraphics[width=1\columnwidth]{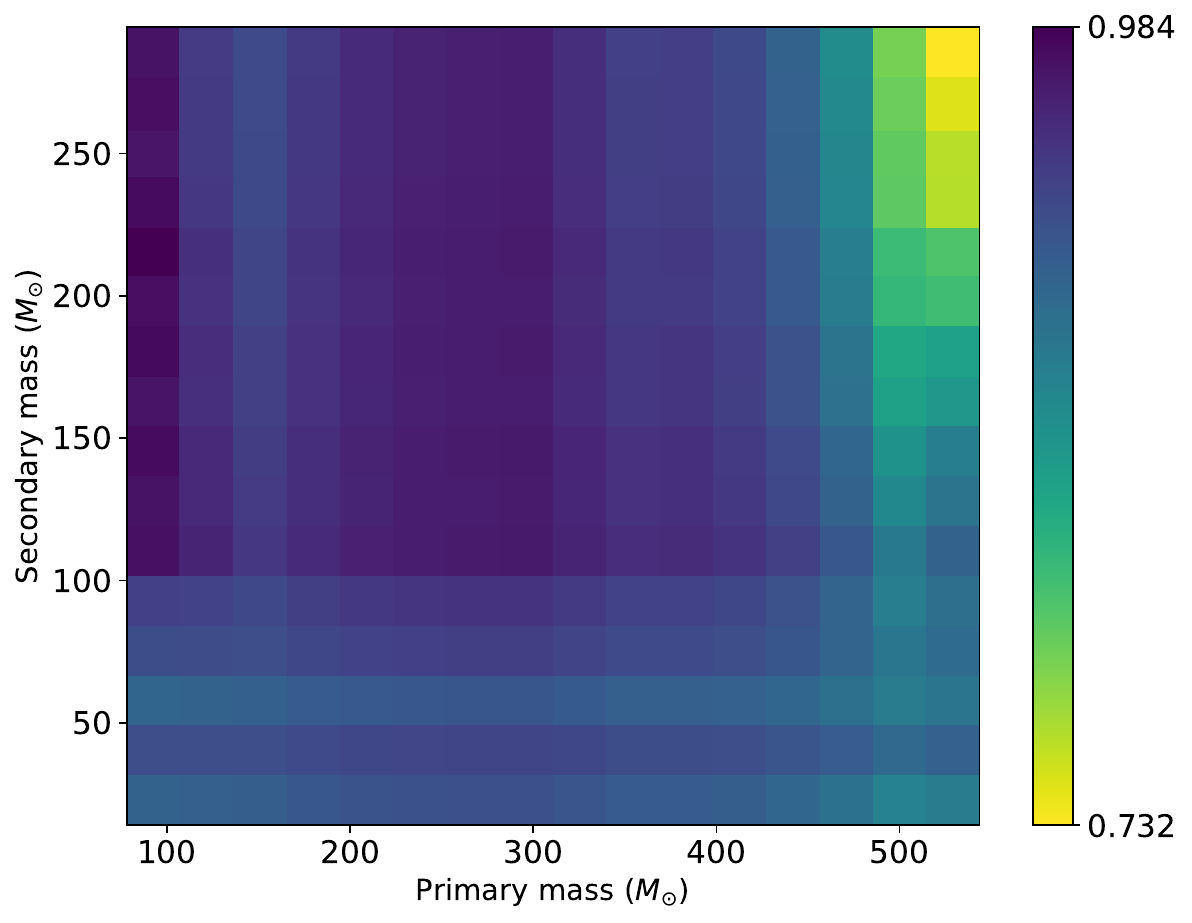}%
    \caption{A heatmap of the outputs from the curriculum classifier on the grid of the primary and secondary masses. This heatmap reveals a sensitivity to higher total masses in the top right corner.}
    \label{fig:c_m1_m2}
\end{figure}

One way to search for outliers as well as to detect possible relations between parameters is to plot heatmaps, in which bins are created for the combinations of the values on the axes, and the average of the classifier outputs within these bins is shown. Creating heatmaps involving the inclination, the first parameter of interest found in the previous subsection, revealed no new patterns. For the primary and secondary masses, a heatmap is shown in Fig.~\ref{fig:c_m1_m2}. Although this figure shows that the curriculum classifier is on average correct and fairly confident (with a minimum cell average of approximately $0.73$), uncertainty sets in in the highest region of total mass, or as the primary mass increases. This reflects the correlation of matched filtering sensitivty with the primary mass shown in \cite{Davis_2020}, although the limitations of the curriculum classifier set in for total masses that are considerably higher. A direct comparison is made difficult by the difference in analysis, as \cite{Davis_2020} studies the probability of matched-filtering triggers being produced in these ranges.

\begin{figure}[!]
    \includegraphics[width=1\columnwidth]{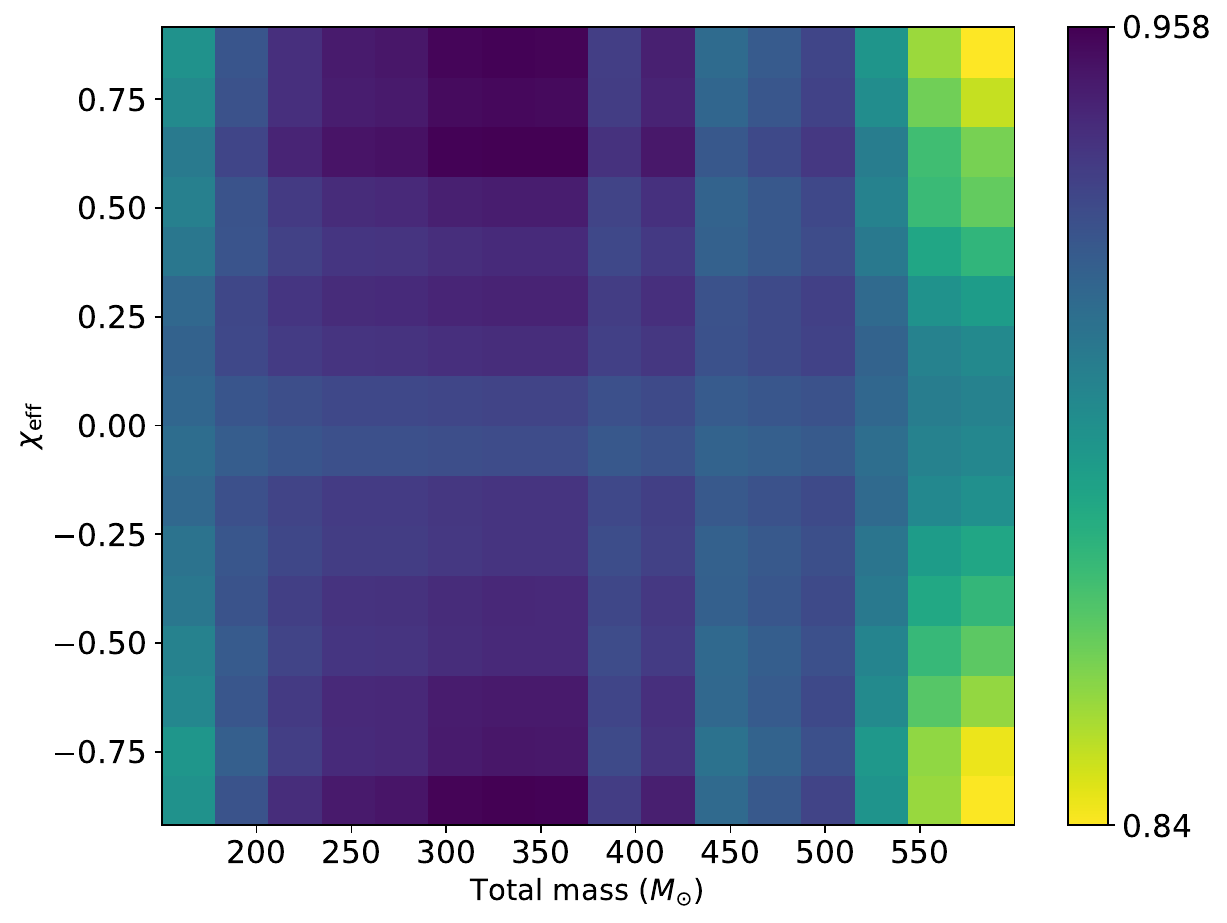}%
    \caption{A heatmap of the outputs from the curriculum classifier on the grid of the total mass and $\chi_{\textup{eff}}$. It can be seen that there is a natural symmetry on the effective spin, and the classifier is proportionally sensitive to higher absolute values of this parameter.}
    \label{fig:c_total_chieff}
\end{figure}

Resuming the analysis on the total mass with the addition of the effective spin $\chi_{\textup{eff}}$, a similar heatmap is shown in Fig.~\ref{fig:c_total_chieff}. As expected, this heatmap reveals regions of higher uncertainty at high total mass, but also at the extreme ends of $\chi_{\textup{eff}}$. This is again in line with the findings of \cite{Davis_2020}, and together these observations show that to a certain extent the curriculum classifier qualitatively behaves similar to how matched filtering was found to behave in earlier work, but with high confidence even at extreme parameter values. It is possible that convolutional neural networks to some degree learn the same filters as are used in matched filtering, as for some applications convolutional neural networks tend to repeatedly learn similar filters \cite{10.5555/2969033.2969197, pmlr-v44-li15convergent}. This might suggest an interesting connection between the convolutional filters used in matched filtering, and the filters learned by the convolutional neural networks. Further investigation is deferred to future work. Due to the horizontal symmetry in the heatmap, no further conclusions for $\chi_{\textup{eff}}$ can be drawn from this plot.

\begin{figure}[!]
    \includegraphics[width=1\columnwidth]{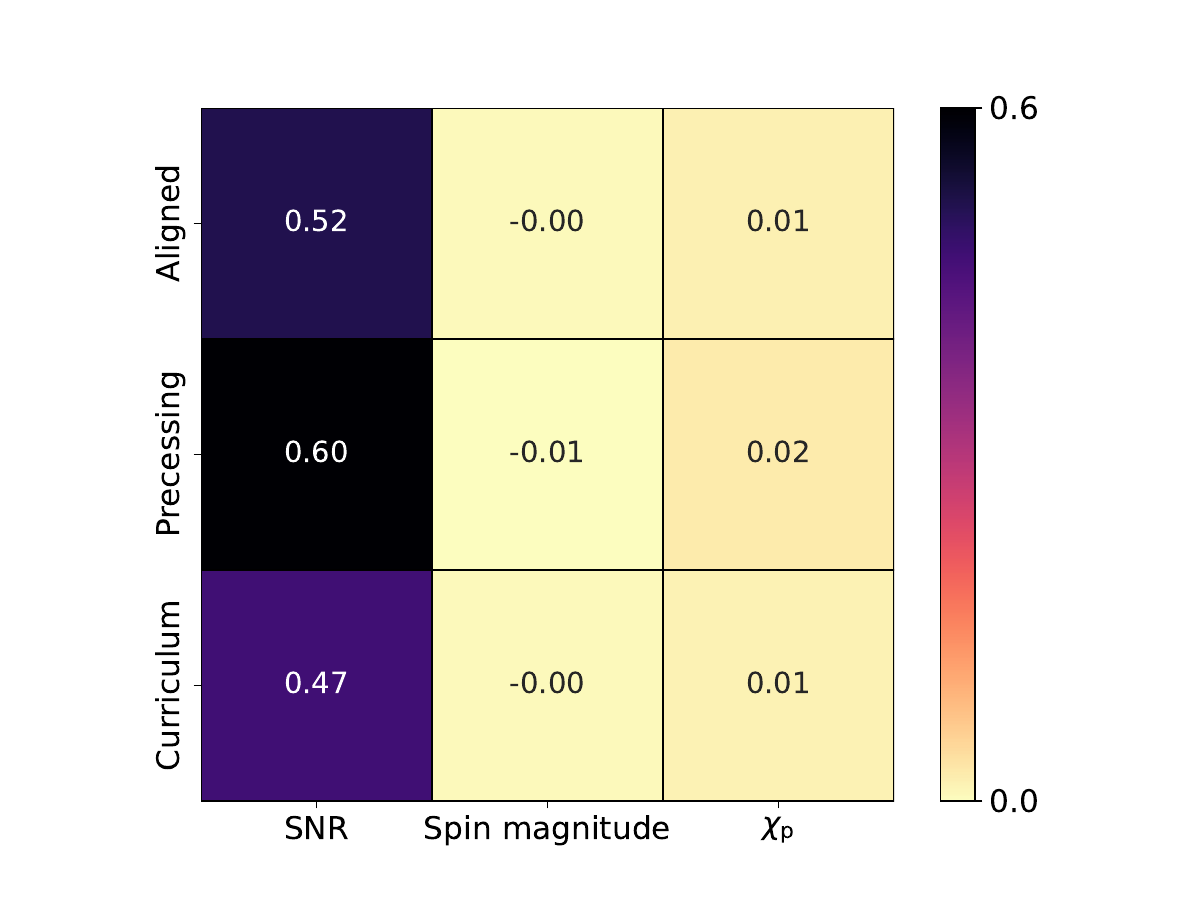}%
    \caption{The values for the Spearman bivariate correlation coefficient with the classifiers on the vertical axis and quantities on the horizontal axis. This matrix shows that of the variables with the strongest correlation to the classifier outputs, the SNR is by far the best predictor.}
    \label{fig:correlation}
\end{figure}

In order to quantify by how much the outputs of the classifiers are influenced by the injection parameters and quantities derived thereof, bivariate correlation coefficients were computed for all three classifiers. In all cases the Spearman correlation coefficient yielded stronger evidence than the Pearson coefficient did, but with similar trends. The strongest correlations are shown in Fig.~\ref{fig:correlation}, clearly showing a moderate correlation of the classifier outputs with the SNR. This shows that the classifiers are mostly hindered by the visibility of the signals within the noise, as the correlation coefficients are only computed for signals, and therefore should never be assigned $0$ by a perfect classifier. It is worthy of note that the correlation of $\chi_{\textup{p}}$ with the output is higher for the precessing-spin classifier than it is for the aligned-spin and curriculum classifiers. This means the precessing-spin classifier is more likely to assign a higher output to examples that show more precession.

From the analysis in this subsection it appears that there is no separate region of the parameter space to which the classifiers are particularly sensitive, and that the binary black hole properties that trouble the classifiers most are akin to those hindering matched filtering searches. The values of the correlation coefficients suggest that beyond a certain point, obtaining correct classification simply becomes a problem of visibility, meaning there is not one specific parameter or statistic that can be appointed as a predictor for misclassification. In terms of robustness to precession, although not strongly correlating, the precessing-spin classifier does respond better to examples with a higher value of $\chi_{\textup{p}}$ than the other classifiers do. This suggests that there are properties of precessing examples that the aligned-spin classifier and curriculum classifier have not fully learned.

\section{Conclusions}\label{sec:conclusions}

Three different neural network classifiers of the same architecture were introduced, each with a different training regime that involves different levels of exposure to precession. It was shown that these classifiers perform exceptionally well for the task of telling signals from glitches, with the curriculum classifier in particular obtaining an accuracy as high as $95\%$ on the held-out test sets. It was also shown that in a simulation on a test set of signals and glitches, all three classifiers outperform the matched filtering algorithm for the task of separating signals from glitches. Matched filtering was in particular hindered by a class of glitches with high values on the Wasserstein metric. The sensitivities of the classifiers to waveform parameters and statistics were investigated, finding that there are no particular combinations of these values that can accurately predict misclassification, although the three classifiers will become less confident as system total mass increases. 

Through comparison of the three classifiers, the robustness of the neural network architecture to the effects of precession was tested. Although all three classifiers perform extremely well, it appears that the three classifiers handle precession differently and in accordance with expectations based on their provided training sets. This is measured through the effective precessing spin statistic, showing that the architecture is not fully robust to precession in the sense that the insertion of precession into the training phase will alter classifier behaviour. This is in part mitigated by the use of curriculum learning, which was shown to be able to improve classifier performance.

Future work could see the inclusion of additional waveform effects such as eccentricity or higher-order modes \cite{10.1093/acprof:oso/9780198570745.001.0001}. Alternatively, the architecture can be modified further. The early overfitting of the classifiers suggests that the architecture could benefit from the addition of max pooling layers to reduce the number of model parameters. There are other opportunities that can be found within the training regimes. One example can be the extension of curriculum learning to a curriculum that is strongly segmented over a larger number of phases, with examples of increasing difficulty being fed as the training progresses. This difficulty can be measured by for instance the effective precessing spin parameter, as was done in this work using a hard cut-off at zero.

This work has demonstrated that neural networks are efficient candidate models in the search for intermediate-mass black hole binaries in the presence of glitches. A baseline for the performance on this task has been established, while at the same time showing that neural networks may serve as a new medium to observe the sometimes minute differences between different classes of gravitational-wave signals and glitches. This exploration is important for future analyses, as such tasks have proven difficult for traditional methods. As these differences become better understood, the quality and quantity of gravitational-wave analyses should benefit directly, enabling a deeper understanding of the cosmos and the many objects residing in it.

\section*{Acknowledgements}

With thanks to Stefano Schmidt, Melissa Lopez, Bhooshan Gadre, Tom Dooney, Vasileios Skliris and Hugo Einsle. Q.M and M.vd.S are supported by the research program of the Netherlands Organisation for Scientific Research (NWO). M.vd.S acknowledges support from the European Union, AHEAD 2020 (grant number 871158) and of the project Cortex with project number 00686766 of the research programme NWA which is (partly) financed by the Dutch Research Council (NWO). S.C is supported by the National Science Foundation under Grant No. PHY-2309332. The authors are grateful for computational resources provided by the LIGO Laboratory and supported by the National Science Foundation Grants No. PHY-0757058 and No. PHY-0823459. This material is based upon work supported by NSF's LIGO Laboratory which is a major facility fully funded by the National Science Foundation.

\bibliography{references}

\begin{thebibliography}{98}%
\makeatletter
\providecommand \@ifxundefined [1]{%
 \@ifx{#1\undefined}
}%
\providecommand \@ifnum [1]{%
 \ifnum #1\expandafter \@firstoftwo
 \else \expandafter \@secondoftwo
 \fi
}%
\providecommand \@ifx [1]{%
 \ifx #1\expandafter \@firstoftwo
 \else \expandafter \@secondoftwo
 \fi
}%
\providecommand \natexlab [1]{#1}%
\providecommand \enquote  [1]{``#1''}%
\providecommand \bibnamefont  [1]{#1}%
\providecommand \bibfnamefont [1]{#1}%
\providecommand \citenamefont [1]{#1}%
\providecommand \href@noop [0]{\@secondoftwo}%
\providecommand \href [0]{\begingroup \@sanitize@url \@href}%
\providecommand \@href[1]{\@@startlink{#1}\@@href}%
\providecommand \@@href[1]{\endgroup#1\@@endlink}%
\providecommand \@sanitize@url [0]{\catcode `\\12\catcode `\$12\catcode `\&12\catcode `\#12\catcode `\^12\catcode `\_12\catcode `\%12\relax}%
\providecommand \@@startlink[1]{}%
\providecommand \@@endlink[0]{}%
\providecommand \url  [0]{\begingroup\@sanitize@url \@url }%
\providecommand \@url [1]{\endgroup\@href {#1}{\urlprefix }}%
\providecommand \urlprefix  [0]{URL }%
\providecommand \Eprint [0]{\href }%
\providecommand \doibase [0]{http://dx.doi.org/}%
\providecommand \selectlanguage [0]{\@gobble}%
\providecommand \bibinfo  [0]{\@secondoftwo}%
\providecommand \bibfield  [0]{\@secondoftwo}%
\providecommand \translation [1]{[#1]}%
\providecommand \BibitemOpen [0]{}%
\providecommand \bibitemStop [0]{}%
\providecommand \bibitemNoStop [0]{.\EOS\space}%
\providecommand \EOS [0]{\spacefactor3000\relax}%
\providecommand \BibitemShut  [1]{\csname bibitem#1\endcsname}%
\let\auto@bib@innerbib\@empty
\bibitem [{\citenamefont {Wald}(2010)}]{wald2010general}%
  \BibitemOpen
  \bibfield  {author} {\bibinfo {author} {\bibfnamefont {R.M.}\ \bibnamefont {Wald}},\ }\href {https://books.google.nl/books?id=9S-hzg6-moYC} {\emph {\bibinfo {title} {General Relativity}}}\ (\bibinfo  {publisher} {University of Chicago Press},\ \bibinfo {year} {2010})\BibitemShut {NoStop}%
\bibitem [{\citenamefont {Gillessen}\ \emph {et~al.}(2009)\citenamefont {Gillessen}, \citenamefont {Eisenhauer}, \citenamefont {Trippe}, \citenamefont {Alexander}, \citenamefont {Genzel}, \citenamefont {Martins},\ and\ \citenamefont {Ott}}]{Gillessen_2009}%
  \BibitemOpen
  \bibfield  {author} {\bibinfo {author} {\bibfnamefont {S.}~\bibnamefont {Gillessen}}, \bibinfo {author} {\bibfnamefont {F.}~\bibnamefont {Eisenhauer}}, \bibinfo {author} {\bibfnamefont {S.}~\bibnamefont {Trippe}}, \bibinfo {author} {\bibfnamefont {T.}~\bibnamefont {Alexander}}, \bibinfo {author} {\bibfnamefont {R.}~\bibnamefont {Genzel}}, \bibinfo {author} {\bibfnamefont {F.}~\bibnamefont {Martins}}, \ and\ \bibinfo {author} {\bibfnamefont {T.}~\bibnamefont {Ott}},\ }\bibfield  {title} {\enquote {\bibinfo {title} {Monitoring stellar orbits around the massive black hole in the galactic center},}\ }\href {\doibase 10.1088/0004-637X/692/2/1075} {\bibfield  {journal} {\bibinfo  {journal} {The Astrophysical Journal}\ }\textbf {\bibinfo {volume} {692}},\ \bibinfo {pages} {1075} (\bibinfo {year} {2009})}\BibitemShut {NoStop}%
\bibitem [{\citenamefont {Ghez}\ \emph {et~al.}(1998)\citenamefont {Ghez}, \citenamefont {Klein}, \citenamefont {Morris},\ and\ \citenamefont {Becklin}}]{Ghez_1998}%
  \BibitemOpen
  \bibfield  {author} {\bibinfo {author} {\bibfnamefont {A.~M.}\ \bibnamefont {Ghez}}, \bibinfo {author} {\bibfnamefont {B.~L.}\ \bibnamefont {Klein}}, \bibinfo {author} {\bibfnamefont {M.}~\bibnamefont {Morris}}, \ and\ \bibinfo {author} {\bibfnamefont {E.~E.}\ \bibnamefont {Becklin}},\ }\bibfield  {title} {\enquote {\bibinfo {title} {High proper-motion stars in the vicinity of sagittarius a*: Evidence for a supermassive black hole at the center of our galaxy},}\ }\href {\doibase 10.1086/306528} {\bibfield  {journal} {\bibinfo  {journal} {The Astrophysical Journal}\ }\textbf {\bibinfo {volume} {509}},\ \bibinfo {pages} {678} (\bibinfo {year} {1998})}\BibitemShut {NoStop}%
\bibitem [{\citenamefont {Grumiller}\ and\ \citenamefont {Sheikh-Jabbari}(2022)}]{Grumiller:2022qhx}%
  \BibitemOpen
  \bibfield  {author} {\bibinfo {author} {\bibfnamefont {Daniel}\ \bibnamefont {Grumiller}}\ and\ \bibinfo {author} {\bibfnamefont {Mohammad~Mehdi}\ \bibnamefont {Sheikh-Jabbari}},\ }\href {\doibase 10.1007/978-3-031-10343-8} {\emph {\bibinfo {title} {{Black Hole Physics: From Collapse to Evaporation}}}},\ Grad.Texts Math.\ (\bibinfo  {publisher} {Springer},\ \bibinfo {year} {2022})\BibitemShut {NoStop}%
\bibitem [{\citenamefont {Bowyer}\ \emph {et~al.}(1965)\citenamefont {Bowyer}, \citenamefont {Byram}, \citenamefont {Chubb},\ and\ \citenamefont {Friedman}}]{doi:10.1126/science.147.3656.394}%
  \BibitemOpen
  \bibfield  {author} {\bibinfo {author} {\bibfnamefont {S.}~\bibnamefont {Bowyer}}, \bibinfo {author} {\bibfnamefont {E.~T.}\ \bibnamefont {Byram}}, \bibinfo {author} {\bibfnamefont {T.~A.}\ \bibnamefont {Chubb}}, \ and\ \bibinfo {author} {\bibfnamefont {H.}~\bibnamefont {Friedman}},\ }\bibfield  {title} {\enquote {\bibinfo {title} {Cosmic x-ray sources},}\ }\href {\doibase 10.1126/science.147.3656.394} {\bibfield  {journal} {\bibinfo  {journal} {Science}\ }\textbf {\bibinfo {volume} {147}},\ \bibinfo {pages} {394--398} (\bibinfo {year} {1965})},\ \Eprint {http://arxiv.org/abs/https://www.science.org/doi/pdf/10.1126/science.147.3656.394} {https://www.science.org/doi/pdf/10.1126/science.147.3656.394} \BibitemShut {NoStop}%
\bibitem [{\citenamefont {{Goss}}\ and\ \citenamefont {{McGee}}(1996)}]{1996ASPC..102..369G}%
  \BibitemOpen
  \bibfield  {author} {\bibinfo {author} {\bibfnamefont {W.~M.}\ \bibnamefont {{Goss}}}\ and\ \bibinfo {author} {\bibfnamefont {R.~X.}\ \bibnamefont {{McGee}}},\ }\bibfield  {title} {\enquote {\bibinfo {title} {{The Discovery of the Radio Source Sagittarius A (Sgr A)}},}\ }in\ \href@noop {} {\emph {\bibinfo {booktitle} {The Galactic Center}}},\ \bibinfo {series} {Astronomical Society of the Pacific Conference Series}, Vol.\ \bibinfo {volume} {102},\ \bibinfo {editor} {edited by\ \bibinfo {editor} {\bibfnamefont {Roland}\ \bibnamefont {{Gredel}}}}\ (\bibinfo {year} {1996})\ p.\ \bibinfo {pages} {369}\BibitemShut {NoStop}%
\bibitem [{\citenamefont {{Genzel}}\ \emph {et~al.}(1994)\citenamefont {{Genzel}}, \citenamefont {{Hollenbach}},\ and\ \citenamefont {{Townes}}}]{1994RPPh...57..417G}%
  \BibitemOpen
  \bibfield  {author} {\bibinfo {author} {\bibfnamefont {R.}~\bibnamefont {{Genzel}}}, \bibinfo {author} {\bibfnamefont {D.}~\bibnamefont {{Hollenbach}}}, \ and\ \bibinfo {author} {\bibfnamefont {C.~H.}\ \bibnamefont {{Townes}}},\ }\bibfield  {title} {\enquote {\bibinfo {title} {{The nucleus of our Galaxy}},}\ }\href {\doibase 10.1088/0034-4885/57/5/001} {\bibfield  {journal} {\bibinfo  {journal} {Reports on Progress in Physics}\ }\textbf {\bibinfo {volume} {57}},\ \bibinfo {pages} {417--479} (\bibinfo {year} {1994})}\BibitemShut {NoStop}%
\bibitem [{\citenamefont {{Shipman}}(1975)}]{1975ApL....16....9S}%
  \BibitemOpen
  \bibfield  {author} {\bibinfo {author} {\bibfnamefont {H.L.}\ \bibnamefont {{Shipman}}},\ }\bibfield  {title} {\enquote {\bibinfo {title} {{The Implausible History of Triple Star Models for Cygnus X-1: Evidence for a Black Hole}},}\ }\href@noop {} {\bibfield  {journal} {\bibinfo  {journal} {Astrophysical Letters}\ }\textbf {\bibinfo {volume} {16}},\ \bibinfo {pages} {9} (\bibinfo {year} {1975})}\BibitemShut {NoStop}%
\bibitem [{\citenamefont {Abbott}\ \emph {et~al.}(2016)\citenamefont {Abbott}, \citenamefont {Abbott}, \citenamefont {Abbott}, \citenamefont {Abernathy}, \citenamefont {Acernese}, \citenamefont {Ackley}, \citenamefont {Adams}, \citenamefont {Adams},\ and\ \citenamefont {Addesso}}]{PhysRevLett.116.061102}%
  \BibitemOpen
  \bibfield  {author} {\bibinfo {author} {\bibfnamefont {B.~P.}\ \bibnamefont {Abbott}}, \bibinfo {author} {\bibfnamefont {R.}~\bibnamefont {Abbott}}, \bibinfo {author} {\bibfnamefont {T.~D.}\ \bibnamefont {Abbott}}, \bibinfo {author} {\bibfnamefont {M.~R.}\ \bibnamefont {Abernathy}}, \bibinfo {author} {\bibfnamefont {F.}~\bibnamefont {Acernese}}, \bibinfo {author} {\bibfnamefont {K.}~\bibnamefont {Ackley}}, \bibinfo {author} {\bibfnamefont {C.}~\bibnamefont {Adams}}, \bibinfo {author} {\bibfnamefont {T.}~\bibnamefont {Adams}}, \ and\ \bibinfo {author} {\bibfnamefont {P.}~\bibnamefont {Addesso}} (\bibinfo {collaboration} {LIGO Scientific Collaboration and Virgo Collaboration}),\ }\bibfield  {title} {\enquote {\bibinfo {title} {Observation of gravitational waves from a binary black hole merger},}\ }\href {\doibase 10.1103/PhysRevLett.116.061102} {\bibfield  {journal} {\bibinfo  {journal} {Phys. Rev. Lett.}\ }\textbf {\bibinfo {volume} {116}},\ \bibinfo {pages} {061102} (\bibinfo {year} {2016})}\BibitemShut
  {NoStop}%
\bibitem [{\citenamefont {Collaboration}\ and\ \citenamefont {the Virgo~Collaboration}(2022)}]{gwtc21}%
  \BibitemOpen
  \bibfield  {author} {\bibinfo {author} {\bibfnamefont {The LIGO~Scientific}\ \bibnamefont {Collaboration}}\ and\ \bibinfo {author} {\bibnamefont {the Virgo~Collaboration}},\ }\bibfield  {title} {\enquote {\bibinfo {title} {Gwtc-2.1: Deep extended catalog of compact binary coalescences observed by ligo and virgo during the first half of the third observing run},}\ }\href@noop {} {\  (\bibinfo {year} {2022})},\ \Eprint {http://arxiv.org/abs/2108.01045} {arXiv:2108.01045 [gr-qc]} \BibitemShut {NoStop}%
\bibitem [{\citenamefont {et~al.}(2019{\natexlab{a}})}]{PhysRevX.9.031040}%
  \BibitemOpen
  \bibfield  {author} {\bibinfo {author} {\bibfnamefont {B.P.~Abbott}\ \bibnamefont {et~al.}} (\bibinfo {collaboration} {LIGO Scientific Collaboration and Virgo Collaboration}),\ }\bibfield  {title} {\enquote {\bibinfo {title} {Gwtc-1: A gravitational-wave transient catalog of compact binary mergers observed by ligo and virgo during the first and second observing runs},}\ }\href {\doibase 10.1103/PhysRevX.9.031040} {\bibfield  {journal} {\bibinfo  {journal} {Phys. Rev. X}\ }\textbf {\bibinfo {volume} {9}},\ \bibinfo {pages} {031040} (\bibinfo {year} {2019}{\natexlab{a}})}\BibitemShut {NoStop}%
\bibitem [{\citenamefont {et~al.}(2021{\natexlab{a}})}]{PhysRevX.11.021053}%
  \BibitemOpen
  \bibfield  {author} {\bibinfo {author} {\bibfnamefont {R.~Abbott}\ \bibnamefont {et~al.}} (\bibinfo {collaboration} {LIGO Scientific Collaboration and Virgo Collaboration}),\ }\bibfield  {title} {\enquote {\bibinfo {title} {Gwtc-2: Compact binary coalescences observed by ligo and virgo during the first half of the third observing run},}\ }\href {\doibase 10.1103/PhysRevX.11.021053} {\bibfield  {journal} {\bibinfo  {journal} {Phys. Rev. X}\ }\textbf {\bibinfo {volume} {11}},\ \bibinfo {pages} {021053} (\bibinfo {year} {2021}{\natexlab{a}})}\BibitemShut {NoStop}%
\bibitem [{\citenamefont {et~al.}(2021{\natexlab{b}})}]{theligoscientificcollaboration2021gwtc3}%
  \BibitemOpen
  \bibfield  {author} {\bibinfo {author} {\bibfnamefont {R.~Abbott}\ \bibnamefont {et~al.}},\ }\href@noop {} {\enquote {\bibinfo {title} {Gwtc-3: Compact binary coalescences observed by ligo and virgo during the second part of the third observing run},}\ } (\bibinfo {year} {2021}{\natexlab{b}}),\ \Eprint {http://arxiv.org/abs/2111.03606} {arXiv:2111.03606 [gr-qc]} \BibitemShut {NoStop}%
\bibitem [{\citenamefont {Abbott}\ \emph {et~al.}(2020{\natexlab{a}})\citenamefont {Abbott}, \citenamefont {Abbott}, \citenamefont {Abraham}, \citenamefont {Acernese}, \citenamefont {Ackley}, \citenamefont {Adams}, \citenamefont {Adhikari}, \citenamefont {Adya}, \citenamefont {Affeldt}, \citenamefont {Agathos}, \citenamefont {Agatsuma}, \citenamefont {Aggarwal}, \citenamefont {Aguiar}, \citenamefont {Aich}, \citenamefont {Aiello}, \citenamefont {Ain}, \citenamefont {Ajith}, \citenamefont {Akcay}, \citenamefont {Allen}, \citenamefont {Allocca}, \citenamefont {Altin}, \citenamefont {Amato}, \citenamefont {Anand}, \citenamefont {Ananyeva},\ and\ \citenamefont {Anderson}}]{Abbott_2020}%
  \BibitemOpen
  \bibfield  {author} {\bibinfo {author} {\bibfnamefont {R.}~\bibnamefont {Abbott}}, \bibinfo {author} {\bibfnamefont {T.D.}\ \bibnamefont {Abbott}}, \bibinfo {author} {\bibfnamefont {S.}~\bibnamefont {Abraham}}, \bibinfo {author} {\bibfnamefont {F.}~\bibnamefont {Acernese}}, \bibinfo {author} {\bibfnamefont {K.}~\bibnamefont {Ackley}}, \bibinfo {author} {\bibfnamefont {C.}~\bibnamefont {Adams}}, \bibinfo {author} {\bibfnamefont {R.X.}\ \bibnamefont {Adhikari}}, \bibinfo {author} {\bibfnamefont {V.B.}\ \bibnamefont {Adya}}, \bibinfo {author} {\bibfnamefont {C.}~\bibnamefont {Affeldt}}, \bibinfo {author} {\bibfnamefont {M.}~\bibnamefont {Agathos}}, \bibinfo {author} {\bibfnamefont {K.}~\bibnamefont {Agatsuma}}, \bibinfo {author} {\bibfnamefont {N.}~\bibnamefont {Aggarwal}}, \bibinfo {author} {\bibfnamefont {O.D.}\ \bibnamefont {Aguiar}}, \bibinfo {author} {\bibfnamefont {A.}~\bibnamefont {Aich}}, \bibinfo {author} {\bibfnamefont {L.}~\bibnamefont {Aiello}}, \bibinfo {author} {\bibfnamefont {A.}~\bibnamefont
  {Ain}}, \bibinfo {author} {\bibfnamefont {P.}~\bibnamefont {Ajith}}, \bibinfo {author} {\bibfnamefont {S.}~\bibnamefont {Akcay}}, \bibinfo {author} {\bibfnamefont {G.}~\bibnamefont {Allen}}, \bibinfo {author} {\bibfnamefont {A.}~\bibnamefont {Allocca}}, \bibinfo {author} {\bibfnamefont {P.A.}\ \bibnamefont {Altin}}, \bibinfo {author} {\bibfnamefont {A.}~\bibnamefont {Amato}}, \bibinfo {author} {\bibfnamefont {S.}~\bibnamefont {Anand}}, \bibinfo {author} {\bibfnamefont {A.}~\bibnamefont {Ananyeva}}, \ and\ \bibinfo {author} {\bibfnamefont {S.B.}\ \bibnamefont {Anderson}},\ }\bibfield  {title} {\enquote {\bibinfo {title} {Gw190521: A binary black hole merger with a total mass of 150$m_{\odot}$},}\ }\href {\doibase 10.1103/physrevlett.125.101102} {\bibfield  {journal} {\bibinfo  {journal} {Physical Review Letters}\ }\textbf {\bibinfo {volume} {125}} (\bibinfo {year} {2020}{\natexlab{a}}),\ 10.1103/physrevlett.125.101102}\BibitemShut {NoStop}%
\bibitem [{\citenamefont {Greene}\ \emph {et~al.}(2020)\citenamefont {Greene}, \citenamefont {Strader},\ and\ \citenamefont {Ho}}]{Greene_2020}%
  \BibitemOpen
  \bibfield  {author} {\bibinfo {author} {\bibfnamefont {Jenny~E.}\ \bibnamefont {Greene}}, \bibinfo {author} {\bibfnamefont {Jay}\ \bibnamefont {Strader}}, \ and\ \bibinfo {author} {\bibfnamefont {Luis~C.}\ \bibnamefont {Ho}},\ }\bibfield  {title} {\enquote {\bibinfo {title} {Intermediate-mass black holes},}\ }\href {\doibase 10.1146/annurev-astro-032620-021835} {\bibfield  {journal} {\bibinfo  {journal} {Annual Review of Astronomy and Astrophysics}\ }\textbf {\bibinfo {volume} {58}},\ \bibinfo {pages} {257–312} (\bibinfo {year} {2020})}\BibitemShut {NoStop}%
\bibitem [{\citenamefont {Abbott}\ \emph {et~al.}(2022)\citenamefont {Abbott}, \citenamefont {Abbott}, \citenamefont {Acernese}, \citenamefont {Ackley}, \citenamefont {Adams}, \citenamefont {Adhikari}, \citenamefont {Adhikari}, \citenamefont {Adya}, \citenamefont {Affeldt}, \citenamefont {Agarwal}, \citenamefont {Agathos}, \citenamefont {Agatsuma}, \citenamefont {Aggarwal}, \citenamefont {Aguiar}, \citenamefont {Aiello}, \citenamefont {Ain}, \citenamefont {Ajith},\ and\ \citenamefont {Akutsu}}]{Abbott_2022}%
  \BibitemOpen
  \bibfield  {author} {\bibinfo {author} {\bibfnamefont {R.}~\bibnamefont {Abbott}}, \bibinfo {author} {\bibfnamefont {T.~D.}\ \bibnamefont {Abbott}}, \bibinfo {author} {\bibfnamefont {F.}~\bibnamefont {Acernese}}, \bibinfo {author} {\bibfnamefont {K.}~\bibnamefont {Ackley}}, \bibinfo {author} {\bibfnamefont {C.}~\bibnamefont {Adams}}, \bibinfo {author} {\bibfnamefont {N.}~\bibnamefont {Adhikari}}, \bibinfo {author} {\bibfnamefont {R.~X.}\ \bibnamefont {Adhikari}}, \bibinfo {author} {\bibfnamefont {V.~B.}\ \bibnamefont {Adya}}, \bibinfo {author} {\bibfnamefont {C.}~\bibnamefont {Affeldt}}, \bibinfo {author} {\bibfnamefont {D.}~\bibnamefont {Agarwal}}, \bibinfo {author} {\bibfnamefont {M.}~\bibnamefont {Agathos}}, \bibinfo {author} {\bibfnamefont {K.}~\bibnamefont {Agatsuma}}, \bibinfo {author} {\bibfnamefont {N.}~\bibnamefont {Aggarwal}}, \bibinfo {author} {\bibfnamefont {O.~D.}\ \bibnamefont {Aguiar}}, \bibinfo {author} {\bibfnamefont {L.}~\bibnamefont {Aiello}}, \bibinfo {author} {\bibfnamefont
  {A.}~\bibnamefont {Ain}}, \bibinfo {author} {\bibfnamefont {P.}~\bibnamefont {Ajith}}, \ and\ \bibinfo {author} {\bibfnamefont {T.}~\bibnamefont {Akutsu}},\ }\bibfield  {title} {\enquote {\bibinfo {title} {Search for intermediate-mass black hole binaries in the third observing run of advanced ligo and advanced virgo},}\ }\href {\doibase 10.1051/0004-6361/202141452} {\bibfield  {journal} {\bibinfo  {journal} {Astronomy \& Astrophysics}\ }\textbf {\bibinfo {volume} {659}},\ \bibinfo {pages} {A84} (\bibinfo {year} {2022})}\BibitemShut {NoStop}%
\bibitem [{\citenamefont {{Haiman}}\ and\ \citenamefont {{Loeb}}(2001)}]{2001ApJ...552..459H}%
  \BibitemOpen
  \bibfield  {author} {\bibinfo {author} {\bibfnamefont {Zolt{\'a}n}\ \bibnamefont {{Haiman}}}\ and\ \bibinfo {author} {\bibfnamefont {Abraham}\ \bibnamefont {{Loeb}}},\ }\bibfield  {title} {\enquote {\bibinfo {title} {{What Is the Highest Plausible Redshift of Luminous Quasars?}}}\ }\href {\doibase 10.1086/320586} {\bibfield  {journal} {\bibinfo  {journal} {\apj}\ }\textbf {\bibinfo {volume} {552}},\ \bibinfo {pages} {459--463} (\bibinfo {year} {2001})},\ \Eprint {http://arxiv.org/abs/astro-ph/0011529} {arXiv:astro-ph/0011529 [astro-ph]} \BibitemShut {NoStop}%
\bibitem [{\citenamefont {{Volonteri}}\ \emph {et~al.}(2003)\citenamefont {{Volonteri}}, \citenamefont {{Haardt}},\ and\ \citenamefont {{Madau}}}]{2003ApJ...582..559V}%
  \BibitemOpen
  \bibfield  {author} {\bibinfo {author} {\bibfnamefont {Marta}\ \bibnamefont {{Volonteri}}}, \bibinfo {author} {\bibfnamefont {Francesco}\ \bibnamefont {{Haardt}}}, \ and\ \bibinfo {author} {\bibfnamefont {Piero}\ \bibnamefont {{Madau}}},\ }\bibfield  {title} {\enquote {\bibinfo {title} {{The Assembly and Merging History of Supermassive Black Holes in Hierarchical Models of Galaxy Formation}},}\ }\href {\doibase 10.1086/344675} {\bibfield  {journal} {\bibinfo  {journal} {\apj}\ }\textbf {\bibinfo {volume} {582}},\ \bibinfo {pages} {559--573} (\bibinfo {year} {2003})},\ \Eprint {http://arxiv.org/abs/astro-ph/0207276} {arXiv:astro-ph/0207276 [astro-ph]} \BibitemShut {NoStop}%
\bibitem [{\citenamefont {Gerosa}\ and\ \citenamefont {Fishbach}(2021)}]{Gerosa_2021}%
  \BibitemOpen
  \bibfield  {author} {\bibinfo {author} {\bibfnamefont {Davide}\ \bibnamefont {Gerosa}}\ and\ \bibinfo {author} {\bibfnamefont {Maya}\ \bibnamefont {Fishbach}},\ }\bibfield  {title} {\enquote {\bibinfo {title} {Hierarchical mergers of stellar-mass black holes and their gravitational-wave signatures},}\ }\href {\doibase 10.1038/s41550-021-01398-w} {\bibfield  {journal} {\bibinfo  {journal} {Nature Astronomy}\ }\textbf {\bibinfo {volume} {5}},\ \bibinfo {pages} {749–760} (\bibinfo {year} {2021})}\BibitemShut {NoStop}%
\bibitem [{\citenamefont {Dewangan}\ \emph {et~al.}(2006)\citenamefont {Dewangan}, \citenamefont {Titarchuk},\ and\ \citenamefont {Griffiths}}]{Dewangan_2006}%
  \BibitemOpen
  \bibfield  {author} {\bibinfo {author} {\bibfnamefont {Gulab~C.}\ \bibnamefont {Dewangan}}, \bibinfo {author} {\bibfnamefont {Lev}\ \bibnamefont {Titarchuk}}, \ and\ \bibinfo {author} {\bibfnamefont {Richard~E.}\ \bibnamefont {Griffiths}},\ }\bibfield  {title} {\enquote {\bibinfo {title} {Black hole mass of the ultraluminous x-ray source m82 x-1},}\ }\href {\doibase 10.1086/499235} {\bibfield  {journal} {\bibinfo  {journal} {The Astrophysical Journal}\ }\textbf {\bibinfo {volume} {637}},\ \bibinfo {pages} {L21} (\bibinfo {year} {2006})}\BibitemShut {NoStop}%
\bibitem [{\citenamefont {Madhusudhan}\ \emph {et~al.}(2006)\citenamefont {Madhusudhan}, \citenamefont {Justham}, \citenamefont {Nelson}, \citenamefont {Paxton}, \citenamefont {Pfahl}, \citenamefont {Podsiadlowski},\ and\ \citenamefont {Rappaport}}]{Madhusudhan_2006}%
  \BibitemOpen
  \bibfield  {author} {\bibinfo {author} {\bibfnamefont {N.}~\bibnamefont {Madhusudhan}}, \bibinfo {author} {\bibfnamefont {S.}~\bibnamefont {Justham}}, \bibinfo {author} {\bibfnamefont {L.}~\bibnamefont {Nelson}}, \bibinfo {author} {\bibfnamefont {B.}~\bibnamefont {Paxton}}, \bibinfo {author} {\bibfnamefont {E.}~\bibnamefont {Pfahl}}, \bibinfo {author} {\bibfnamefont {Ph.}\ \bibnamefont {Podsiadlowski}}, \ and\ \bibinfo {author} {\bibfnamefont {S.}~\bibnamefont {Rappaport}},\ }\bibfield  {title} {\enquote {\bibinfo {title} {Models of ultraluminous x-ray sources with intermediate-mass black holes},}\ }\href {\doibase 10.1086/500238} {\bibfield  {journal} {\bibinfo  {journal} {The Astrophysical Journal}\ }\textbf {\bibinfo {volume} {640}},\ \bibinfo {pages} {918} (\bibinfo {year} {2006})}\BibitemShut {NoStop}%
\bibitem [{\citenamefont {Cabero}\ \emph {et~al.}(2019)\citenamefont {Cabero}, \citenamefont {Lundgren}, \citenamefont {Nitz}, \citenamefont {Dent}, \citenamefont {Barker}, \citenamefont {Goetz}, \citenamefont {Kissel}, \citenamefont {Nuttall}, \citenamefont {Schale}, \citenamefont {Schofield},\ and\ \citenamefont {Davis}}]{Cabero_2019}%
  \BibitemOpen
  \bibfield  {author} {\bibinfo {author} {\bibfnamefont {M}~\bibnamefont {Cabero}}, \bibinfo {author} {\bibfnamefont {A}~\bibnamefont {Lundgren}}, \bibinfo {author} {\bibfnamefont {A~H}\ \bibnamefont {Nitz}}, \bibinfo {author} {\bibfnamefont {T}~\bibnamefont {Dent}}, \bibinfo {author} {\bibfnamefont {D}~\bibnamefont {Barker}}, \bibinfo {author} {\bibfnamefont {E}~\bibnamefont {Goetz}}, \bibinfo {author} {\bibfnamefont {J~S}\ \bibnamefont {Kissel}}, \bibinfo {author} {\bibfnamefont {L~K}\ \bibnamefont {Nuttall}}, \bibinfo {author} {\bibfnamefont {P}~\bibnamefont {Schale}}, \bibinfo {author} {\bibfnamefont {R}~\bibnamefont {Schofield}}, \ and\ \bibinfo {author} {\bibfnamefont {D}~\bibnamefont {Davis}},\ }\bibfield  {title} {\enquote {\bibinfo {title} {Blip glitches in advanced ligo data},}\ }\href {\doibase 10.1088/1361-6382/ab2e14} {\bibfield  {journal} {\bibinfo  {journal} {Classical and Quantum Gravity}\ }\textbf {\bibinfo {volume} {36}},\ \bibinfo {pages} {155010} (\bibinfo {year} {2019})}\BibitemShut
  {NoStop}%
\bibitem [{\citenamefont {Ghosh}\ \emph {et~al.}(2023)\citenamefont {Ghosh}, \citenamefont {Chandra},\ and\ \citenamefont {Pai}}]{ghosh2023unmasking}%
  \BibitemOpen
  \bibfield  {author} {\bibinfo {author} {\bibfnamefont {Sayantan}\ \bibnamefont {Ghosh}}, \bibinfo {author} {\bibfnamefont {Koustav}\ \bibnamefont {Chandra}}, \ and\ \bibinfo {author} {\bibfnamefont {Archana}\ \bibnamefont {Pai}},\ }\href@noop {} {\enquote {\bibinfo {title} {Unmasking noise transients masquerading as intermediate-mass black hole binaries},}\ } (\bibinfo {year} {2023}),\ \Eprint {http://arxiv.org/abs/2312.01211} {arXiv:2312.01211 [gr-qc]} \BibitemShut {NoStop}%
\bibitem [{\citenamefont {Chandra}\ \emph {et~al.}(2022)\citenamefont {Chandra}, \citenamefont {Bustillo}, \citenamefont {Pai},\ and\ \citenamefont {Harry}}]{Chandra_2022}%
  \BibitemOpen
  \bibfield  {author} {\bibinfo {author} {\bibfnamefont {Koustav}\ \bibnamefont {Chandra}}, \bibinfo {author} {\bibfnamefont {Juan~Calderón}\ \bibnamefont {Bustillo}}, \bibinfo {author} {\bibfnamefont {Archana}\ \bibnamefont {Pai}}, \ and\ \bibinfo {author} {\bibfnamefont {I.W.}\ \bibnamefont {Harry}},\ }\bibfield  {title} {\enquote {\bibinfo {title} {First gravitational-wave search for intermediate-mass black hole mergers with higher-order harmonics},}\ }\href {\doibase 10.1103/physrevd.106.123003} {\bibfield  {journal} {\bibinfo  {journal} {Physical Review D}\ }\textbf {\bibinfo {volume} {106}} (\bibinfo {year} {2022}),\ 10.1103/physrevd.106.123003}\BibitemShut {NoStop}%
\bibitem [{\citenamefont {Bishop}(2006)}]{10.5555/1162264}%
  \BibitemOpen
  \bibfield  {author} {\bibinfo {author} {\bibfnamefont {Christopher~M.}\ \bibnamefont {Bishop}},\ }\href@noop {} {\emph {\bibinfo {title} {Pattern Recognition and Machine Learning (Information Science and Statistics)}}}\ (\bibinfo  {publisher} {Springer-Verlag},\ \bibinfo {address} {Berlin, Heidelberg},\ \bibinfo {year} {2006})\BibitemShut {NoStop}%
\bibitem [{\citenamefont {Wei}\ \emph {et~al.}(2021)\citenamefont {Wei}, \citenamefont {Khan}, \citenamefont {Huerta}, \citenamefont {Huang},\ and\ \citenamefont {Tian}}]{Wei_2021}%
  \BibitemOpen
  \bibfield  {author} {\bibinfo {author} {\bibfnamefont {Wei}\ \bibnamefont {Wei}}, \bibinfo {author} {\bibfnamefont {Asad}\ \bibnamefont {Khan}}, \bibinfo {author} {\bibfnamefont {E.A.}\ \bibnamefont {Huerta}}, \bibinfo {author} {\bibfnamefont {Xiaobo}\ \bibnamefont {Huang}}, \ and\ \bibinfo {author} {\bibfnamefont {Minyang}\ \bibnamefont {Tian}},\ }\bibfield  {title} {\enquote {\bibinfo {title} {Deep learning ensemble for real-time gravitational wave detection of spinning binary black hole mergers},}\ }\href {\doibase 10.1016/j.physletb.2020.136029} {\bibfield  {journal} {\bibinfo  {journal} {Physics Letters B}\ }\textbf {\bibinfo {volume} {812}},\ \bibinfo {pages} {136029} (\bibinfo {year} {2021})}\BibitemShut {NoStop}%
\bibitem [{\citenamefont {Wang}\ \emph {et~al.}(2020)\citenamefont {Wang}, \citenamefont {Wu}, \citenamefont {Cao}, \citenamefont {Liu},\ and\ \citenamefont {Zhu}}]{PhysRevD.101.104003}%
  \BibitemOpen
  \bibfield  {author} {\bibinfo {author} {\bibfnamefont {He}~\bibnamefont {Wang}}, \bibinfo {author} {\bibfnamefont {Shichao}\ \bibnamefont {Wu}}, \bibinfo {author} {\bibfnamefont {Zhoujian}\ \bibnamefont {Cao}}, \bibinfo {author} {\bibfnamefont {Xiaolin}\ \bibnamefont {Liu}}, \ and\ \bibinfo {author} {\bibfnamefont {Jian-Yang}\ \bibnamefont {Zhu}},\ }\bibfield  {title} {\enquote {\bibinfo {title} {Gravitational-wave signal recognition of ligo data by deep learning},}\ }\href {\doibase 10.1103/PhysRevD.101.104003} {\bibfield  {journal} {\bibinfo  {journal} {Phys. Rev. D}\ }\textbf {\bibinfo {volume} {101}},\ \bibinfo {pages} {104003} (\bibinfo {year} {2020})}\BibitemShut {NoStop}%
\bibitem [{\citenamefont {George}\ and\ \citenamefont {Huerta}(2017)}]{george2017deeplearningrealtimegravitational}%
  \BibitemOpen
  \bibfield  {author} {\bibinfo {author} {\bibfnamefont {Daniel}\ \bibnamefont {George}}\ and\ \bibinfo {author} {\bibfnamefont {E.~A.}\ \bibnamefont {Huerta}},\ }\href {https://arxiv.org/abs/1711.07966} {\enquote {\bibinfo {title} {Deep learning for real-time gravitational wave detection and parameter estimation with ligo data},}\ } (\bibinfo {year} {2017}),\ \Eprint {http://arxiv.org/abs/1711.07966} {arXiv:1711.07966 [gr-qc]} \BibitemShut {NoStop}%
\bibitem [{\citenamefont {George}\ and\ \citenamefont {Huerta}(2018)}]{GEORGE201864}%
  \BibitemOpen
  \bibfield  {author} {\bibinfo {author} {\bibfnamefont {Daniel}\ \bibnamefont {George}}\ and\ \bibinfo {author} {\bibfnamefont {E.A.}\ \bibnamefont {Huerta}},\ }\bibfield  {title} {\enquote {\bibinfo {title} {Deep learning for real-time gravitational wave detection and parameter estimation: Results with advanced ligo data},}\ }\href {\doibase https://doi.org/10.1016/j.physletb.2017.12.053} {\bibfield  {journal} {\bibinfo  {journal} {Physics Letters B}\ }\textbf {\bibinfo {volume} {778}},\ \bibinfo {pages} {64--70} (\bibinfo {year} {2018})}\BibitemShut {NoStop}%
\bibitem [{\citenamefont {Gebhard}\ \emph {et~al.}(2019)\citenamefont {Gebhard}, \citenamefont {Kilbertus}, \citenamefont {Harry},\ and\ \citenamefont {Sch\"olkopf}}]{PhysRevD.100.063015}%
  \BibitemOpen
  \bibfield  {author} {\bibinfo {author} {\bibfnamefont {Timothy~D.}\ \bibnamefont {Gebhard}}, \bibinfo {author} {\bibfnamefont {Niki}\ \bibnamefont {Kilbertus}}, \bibinfo {author} {\bibfnamefont {Ian}\ \bibnamefont {Harry}}, \ and\ \bibinfo {author} {\bibfnamefont {Bernhard}\ \bibnamefont {Sch\"olkopf}},\ }\bibfield  {title} {\enquote {\bibinfo {title} {Convolutional neural networks: A magic bullet for gravitational-wave detection?}}\ }\href {\doibase 10.1103/PhysRevD.100.063015} {\bibfield  {journal} {\bibinfo  {journal} {Phys. Rev. D}\ }\textbf {\bibinfo {volume} {100}},\ \bibinfo {pages} {063015} (\bibinfo {year} {2019})}\BibitemShut {NoStop}%
\bibitem [{\citenamefont {Fan}\ \emph {et~al.}(2019)\citenamefont {Fan}, \citenamefont {Li}, \citenamefont {Li}, \citenamefont {Zhong},\ and\ \citenamefont {Cao}}]{Fan_2019}%
  \BibitemOpen
  \bibfield  {author} {\bibinfo {author} {\bibfnamefont {XiLong}\ \bibnamefont {Fan}}, \bibinfo {author} {\bibfnamefont {Jin}\ \bibnamefont {Li}}, \bibinfo {author} {\bibfnamefont {Xin}\ \bibnamefont {Li}}, \bibinfo {author} {\bibfnamefont {YuanHong}\ \bibnamefont {Zhong}}, \ and\ \bibinfo {author} {\bibfnamefont {JunWei}\ \bibnamefont {Cao}},\ }\bibfield  {title} {\enquote {\bibinfo {title} {Applying deep neural networks to the detection and space parameter estimation of compact binary coalescence with a network of gravitational wave detectors},}\ }\href {\doibase 10.1007/s11433-018-9321-7} {\bibfield  {journal} {\bibinfo  {journal} {Science China Physics, Mechanics \& Astronomy}\ }\textbf {\bibinfo {volume} {62}} (\bibinfo {year} {2019}),\ 10.1007/s11433-018-9321-7}\BibitemShut {NoStop}%
\bibitem [{\citenamefont {Verma}\ \emph {et~al.}(2022)\citenamefont {Verma}, \citenamefont {Reza}, \citenamefont {Krishnaswamy}, \citenamefont {Caudill},\ and\ \citenamefont {Gaur}}]{Verma_2022}%
  \BibitemOpen
  \bibfield  {author} {\bibinfo {author} {\bibfnamefont {Chetan}\ \bibnamefont {Verma}}, \bibinfo {author} {\bibfnamefont {Amit}\ \bibnamefont {Reza}}, \bibinfo {author} {\bibfnamefont {Dilip}\ \bibnamefont {Krishnaswamy}}, \bibinfo {author} {\bibfnamefont {Sarah}\ \bibnamefont {Caudill}}, \ and\ \bibinfo {author} {\bibfnamefont {Gurudatt}\ \bibnamefont {Gaur}},\ }\bibfield  {title} {\enquote {\bibinfo {title} {Employing deep learning for detection of gravitational waves from compact binary coalescences},}\ }in\ \href {\doibase 10.1063/5.0108682} {\emph {\bibinfo {booktitle} {THE 6TH INTERNATIONAL CONFERENCE ON SCIENCE AND TECHNOLOGY (ICST21): Challenges and Opportunities for Innovation Research on Science Materials, and Technology in the Covid-19 Era}}}\ (\bibinfo  {publisher} {AIP Publishing},\ \bibinfo {year} {2022})\BibitemShut {NoStop}%
\bibitem [{\citenamefont {Men\'endez-V\'azquez}\ \emph {et~al.}(2021)\citenamefont {Men\'endez-V\'azquez}, \citenamefont {Kolstein}, \citenamefont {Mart\'{\i}nez},\ and\ \citenamefont {Mir}}]{PhysRevD.103.062004}%
  \BibitemOpen
  \bibfield  {author} {\bibinfo {author} {\bibfnamefont {A.}~\bibnamefont {Men\'endez-V\'azquez}}, \bibinfo {author} {\bibfnamefont {M.}~\bibnamefont {Kolstein}}, \bibinfo {author} {\bibfnamefont {M.}~\bibnamefont {Mart\'{\i}nez}}, \ and\ \bibinfo {author} {\bibfnamefont {Ll.~M.}\ \bibnamefont {Mir}},\ }\bibfield  {title} {\enquote {\bibinfo {title} {Searches for compact binary coalescence events using neural networks in the ligo/virgo second observation period},}\ }\href {\doibase 10.1103/PhysRevD.103.062004} {\bibfield  {journal} {\bibinfo  {journal} {Phys. Rev. D}\ }\textbf {\bibinfo {volume} {103}},\ \bibinfo {pages} {062004} (\bibinfo {year} {2021})}\BibitemShut {NoStop}%
\bibitem [{\citenamefont {Jadhav}\ \emph {et~al.}(2023)\citenamefont {Jadhav}, \citenamefont {Shrivastava},\ and\ \citenamefont {Mitra}}]{Jadhav_2023}%
  \BibitemOpen
  \bibfield  {author} {\bibinfo {author} {\bibfnamefont {Shreejit}\ \bibnamefont {Jadhav}}, \bibinfo {author} {\bibfnamefont {Mihir}\ \bibnamefont {Shrivastava}}, \ and\ \bibinfo {author} {\bibfnamefont {Sanjit}\ \bibnamefont {Mitra}},\ }\bibfield  {title} {\enquote {\bibinfo {title} {Towards a robust and reliable deep learning approach for detection of compact binary mergers in gravitational wave data},}\ }\href {\doibase 10.1088/2632-2153/ad0938} {\bibfield  {journal} {\bibinfo  {journal} {Machine Learning: Science and Technology}\ }\textbf {\bibinfo {volume} {4}},\ \bibinfo {pages} {045028} (\bibinfo {year} {2023})}\BibitemShut {NoStop}%
\bibitem [{\citenamefont {Qiu}\ \emph {et~al.}(2023)\citenamefont {Qiu}, \citenamefont {Krastev}, \citenamefont {Gill},\ and\ \citenamefont {Berger}}]{QIU2023137850}%
  \BibitemOpen
  \bibfield  {author} {\bibinfo {author} {\bibfnamefont {Richard}\ \bibnamefont {Qiu}}, \bibinfo {author} {\bibfnamefont {Plamen~G.}\ \bibnamefont {Krastev}}, \bibinfo {author} {\bibfnamefont {Kiranjyot}\ \bibnamefont {Gill}}, \ and\ \bibinfo {author} {\bibfnamefont {Edo}\ \bibnamefont {Berger}},\ }\bibfield  {title} {\enquote {\bibinfo {title} {Deep learning detection and classification of gravitational waves from neutron star-black hole mergers},}\ }\href {\doibase https://doi.org/10.1016/j.physletb.2023.137850} {\bibfield  {journal} {\bibinfo  {journal} {Physics Letters B}\ }\textbf {\bibinfo {volume} {840}},\ \bibinfo {pages} {137850} (\bibinfo {year} {2023})}\BibitemShut {NoStop}%
\bibitem [{\citenamefont {Meijer}\ \emph {et~al.}(2024)\citenamefont {Meijer}, \citenamefont {Lopez}, \citenamefont {Tsuna},\ and\ \citenamefont {Caudill}}]{PhysRevD.109.022006}%
  \BibitemOpen
  \bibfield  {author} {\bibinfo {author} {\bibfnamefont {Quirijn}\ \bibnamefont {Meijer}}, \bibinfo {author} {\bibfnamefont {Melissa}\ \bibnamefont {Lopez}}, \bibinfo {author} {\bibfnamefont {Daichi}\ \bibnamefont {Tsuna}}, \ and\ \bibinfo {author} {\bibfnamefont {Sarah}\ \bibnamefont {Caudill}},\ }\bibfield  {title} {\enquote {\bibinfo {title} {Gravitational-wave searches for cosmic string cusps in einstein telescope data using deep learning},}\ }\href {\doibase 10.1103/PhysRevD.109.022006} {\bibfield  {journal} {\bibinfo  {journal} {Phys. Rev. D}\ }\textbf {\bibinfo {volume} {109}},\ \bibinfo {pages} {022006} (\bibinfo {year} {2024})}\BibitemShut {NoStop}%
\bibitem [{\citenamefont {Chatterjee}\ \emph {et~al.}(2021)\citenamefont {Chatterjee}, \citenamefont {Wen}, \citenamefont {Diakogiannis},\ and\ \citenamefont {Vinsen}}]{PhysRevD.104.064046}%
  \BibitemOpen
  \bibfield  {author} {\bibinfo {author} {\bibfnamefont {Chayan}\ \bibnamefont {Chatterjee}}, \bibinfo {author} {\bibfnamefont {Linqing}\ \bibnamefont {Wen}}, \bibinfo {author} {\bibfnamefont {Foivos}\ \bibnamefont {Diakogiannis}}, \ and\ \bibinfo {author} {\bibfnamefont {Kevin}\ \bibnamefont {Vinsen}},\ }\bibfield  {title} {\enquote {\bibinfo {title} {Extraction of binary black hole gravitational wave signals from detector data using deep learning},}\ }\href {\doibase 10.1103/PhysRevD.104.064046} {\bibfield  {journal} {\bibinfo  {journal} {Phys. Rev. D}\ }\textbf {\bibinfo {volume} {104}},\ \bibinfo {pages} {064046} (\bibinfo {year} {2021})}\BibitemShut {NoStop}%
\bibitem [{\citenamefont {Boudart}(2023)}]{PhysRevD.107.024007}%
  \BibitemOpen
  \bibfield  {author} {\bibinfo {author} {\bibfnamefont {Vincent}\ \bibnamefont {Boudart}},\ }\bibfield  {title} {\enquote {\bibinfo {title} {Convolutional neural network to distinguish glitches from minute-long gravitational wave transients},}\ }\href {\doibase 10.1103/PhysRevD.107.024007} {\bibfield  {journal} {\bibinfo  {journal} {Phys. Rev. D}\ }\textbf {\bibinfo {volume} {107}},\ \bibinfo {pages} {024007} (\bibinfo {year} {2023})}\BibitemShut {NoStop}%
\bibitem [{\citenamefont {Koyama}\ \emph {et~al.}(2024)\citenamefont {Koyama}, \citenamefont {Sakai}, \citenamefont {Sasaoka}, \citenamefont {Dominguez}, \citenamefont {Somiya}, \citenamefont {Omae}, \citenamefont {Terada}, \citenamefont {Meyer-Conde},\ and\ \citenamefont {Takahashi}}]{Koyama:2024zos}%
  \BibitemOpen
  \bibfield  {author} {\bibinfo {author} {\bibfnamefont {Naoki}\ \bibnamefont {Koyama}}, \bibinfo {author} {\bibfnamefont {Yusuke}\ \bibnamefont {Sakai}}, \bibinfo {author} {\bibfnamefont {Seiya}\ \bibnamefont {Sasaoka}}, \bibinfo {author} {\bibfnamefont {Diego}\ \bibnamefont {Dominguez}}, \bibinfo {author} {\bibfnamefont {Kentaro}\ \bibnamefont {Somiya}}, \bibinfo {author} {\bibfnamefont {Yuto}\ \bibnamefont {Omae}}, \bibinfo {author} {\bibfnamefont {Yoshikazu}\ \bibnamefont {Terada}}, \bibinfo {author} {\bibfnamefont {Marco}\ \bibnamefont {Meyer-Conde}}, \ and\ \bibinfo {author} {\bibfnamefont {Hirotaka}\ \bibnamefont {Takahashi}},\ }\bibfield  {title} {\enquote {\bibinfo {title} {{Enhancing the rationale of convolutional neural networks for glitch classification in gravitational wave detectors: a visual explanation}},}\ }\href {\doibase 10.1088/2632-2153/ad6391} {\bibfield  {journal} {\bibinfo  {journal} {Mach. Learn. Sci. Tech.}\ }\textbf {\bibinfo {volume} {5}},\ \bibinfo {pages} {035028} (\bibinfo
  {year} {2024})}\BibitemShut {NoStop}%
\bibitem [{\citenamefont {George}\ \emph {et~al.}(2018)\citenamefont {George}, \citenamefont {Shen},\ and\ \citenamefont {Huerta}}]{PhysRevD.97.101501}%
  \BibitemOpen
  \bibfield  {author} {\bibinfo {author} {\bibfnamefont {Daniel}\ \bibnamefont {George}}, \bibinfo {author} {\bibfnamefont {Hongyu}\ \bibnamefont {Shen}}, \ and\ \bibinfo {author} {\bibfnamefont {E.~A.}\ \bibnamefont {Huerta}},\ }\bibfield  {title} {\enquote {\bibinfo {title} {Classification and unsupervised clustering of ligo data with deep transfer learning},}\ }\href {\doibase 10.1103/PhysRevD.97.101501} {\bibfield  {journal} {\bibinfo  {journal} {Phys. Rev. D}\ }\textbf {\bibinfo {volume} {97}},\ \bibinfo {pages} {101501} (\bibinfo {year} {2018})}\BibitemShut {NoStop}%
\bibitem [{\citenamefont {Lopez}\ \emph {et~al.}(2022{\natexlab{a}})\citenamefont {Lopez}, \citenamefont {Boudart}, \citenamefont {Buijsman}, \citenamefont {Reza},\ and\ \citenamefont {Caudill}}]{PhysRevD.106.023027}%
  \BibitemOpen
  \bibfield  {author} {\bibinfo {author} {\bibfnamefont {Melissa}\ \bibnamefont {Lopez}}, \bibinfo {author} {\bibfnamefont {Vincent}\ \bibnamefont {Boudart}}, \bibinfo {author} {\bibfnamefont {Kerwin}\ \bibnamefont {Buijsman}}, \bibinfo {author} {\bibfnamefont {Amit}\ \bibnamefont {Reza}}, \ and\ \bibinfo {author} {\bibfnamefont {Sarah}\ \bibnamefont {Caudill}},\ }\bibfield  {title} {\enquote {\bibinfo {title} {Simulating transient noise bursts in ligo with generative adversarial networks},}\ }\href {\doibase 10.1103/PhysRevD.106.023027} {\bibfield  {journal} {\bibinfo  {journal} {Phys. Rev. D}\ }\textbf {\bibinfo {volume} {106}},\ \bibinfo {pages} {023027} (\bibinfo {year} {2022}{\natexlab{a}})}\BibitemShut {NoStop}%
\bibitem [{\citenamefont {Lopez}\ \emph {et~al.}(2022{\natexlab{b}})\citenamefont {Lopez}, \citenamefont {Boudart}, \citenamefont {Schmidt},\ and\ \citenamefont {Caudill}}]{lopez2022simulating}%
  \BibitemOpen
  \bibfield  {author} {\bibinfo {author} {\bibfnamefont {Melissa}\ \bibnamefont {Lopez}}, \bibinfo {author} {\bibfnamefont {Vincent}\ \bibnamefont {Boudart}}, \bibinfo {author} {\bibfnamefont {Stefano}\ \bibnamefont {Schmidt}}, \ and\ \bibinfo {author} {\bibfnamefont {Sarah}\ \bibnamefont {Caudill}},\ }\href@noop {} {\enquote {\bibinfo {title} {Simulating transient noise bursts in ligo with gengli},}\ } (\bibinfo {year} {2022}{\natexlab{b}}),\ \Eprint {http://arxiv.org/abs/2205.09204} {arXiv:2205.09204 [astro-ph.IM]} \BibitemShut {NoStop}%
\bibitem [{\citenamefont {Apostolatos}\ \emph {et~al.}(1994)\citenamefont {Apostolatos}, \citenamefont {Cutler}, \citenamefont {Sussman},\ and\ \citenamefont {Thorne}}]{PhysRevD.49.6274}%
  \BibitemOpen
  \bibfield  {author} {\bibinfo {author} {\bibfnamefont {Theocharis~A.}\ \bibnamefont {Apostolatos}}, \bibinfo {author} {\bibfnamefont {Curt}\ \bibnamefont {Cutler}}, \bibinfo {author} {\bibfnamefont {Gerald~J.}\ \bibnamefont {Sussman}}, \ and\ \bibinfo {author} {\bibfnamefont {Kip~S.}\ \bibnamefont {Thorne}},\ }\bibfield  {title} {\enquote {\bibinfo {title} {Spin-induced orbital precession and its modulation of the gravitational waveforms from merging binaries},}\ }\href {\doibase 10.1103/PhysRevD.49.6274} {\bibfield  {journal} {\bibinfo  {journal} {Phys. Rev. D}\ }\textbf {\bibinfo {volume} {49}},\ \bibinfo {pages} {6274--6297} (\bibinfo {year} {1994})}\BibitemShut {NoStop}%
\bibitem [{\citenamefont {Schmidt}\ \emph {et~al.}(2012)\citenamefont {Schmidt}, \citenamefont {Hannam},\ and\ \citenamefont {Husa}}]{PhysRevD.86.104063}%
  \BibitemOpen
  \bibfield  {author} {\bibinfo {author} {\bibfnamefont {Patricia}\ \bibnamefont {Schmidt}}, \bibinfo {author} {\bibfnamefont {Mark}\ \bibnamefont {Hannam}}, \ and\ \bibinfo {author} {\bibfnamefont {Sascha}\ \bibnamefont {Husa}},\ }\bibfield  {title} {\enquote {\bibinfo {title} {Towards models of gravitational waveforms from generic binaries: A simple approximate mapping between precessing and nonprecessing inspiral signals},}\ }\href {\doibase 10.1103/PhysRevD.86.104063} {\bibfield  {journal} {\bibinfo  {journal} {Phys. Rev. D}\ }\textbf {\bibinfo {volume} {86}},\ \bibinfo {pages} {104063} (\bibinfo {year} {2012})}\BibitemShut {NoStop}%
\bibitem [{\citenamefont {Helstrom}(1960)}]{helstrom1960statistical}%
  \BibitemOpen
  \bibfield  {author} {\bibinfo {author} {\bibfnamefont {C.W.}\ \bibnamefont {Helstrom}},\ }\href {https://books.google.nl/books?id=RNAyAAAAMAAJ} {\emph {\bibinfo {title} {Statistical Theory of Signal Detection}}},\ International series of monographs on electronics and instrumentation\ (\bibinfo  {publisher} {Pergamon Press},\ \bibinfo {year} {1960})\BibitemShut {NoStop}%
\bibitem [{\citenamefont {Maggiore}(2007)}]{10.1093/acprof:oso/9780198570745.001.0001}%
  \BibitemOpen
  \bibfield  {author} {\bibinfo {author} {\bibfnamefont {Michele}\ \bibnamefont {Maggiore}},\ }\href@noop {} {\enquote {\bibinfo {title} {Gravitational waves: Volume 1: Theory and experiments},}\ } (\bibinfo {year} {2007})\BibitemShut {NoStop}%
\bibitem [{\citenamefont {Goddi}\ \emph {et~al.}(2019)\citenamefont {Goddi}, \citenamefont {Crew}, \citenamefont {Impellizzeri}, \citenamefont {Martí-Vidal}, \citenamefont {Matthews}, \citenamefont {Messias}, \citenamefont {Rottmann}, \citenamefont {Alef}, \citenamefont {Blackburn}, \citenamefont {Bronzwaer}, \citenamefont {Chan}, \citenamefont {Davelaar}, \citenamefont {Deane}, \citenamefont {Dexter}, \citenamefont {Doeleman}, \citenamefont {Falcke}, \citenamefont {Fish}, \citenamefont {Fraga-Encinas}, \citenamefont {Fromm}, \citenamefont {Herrero-Illana}, \citenamefont {Issaoun}, \citenamefont {James}, \citenamefont {Janssen}, \citenamefont {Kramer}, \citenamefont {Krichbaum}, \citenamefont {De~Laurentis}, \citenamefont {Liuzzo}, \citenamefont {Mizuno}, \citenamefont {Moscibrodzka}, \citenamefont {Natarajan}, \citenamefont {Porth}, \citenamefont {Rezzolla}, \citenamefont {Rygl}, \citenamefont {Roelofs}, \citenamefont {Ros}, \citenamefont {Roy}, \citenamefont {Shao}, \citenamefont {Van~Langevelde},
  \citenamefont {Van~Bemmel}, \citenamefont {Tilanus}, \citenamefont {Torne}, \citenamefont {Wielgus}, \citenamefont {Younsi}, \citenamefont {Zensus},\ and\ \citenamefont {Collaboration}}]{https://doi.org/10.18727/0722-6691/5150}%
  \BibitemOpen
  \bibfield  {author} {\bibinfo {author} {\bibfnamefont {Ciriaco}\ \bibnamefont {Goddi}}, \bibinfo {author} {\bibfnamefont {Geoff}\ \bibnamefont {Crew}}, \bibinfo {author} {\bibfnamefont {Violette}\ \bibnamefont {Impellizzeri}}, \bibinfo {author} {\bibfnamefont {Iván}\ \bibnamefont {Martí-Vidal}}, \bibinfo {author} {\bibfnamefont {Lynn~D.}\ \bibnamefont {Matthews}}, \bibinfo {author} {\bibfnamefont {Hugo}\ \bibnamefont {Messias}}, \bibinfo {author} {\bibfnamefont {Helge}\ \bibnamefont {Rottmann}}, \bibinfo {author} {\bibfnamefont {Walter}\ \bibnamefont {Alef}}, \bibinfo {author} {\bibfnamefont {Lindy}\ \bibnamefont {Blackburn}}, \bibinfo {author} {\bibfnamefont {Thomas}\ \bibnamefont {Bronzwaer}}, \bibinfo {author} {\bibfnamefont {Chi-Kwan}\ \bibnamefont {Chan}}, \bibinfo {author} {\bibfnamefont {Jordy}\ \bibnamefont {Davelaar}}, \bibinfo {author} {\bibfnamefont {Roger}\ \bibnamefont {Deane}}, \bibinfo {author} {\bibfnamefont {Jason}\ \bibnamefont {Dexter}}, \bibinfo {author} {\bibfnamefont {Shep}\
  \bibnamefont {Doeleman}}, \bibinfo {author} {\bibfnamefont {Heino}\ \bibnamefont {Falcke}}, \bibinfo {author} {\bibfnamefont {Vincent~L.}\ \bibnamefont {Fish}}, \bibinfo {author} {\bibfnamefont {Raquel}\ \bibnamefont {Fraga-Encinas}}, \bibinfo {author} {\bibfnamefont {Christian~M.}\ \bibnamefont {Fromm}}, \bibinfo {author} {\bibfnamefont {Ruben}\ \bibnamefont {Herrero-Illana}}, \bibinfo {author} {\bibfnamefont {Sara}\ \bibnamefont {Issaoun}}, \bibinfo {author} {\bibfnamefont {David}\ \bibnamefont {James}}, \bibinfo {author} {\bibfnamefont {Michael}\ \bibnamefont {Janssen}}, \bibinfo {author} {\bibfnamefont {Michael}\ \bibnamefont {Kramer}}, \bibinfo {author} {\bibfnamefont {Thomas~P.}\ \bibnamefont {Krichbaum}}, \bibinfo {author} {\bibfnamefont {Mariafelicia}\ \bibnamefont {De~Laurentis}}, \bibinfo {author} {\bibfnamefont {Elisabetta}\ \bibnamefont {Liuzzo}}, \bibinfo {author} {\bibfnamefont {Yosuke}\ \bibnamefont {Mizuno}}, \bibinfo {author} {\bibfnamefont {Monika}\ \bibnamefont {Moscibrodzka}}, \bibinfo
  {author} {\bibfnamefont {Iniyan}\ \bibnamefont {Natarajan}}, \bibinfo {author} {\bibfnamefont {Oliver}\ \bibnamefont {Porth}}, \bibinfo {author} {\bibfnamefont {Luciano}\ \bibnamefont {Rezzolla}}, \bibinfo {author} {\bibfnamefont {Kazi}\ \bibnamefont {Rygl}}, \bibinfo {author} {\bibfnamefont {Freek}\ \bibnamefont {Roelofs}}, \bibinfo {author} {\bibfnamefont {Eduardo}\ \bibnamefont {Ros}}, \bibinfo {author} {\bibfnamefont {Alan~L.}\ \bibnamefont {Roy}}, \bibinfo {author} {\bibfnamefont {Lijing}\ \bibnamefont {Shao}}, \bibinfo {author} {\bibfnamefont {Huib~Jan}\ \bibnamefont {Van~Langevelde}}, \bibinfo {author} {\bibfnamefont {Ilse}\ \bibnamefont {Van~Bemmel}}, \bibinfo {author} {\bibfnamefont {Remo}\ \bibnamefont {Tilanus}}, \bibinfo {author} {\bibfnamefont {Pablo}\ \bibnamefont {Torne}}, \bibinfo {author} {\bibfnamefont {Maciek}\ \bibnamefont {Wielgus}}, \bibinfo {author} {\bibfnamefont {Ziri}\ \bibnamefont {Younsi}}, \bibinfo {author} {\bibfnamefont {J.~Anton}\ \bibnamefont {Zensus}}, \ and\ \bibinfo
  {author} {\bibfnamefont {The Event Horizon~Telescope}\ \bibnamefont {Collaboration}},\ }\bibfield  {title} {\enquote {\bibinfo {title} {First m87 event horizon telescope results and the role of alma},}\ }\href {\doibase 10.18727/0722-6691/5150} {\bibfield  {journal} {\bibinfo  {journal} {Published in The Messenger vol. 177}\ }\textbf {\bibinfo {volume} {pp. 25-35}},\ \bibinfo {pages} {September 2019.} (\bibinfo {year} {2019})}\BibitemShut {NoStop}%
\bibitem [{\citenamefont {Collaboration}(2024{\natexlab{a}})}]{sagvii}%
  \BibitemOpen
  \bibfield  {author} {\bibinfo {author} {\bibfnamefont {The Event Horizon~Telescope}\ \bibnamefont {Collaboration}},\ }\bibfield  {title} {\enquote {\bibinfo {title} {First sagittarius a* event horizon telescope results. vii. polarization of the ring},}\ }\href {\doibase 10.3847/2041-8213/ad2df0} {\bibfield  {journal} {\bibinfo  {journal} {The Astrophysical Journal Letters}\ }\textbf {\bibinfo {volume} {964}},\ \bibinfo {pages} {L25} (\bibinfo {year} {2024}{\natexlab{a}})}\BibitemShut {NoStop}%
\bibitem [{\citenamefont {Collaboration}(2024{\natexlab{b}})}]{sagviii}%
  \BibitemOpen
  \bibfield  {author} {\bibinfo {author} {\bibfnamefont {The Event Horizon~Telescope}\ \bibnamefont {Collaboration}},\ }\bibfield  {title} {\enquote {\bibinfo {title} {First sagittarius a* event horizon telescope results. viii. physical interpretation of the polarized ring},}\ }\href {\doibase 10.3847/2041-8213/ad2df1} {\bibfield  {journal} {\bibinfo  {journal} {The Astrophysical Journal Letters}\ }\textbf {\bibinfo {volume} {964}},\ \bibinfo {pages} {L26} (\bibinfo {year} {2024}{\natexlab{b}})}\BibitemShut {NoStop}%
\bibitem [{\citenamefont {{Abbott, R.}}\ \emph {et~al.}(2022)\citenamefont {{Abbott, R.}}, \citenamefont {{Abbott, T. D.}}, \citenamefont {{Acernese, F.}}, \citenamefont {{Ackley, K.}}, \citenamefont {{Adams, C.}}, \citenamefont {{Adhikari, N.}}, \citenamefont {{Adhikari, R. X.}}, \citenamefont {{Adya, V. B.}}, \citenamefont {{Affeldt, C.}}, \citenamefont {{Agarwal, D.}}, \citenamefont {{Agathos, M.}}, \citenamefont {{Agatsuma, K.}}, \citenamefont {{Aggarwal, N.}}, \citenamefont {{Aguiar, O. D.}}, \citenamefont {{Aiello, L.}}, \citenamefont {{Ain, A.}}, \citenamefont {{Ajith, P.}}, \citenamefont {{Akutsu, T.}}, \citenamefont {{Albanesi, S.}}, \citenamefont {{Allocca, A.}}, \citenamefont {{Altin, P. A.}},\ and\ \citenamefont {{Amato, A.}}}]{imbho3}%
  \BibitemOpen
  \bibfield  {author} {\bibinfo {author} {\bibnamefont {{Abbott, R.}}}, \bibinfo {author} {\bibnamefont {{Abbott, T. D.}}}, \bibinfo {author} {\bibnamefont {{Acernese, F.}}}, \bibinfo {author} {\bibnamefont {{Ackley, K.}}}, \bibinfo {author} {\bibnamefont {{Adams, C.}}}, \bibinfo {author} {\bibnamefont {{Adhikari, N.}}}, \bibinfo {author} {\bibnamefont {{Adhikari, R. X.}}}, \bibinfo {author} {\bibnamefont {{Adya, V. B.}}}, \bibinfo {author} {\bibnamefont {{Affeldt, C.}}}, \bibinfo {author} {\bibnamefont {{Agarwal, D.}}}, \bibinfo {author} {\bibnamefont {{Agathos, M.}}}, \bibinfo {author} {\bibnamefont {{Agatsuma, K.}}}, \bibinfo {author} {\bibnamefont {{Aggarwal, N.}}}, \bibinfo {author} {\bibnamefont {{Aguiar, O. D.}}}, \bibinfo {author} {\bibnamefont {{Aiello, L.}}}, \bibinfo {author} {\bibnamefont {{Ain, A.}}}, \bibinfo {author} {\bibnamefont {{Ajith, P.}}}, \bibinfo {author} {\bibnamefont {{Akutsu, T.}}}, \bibinfo {author} {\bibnamefont {{Albanesi, S.}}}, \bibinfo {author} {\bibnamefont {{Allocca, A.}}},
  \bibinfo {author} {\bibnamefont {{Altin, P. A.}}}, \ and\ \bibinfo {author} {\bibnamefont {{Amato, A.}}},\ }\bibfield  {title} {\enquote {\bibinfo {title} {Search for intermediate-mass black hole binaries in the third observing run of advanced ligo and advanced virgo},}\ }\href {\doibase 10.1051/0004-6361/202141452} {\bibfield  {journal} {\bibinfo  {journal} {Astronomy \& Astrophysics}\ }\textbf {\bibinfo {volume} {659}},\ \bibinfo {pages} {A84} (\bibinfo {year} {2022})}\BibitemShut {NoStop}%
\bibitem [{\citenamefont {{Fowler}}\ and\ \citenamefont {{Hoyle}}(1964)}]{1964ApJS....9..201F}%
  \BibitemOpen
  \bibfield  {author} {\bibinfo {author} {\bibfnamefont {William~A.}\ \bibnamefont {{Fowler}}}\ and\ \bibinfo {author} {\bibfnamefont {F.}~\bibnamefont {{Hoyle}}},\ }\bibfield  {title} {\enquote {\bibinfo {title} {{Neutrino Processes and Pair Formation in Massive Stars and Supernovae.}}}\ }\href {\doibase 10.1086/190103} {\bibfield  {journal} {\bibinfo  {journal} {Astrophys. J. Suppl.}\ }\textbf {\bibinfo {volume} {9}},\ \bibinfo {pages} {201} (\bibinfo {year} {1964})}\BibitemShut {NoStop}%
\bibitem [{\citenamefont {Barkat}\ \emph {et~al.}(1967)\citenamefont {Barkat}, \citenamefont {Rakavy},\ and\ \citenamefont {Sack}}]{PhysRevLett.18.379}%
  \BibitemOpen
  \bibfield  {author} {\bibinfo {author} {\bibfnamefont {Z.}~\bibnamefont {Barkat}}, \bibinfo {author} {\bibfnamefont {G.}~\bibnamefont {Rakavy}}, \ and\ \bibinfo {author} {\bibfnamefont {N.}~\bibnamefont {Sack}},\ }\bibfield  {title} {\enquote {\bibinfo {title} {Dynamics of supernova explosion resulting from pair formation},}\ }\href {\doibase 10.1103/PhysRevLett.18.379} {\bibfield  {journal} {\bibinfo  {journal} {Phys. Rev. Lett.}\ }\textbf {\bibinfo {volume} {18}},\ \bibinfo {pages} {379--381} (\bibinfo {year} {1967})}\BibitemShut {NoStop}%
\bibitem [{\citenamefont {{Bond}}\ \emph {et~al.}(1984)\citenamefont {{Bond}}, \citenamefont {{Arnett}},\ and\ \citenamefont {{Carr}}}]{1984ApJ...280..825B}%
  \BibitemOpen
  \bibfield  {author} {\bibinfo {author} {\bibfnamefont {J.~R.}\ \bibnamefont {{Bond}}}, \bibinfo {author} {\bibfnamefont {W.~D.}\ \bibnamefont {{Arnett}}}, \ and\ \bibinfo {author} {\bibfnamefont {B.~J.}\ \bibnamefont {{Carr}}},\ }\bibfield  {title} {\enquote {\bibinfo {title} {{The evolution and fate of Very Massive Objects}},}\ }\href {\doibase 10.1086/162057} {\bibfield  {journal} {\bibinfo  {journal} {\apj}\ }\textbf {\bibinfo {volume} {280}},\ \bibinfo {pages} {825--847} (\bibinfo {year} {1984})}\BibitemShut {NoStop}%
\bibitem [{\citenamefont {Woosley}\ \emph {et~al.}(2007)\citenamefont {Woosley}, \citenamefont {Blinnikov},\ and\ \citenamefont {Heger}}]{Woosley_2007}%
  \BibitemOpen
  \bibfield  {author} {\bibinfo {author} {\bibfnamefont {S.~E.}\ \bibnamefont {Woosley}}, \bibinfo {author} {\bibfnamefont {S.}~\bibnamefont {Blinnikov}}, \ and\ \bibinfo {author} {\bibfnamefont {Alexander}\ \bibnamefont {Heger}},\ }\bibfield  {title} {\enquote {\bibinfo {title} {Pulsational pair instability as an explanation for the most luminous supernovae},}\ }\href {\doibase 10.1038/nature06333} {\bibfield  {journal} {\bibinfo  {journal} {Nature}\ }\textbf {\bibinfo {volume} {450}},\ \bibinfo {pages} {390–392} (\bibinfo {year} {2007})}\BibitemShut {NoStop}%
\bibitem [{\citenamefont {{Woosley}}\ and\ \citenamefont {{Heger}}(2021)}]{2021ApJ...912L..31W}%
  \BibitemOpen
  \bibfield  {author} {\bibinfo {author} {\bibfnamefont {S.~E.}\ \bibnamefont {{Woosley}}}\ and\ \bibinfo {author} {\bibfnamefont {Alexander}\ \bibnamefont {{Heger}}},\ }\bibfield  {title} {\enquote {\bibinfo {title} {{The Pair-instability Mass Gap for Black Holes}},}\ }\href {\doibase 10.3847/2041-8213/abf2c4} {\bibfield  {journal} {\bibinfo  {journal} {Astrophys. J. Lett.}\ }\textbf {\bibinfo {volume} {912}},\ \bibinfo {eid} {L31} (\bibinfo {year} {2021})},\ \Eprint {http://arxiv.org/abs/2103.07933} {arXiv:2103.07933 [astro-ph.SR]} \BibitemShut {NoStop}%
\bibitem [{\citenamefont {Greene}(2012)}]{Greene_2012}%
  \BibitemOpen
  \bibfield  {author} {\bibinfo {author} {\bibfnamefont {Jenny~E.}\ \bibnamefont {Greene}},\ }\bibfield  {title} {\enquote {\bibinfo {title} {Low-mass black holes as the remnants of primordial black hole formation},}\ }\href {\doibase 10.1038/ncomms2314} {\bibfield  {journal} {\bibinfo  {journal} {Nature Communications}\ }\textbf {\bibinfo {volume} {3}} (\bibinfo {year} {2012}),\ 10.1038/ncomms2314}\BibitemShut {NoStop}%
\bibitem [{\citenamefont {Kawaguchi}\ \emph {et~al.}(2008)\citenamefont {Kawaguchi}, \citenamefont {Kawasaki}, \citenamefont {Takayama}, \citenamefont {Yamaguchi},\ and\ \citenamefont {Yokoyama}}]{10.1111/j.1365-2966.2008.13523.x}%
  \BibitemOpen
  \bibfield  {author} {\bibinfo {author} {\bibfnamefont {Toshihiro}\ \bibnamefont {Kawaguchi}}, \bibinfo {author} {\bibfnamefont {Masahiro}\ \bibnamefont {Kawasaki}}, \bibinfo {author} {\bibfnamefont {Tsutomu}\ \bibnamefont {Takayama}}, \bibinfo {author} {\bibfnamefont {Masahide}\ \bibnamefont {Yamaguchi}}, \ and\ \bibinfo {author} {\bibfnamefont {Jun'ichi}\ \bibnamefont {Yokoyama}},\ }\bibfield  {title} {\enquote {\bibinfo {title} {{Formation of intermediate-mass black holes as primordial black holes in the inflationary cosmology with running spectral index}},}\ }\href {\doibase 10.1111/j.1365-2966.2008.13523.x} {\bibfield  {journal} {\bibinfo  {journal} {Monthly Notices of the Royal Astronomical Society}\ }\textbf {\bibinfo {volume} {388}},\ \bibinfo {pages} {1426--1432} (\bibinfo {year} {2008})},\ \Eprint {http://arxiv.org/abs/https://academic.oup.com/mnras/article-pdf/388/3/1426/2819501/mnras0388-1426.pdf} {https://academic.oup.com/mnras/article-pdf/388/3/1426/2819501/mnras0388-1426.pdf} \BibitemShut
  {NoStop}%
\bibitem [{\citenamefont {Baumgarte}\ and\ \citenamefont {Shapiro}(2010)}]{baumgarte2010numerical}%
  \BibitemOpen
  \bibfield  {author} {\bibinfo {author} {\bibfnamefont {T.W.}\ \bibnamefont {Baumgarte}}\ and\ \bibinfo {author} {\bibfnamefont {S.L.}\ \bibnamefont {Shapiro}},\ }\href {https://books.google.nl/books?id=dxU1OEinvRUC} {\emph {\bibinfo {title} {Numerical Relativity: Solving Einsteins Equations on the Computer}}}\ (\bibinfo  {publisher} {Cambridge University Press},\ \bibinfo {year} {2010})\BibitemShut {NoStop}%
\bibitem [{\citenamefont {Healy}\ \emph {et~al.}(2017)\citenamefont {Healy}, \citenamefont {Lousto}, \citenamefont {Zlochower},\ and\ \citenamefont {Campanelli}}]{Healy_2017}%
  \BibitemOpen
  \bibfield  {author} {\bibinfo {author} {\bibfnamefont {James}\ \bibnamefont {Healy}}, \bibinfo {author} {\bibfnamefont {Carlos~O}\ \bibnamefont {Lousto}}, \bibinfo {author} {\bibfnamefont {Yosef}\ \bibnamefont {Zlochower}}, \ and\ \bibinfo {author} {\bibfnamefont {Manuela}\ \bibnamefont {Campanelli}},\ }\bibfield  {title} {\enquote {\bibinfo {title} {The rit binary black hole simulations catalog},}\ }\href {\doibase 10.1088/1361-6382/aa91b1} {\bibfield  {journal} {\bibinfo  {journal} {Classical and Quantum Gravity}\ }\textbf {\bibinfo {volume} {34}},\ \bibinfo {pages} {224001} (\bibinfo {year} {2017})}\BibitemShut {NoStop}%
\bibitem [{\citenamefont {Buonanno}\ and\ \citenamefont {Damour}(1999)}]{Buonanno_1999}%
  \BibitemOpen
  \bibfield  {author} {\bibinfo {author} {\bibfnamefont {A.}~\bibnamefont {Buonanno}}\ and\ \bibinfo {author} {\bibfnamefont {T.}~\bibnamefont {Damour}},\ }\bibfield  {title} {\enquote {\bibinfo {title} {Effective one-body approach to general relativistic two-body dynamics},}\ }\href {\doibase 10.1103/physrevd.59.084006} {\bibfield  {journal} {\bibinfo  {journal} {Physical Review D}\ }\textbf {\bibinfo {volume} {59}} (\bibinfo {year} {1999}),\ 10.1103/physrevd.59.084006}\BibitemShut {NoStop}%
\bibitem [{\citenamefont {Buonanno}\ \emph {et~al.}(2009)\citenamefont {Buonanno}, \citenamefont {Iyer}, \citenamefont {Ochsner}, \citenamefont {Pan},\ and\ \citenamefont {Sathyaprakash}}]{PhysRevD.80.084043}%
  \BibitemOpen
  \bibfield  {author} {\bibinfo {author} {\bibfnamefont {Alessandra}\ \bibnamefont {Buonanno}}, \bibinfo {author} {\bibfnamefont {Bala~R.}\ \bibnamefont {Iyer}}, \bibinfo {author} {\bibfnamefont {Evan}\ \bibnamefont {Ochsner}}, \bibinfo {author} {\bibfnamefont {Yi}~\bibnamefont {Pan}}, \ and\ \bibinfo {author} {\bibfnamefont {B.~S.}\ \bibnamefont {Sathyaprakash}},\ }\bibfield  {title} {\enquote {\bibinfo {title} {Comparison of post-newtonian templates for compact binary inspiral signals in gravitational-wave detectors},}\ }\href {\doibase 10.1103/PhysRevD.80.084043} {\bibfield  {journal} {\bibinfo  {journal} {Phys. Rev. D}\ }\textbf {\bibinfo {volume} {80}},\ \bibinfo {pages} {084043} (\bibinfo {year} {2009})}\BibitemShut {NoStop}%
\bibitem [{\citenamefont {Pratten}\ \emph {et~al.}(2020)\citenamefont {Pratten}, \citenamefont {Husa}, \citenamefont {Garc\'{\i}a-Quir\'os}, \citenamefont {Colleoni}, \citenamefont {Ramos-Buades}, \citenamefont {Estell\'es},\ and\ \citenamefont {Jaume}}]{PhysRevD.102.064001}%
  \BibitemOpen
  \bibfield  {author} {\bibinfo {author} {\bibfnamefont {Geraint}\ \bibnamefont {Pratten}}, \bibinfo {author} {\bibfnamefont {Sascha}\ \bibnamefont {Husa}}, \bibinfo {author} {\bibfnamefont {Cecilio}\ \bibnamefont {Garc\'{\i}a-Quir\'os}}, \bibinfo {author} {\bibfnamefont {Marta}\ \bibnamefont {Colleoni}}, \bibinfo {author} {\bibfnamefont {Antoni}\ \bibnamefont {Ramos-Buades}}, \bibinfo {author} {\bibfnamefont {H\'ector}\ \bibnamefont {Estell\'es}}, \ and\ \bibinfo {author} {\bibfnamefont {Rafel}\ \bibnamefont {Jaume}},\ }\bibfield  {title} {\enquote {\bibinfo {title} {Setting the cornerstone for a family of models for gravitational waves from compact binaries: The dominant harmonic for nonprecessing quasicircular black holes},}\ }\href {\doibase 10.1103/PhysRevD.102.064001} {\bibfield  {journal} {\bibinfo  {journal} {Phys. Rev. D}\ }\textbf {\bibinfo {volume} {102}},\ \bibinfo {pages} {064001} (\bibinfo {year} {2020})}\BibitemShut {NoStop}%
\bibitem [{\citenamefont {Garc\'\i{}a-Quir\'os}\ \emph {et~al.}(2020)\citenamefont {Garc\'\i{}a-Quir\'os}, \citenamefont {Colleoni}, \citenamefont {Husa}, \citenamefont {Estell\'es}, \citenamefont {Pratten}, \citenamefont {Ramos-Buades}, \citenamefont {Mateu-Lucena},\ and\ \citenamefont {Jaume}}]{Garcia-Quiros:2020qpx}%
  \BibitemOpen
  \bibfield  {author} {\bibinfo {author} {\bibfnamefont {Cecilio}\ \bibnamefont {Garc\'\i{}a-Quir\'os}}, \bibinfo {author} {\bibfnamefont {Marta}\ \bibnamefont {Colleoni}}, \bibinfo {author} {\bibfnamefont {Sascha}\ \bibnamefont {Husa}}, \bibinfo {author} {\bibfnamefont {H\'ector}\ \bibnamefont {Estell\'es}}, \bibinfo {author} {\bibfnamefont {Geraint}\ \bibnamefont {Pratten}}, \bibinfo {author} {\bibfnamefont {Antoni}\ \bibnamefont {Ramos-Buades}}, \bibinfo {author} {\bibfnamefont {Maite}\ \bibnamefont {Mateu-Lucena}}, \ and\ \bibinfo {author} {\bibfnamefont {Rafel}\ \bibnamefont {Jaume}},\ }\bibfield  {title} {\enquote {\bibinfo {title} {{Multimode frequency-domain model for the gravitational wave signal from nonprecessing black-hole binaries}},}\ }\href {\doibase 10.1103/PhysRevD.102.064002} {\bibfield  {journal} {\bibinfo  {journal} {Phys. Rev. D}\ }\textbf {\bibinfo {volume} {102}},\ \bibinfo {pages} {064002} (\bibinfo {year} {2020})},\ \Eprint {http://arxiv.org/abs/2001.10914} {arXiv:2001.10914 [gr-qc]}
  \BibitemShut {NoStop}%
\bibitem [{\citenamefont {Pratten}\ \emph {et~al.}(2021)\citenamefont {Pratten}, \citenamefont {Garc\'{\i}a-Quir\'os}, \citenamefont {Colleoni}, \citenamefont {Ramos-Buades}, \citenamefont {Estell\'es}, \citenamefont {Mateu-Lucena}, \citenamefont {Jaume}, \citenamefont {Haney}, \citenamefont {Keitel}, \citenamefont {Thompson},\ and\ \citenamefont {Husa}}]{PhysRevD.103.104056}%
  \BibitemOpen
  \bibfield  {author} {\bibinfo {author} {\bibfnamefont {Geraint}\ \bibnamefont {Pratten}}, \bibinfo {author} {\bibfnamefont {Cecilio}\ \bibnamefont {Garc\'{\i}a-Quir\'os}}, \bibinfo {author} {\bibfnamefont {Marta}\ \bibnamefont {Colleoni}}, \bibinfo {author} {\bibfnamefont {Antoni}\ \bibnamefont {Ramos-Buades}}, \bibinfo {author} {\bibfnamefont {H\'ector}\ \bibnamefont {Estell\'es}}, \bibinfo {author} {\bibfnamefont {Maite}\ \bibnamefont {Mateu-Lucena}}, \bibinfo {author} {\bibfnamefont {Rafel}\ \bibnamefont {Jaume}}, \bibinfo {author} {\bibfnamefont {Maria}\ \bibnamefont {Haney}}, \bibinfo {author} {\bibfnamefont {David}\ \bibnamefont {Keitel}}, \bibinfo {author} {\bibfnamefont {Jonathan~E.}\ \bibnamefont {Thompson}}, \ and\ \bibinfo {author} {\bibfnamefont {Sascha}\ \bibnamefont {Husa}},\ }\bibfield  {title} {\enquote {\bibinfo {title} {Computationally efficient models for the dominant and subdominant harmonic modes of precessing binary black holes},}\ }\href {\doibase 10.1103/PhysRevD.103.104056}
  {\bibfield  {journal} {\bibinfo  {journal} {Phys. Rev. D}\ }\textbf {\bibinfo {volume} {103}},\ \bibinfo {pages} {104056} (\bibinfo {year} {2021})}\BibitemShut {NoStop}%
\bibitem [{\citenamefont {Messina}\ \emph {et~al.}(2019)\citenamefont {Messina}, \citenamefont {Dudi}, \citenamefont {Nagar},\ and\ \citenamefont {Bernuzzi}}]{PhysRevD.99.124051}%
  \BibitemOpen
  \bibfield  {author} {\bibinfo {author} {\bibfnamefont {Francesco}\ \bibnamefont {Messina}}, \bibinfo {author} {\bibfnamefont {Reetika}\ \bibnamefont {Dudi}}, \bibinfo {author} {\bibfnamefont {Alessandro}\ \bibnamefont {Nagar}}, \ and\ \bibinfo {author} {\bibfnamefont {Sebastiano}\ \bibnamefont {Bernuzzi}},\ }\bibfield  {title} {\enquote {\bibinfo {title} {Quasi-5.5pn taylorf2 approximant for compact binaries: Point-mass phasing and impact on the tidal polarizability inference},}\ }\href {\doibase 10.1103/PhysRevD.99.124051} {\bibfield  {journal} {\bibinfo  {journal} {Phys. Rev. D}\ }\textbf {\bibinfo {volume} {99}},\ \bibinfo {pages} {124051} (\bibinfo {year} {2019})}\BibitemShut {NoStop}%
\bibitem [{\citenamefont {Gamba}\ \emph {et~al.}(2022)\citenamefont {Gamba}, \citenamefont {Ak\ifmmode~\mbox{\c{c}}\else \c{c}\fi{}ay}, \citenamefont {Bernuzzi},\ and\ \citenamefont {Williams}}]{PhysRevD.106.024020}%
  \BibitemOpen
  \bibfield  {author} {\bibinfo {author} {\bibfnamefont {Rossella}\ \bibnamefont {Gamba}}, \bibinfo {author} {\bibfnamefont {Sarp}\ \bibnamefont {Ak\ifmmode~\mbox{\c{c}}\else \c{c}\fi{}ay}}, \bibinfo {author} {\bibfnamefont {Sebastiano}\ \bibnamefont {Bernuzzi}}, \ and\ \bibinfo {author} {\bibfnamefont {Jake}\ \bibnamefont {Williams}},\ }\bibfield  {title} {\enquote {\bibinfo {title} {Effective-one-body waveforms for precessing coalescing compact binaries with post-newtonian twist},}\ }\href {\doibase 10.1103/PhysRevD.106.024020} {\bibfield  {journal} {\bibinfo  {journal} {Phys. Rev. D}\ }\textbf {\bibinfo {volume} {106}},\ \bibinfo {pages} {024020} (\bibinfo {year} {2022})}\BibitemShut {NoStop}%
\bibitem [{\citenamefont {Estellés}\ \emph {et~al.}(2021)\citenamefont {Estellés}, \citenamefont {Ramos-Buades}, \citenamefont {Husa}, \citenamefont {García-Quirós}, \citenamefont {Colleoni}, \citenamefont {Haegel},\ and\ \citenamefont {Jaume}}]{Estell_s_2021}%
  \BibitemOpen
  \bibfield  {author} {\bibinfo {author} {\bibfnamefont {Héctor}\ \bibnamefont {Estellés}}, \bibinfo {author} {\bibfnamefont {Antoni}\ \bibnamefont {Ramos-Buades}}, \bibinfo {author} {\bibfnamefont {Sascha}\ \bibnamefont {Husa}}, \bibinfo {author} {\bibfnamefont {Cecilio}\ \bibnamefont {García-Quirós}}, \bibinfo {author} {\bibfnamefont {Marta}\ \bibnamefont {Colleoni}}, \bibinfo {author} {\bibfnamefont {Leïla}\ \bibnamefont {Haegel}}, \ and\ \bibinfo {author} {\bibfnamefont {Rafel}\ \bibnamefont {Jaume}},\ }\bibfield  {title} {\enquote {\bibinfo {title} {Phenomenological time domain model for dominant quadrupole gravitational wave signal of coalescing binary black holes},}\ }\href {\doibase 10.1103/physrevd.103.124060} {\bibfield  {journal} {\bibinfo  {journal} {Physical Review D}\ }\textbf {\bibinfo {volume} {103}} (\bibinfo {year} {2021}),\ 10.1103/physrevd.103.124060}\BibitemShut {NoStop}%
\bibitem [{\citenamefont {Aasi}\ \emph {et~al.}(2014)\citenamefont {Aasi}, \citenamefont {Abbott}, \citenamefont {Abbott}, \citenamefont {Abbott}, \citenamefont {Abernathy}, \citenamefont {Acernese}, \citenamefont {Ackley}, \citenamefont {Adams}, \citenamefont {Adams}, \citenamefont {Addesso}, \citenamefont {Adhikari}, \citenamefont {Affeldt}, \citenamefont {Agathos}, \citenamefont {Aggarwal}, \citenamefont {Aguiar}, \citenamefont {Ain}, \citenamefont {Ajith}, \citenamefont {Alemic},\ and\ \citenamefont {Allen}}]{Aasi_2014}%
  \BibitemOpen
  \bibfield  {author} {\bibinfo {author} {\bibfnamefont {J.}~\bibnamefont {Aasi}}, \bibinfo {author} {\bibfnamefont {B.P.}\ \bibnamefont {Abbott}}, \bibinfo {author} {\bibfnamefont {R.}~\bibnamefont {Abbott}}, \bibinfo {author} {\bibfnamefont {T.}~\bibnamefont {Abbott}}, \bibinfo {author} {\bibfnamefont {M.R.}\ \bibnamefont {Abernathy}}, \bibinfo {author} {\bibfnamefont {F.}~\bibnamefont {Acernese}}, \bibinfo {author} {\bibfnamefont {K.}~\bibnamefont {Ackley}}, \bibinfo {author} {\bibfnamefont {C.}~\bibnamefont {Adams}}, \bibinfo {author} {\bibfnamefont {T.}~\bibnamefont {Adams}}, \bibinfo {author} {\bibfnamefont {P.}~\bibnamefont {Addesso}}, \bibinfo {author} {\bibfnamefont {R.X.}\ \bibnamefont {Adhikari}}, \bibinfo {author} {\bibfnamefont {C.}~\bibnamefont {Affeldt}}, \bibinfo {author} {\bibfnamefont {M.}~\bibnamefont {Agathos}}, \bibinfo {author} {\bibfnamefont {N.}~\bibnamefont {Aggarwal}}, \bibinfo {author} {\bibfnamefont {O.D.}\ \bibnamefont {Aguiar}}, \bibinfo {author} {\bibfnamefont {A.}~\bibnamefont
  {Ain}}, \bibinfo {author} {\bibfnamefont {P.}~\bibnamefont {Ajith}}, \bibinfo {author} {\bibfnamefont {A.}~\bibnamefont {Alemic}}, \ and\ \bibinfo {author} {\bibfnamefont {B.}~\bibnamefont {Allen}},\ }\bibfield  {title} {\enquote {\bibinfo {title} {Search for gravitational wave ringdowns from perturbed intermediate mass black holes in ligo-virgo data from 2005–2010},}\ }\href {\doibase 10.1103/physrevd.89.102006} {\bibfield  {journal} {\bibinfo  {journal} {Physical Review D}\ }\textbf {\bibinfo {volume} {89}} (\bibinfo {year} {2014}),\ 10.1103/physrevd.89.102006}\BibitemShut {NoStop}%
\bibitem [{\citenamefont {Collaboration}\ and\ \citenamefont {Collaboration}(2016)}]{Abbott_2016}%
  \BibitemOpen
  \bibfield  {author} {\bibinfo {author} {\bibfnamefont {LIGO~Scientific}\ \bibnamefont {Collaboration}}\ and\ \bibinfo {author} {\bibfnamefont {Virgo}\ \bibnamefont {Collaboration}},\ }\bibfield  {title} {\enquote {\bibinfo {title} {Characterization of transient noise in advanced ligo relevant to gravitational wave signal gw150914},}\ }\href {\doibase 10.1088/0264-9381/33/13/134001} {\bibfield  {journal} {\bibinfo  {journal} {Classical and Quantum Gravity}\ }\textbf {\bibinfo {volume} {33}},\ \bibinfo {pages} {134001} (\bibinfo {year} {2016})}\BibitemShut {NoStop}%
\bibitem [{\citenamefont {Drago}\ \emph {et~al.}(2021)\citenamefont {Drago}, \citenamefont {Gayathri}, \citenamefont {Klimenko}, \citenamefont {Lazzaro}, \citenamefont {Milotti}, \citenamefont {Mitselmakher}, \citenamefont {Necula}, \citenamefont {O'Brian}, \citenamefont {Prodi}, \citenamefont {Salemi}, \citenamefont {Szczepanczyk}, \citenamefont {Tiwari}, \citenamefont {Tiwari}, \citenamefont {Vedovato},\ and\ \citenamefont {Yakushin}}]{drago2021coherent}%
  \BibitemOpen
  \bibfield  {author} {\bibinfo {author} {\bibfnamefont {M.}~\bibnamefont {Drago}}, \bibinfo {author} {\bibfnamefont {V.}~\bibnamefont {Gayathri}}, \bibinfo {author} {\bibfnamefont {S.}~\bibnamefont {Klimenko}}, \bibinfo {author} {\bibfnamefont {C.}~\bibnamefont {Lazzaro}}, \bibinfo {author} {\bibfnamefont {E.}~\bibnamefont {Milotti}}, \bibinfo {author} {\bibfnamefont {G.}~\bibnamefont {Mitselmakher}}, \bibinfo {author} {\bibfnamefont {V.}~\bibnamefont {Necula}}, \bibinfo {author} {\bibfnamefont {B.}~\bibnamefont {O'Brian}}, \bibinfo {author} {\bibfnamefont {G.~A.}\ \bibnamefont {Prodi}}, \bibinfo {author} {\bibfnamefont {F.}~\bibnamefont {Salemi}}, \bibinfo {author} {\bibfnamefont {M.}~\bibnamefont {Szczepanczyk}}, \bibinfo {author} {\bibfnamefont {S.}~\bibnamefont {Tiwari}}, \bibinfo {author} {\bibfnamefont {V.}~\bibnamefont {Tiwari}}, \bibinfo {author} {\bibfnamefont {G.}~\bibnamefont {Vedovato}}, \ and\ \bibinfo {author} {\bibfnamefont {I.}~\bibnamefont {Yakushin}},\ }\href@noop {} {\enquote {\bibinfo
  {title} {Coherent waveburst, a pipeline for unmodeled gravitational-wave data analysis},}\ } (\bibinfo {year} {2021}),\ \Eprint {http://arxiv.org/abs/2006.12604} {arXiv:2006.12604 [gr-qc]} \BibitemShut {NoStop}%
\bibitem [{\citenamefont {Cannon}\ \emph {et~al.}(2020)\citenamefont {Cannon}, \citenamefont {Caudill}, \citenamefont {Chan}, \citenamefont {Cousins}, \citenamefont {Creighton}, \citenamefont {Ewing}, \citenamefont {Fong}, \citenamefont {Godwin}, \citenamefont {Hanna}, \citenamefont {Hooper}, \citenamefont {Huxford}, \citenamefont {Magee}, \citenamefont {Meacher}, \citenamefont {Messick}, \citenamefont {Morisaki}, \citenamefont {Mukherjee}, \citenamefont {Ohta}, \citenamefont {Pace}, \citenamefont {Privitera}, \citenamefont {de~Ruiter}, \citenamefont {Sachdev}, \citenamefont {Singer}, \citenamefont {Singh}, \citenamefont {Tapia}, \citenamefont {Tsukada}, \citenamefont {Tsuna}, \citenamefont {Tsutsui}, \citenamefont {Ueno}, \citenamefont {Viets}, \citenamefont {Wade},\ and\ \citenamefont {Wade}}]{cannon2020gstlal}%
  \BibitemOpen
  \bibfield  {author} {\bibinfo {author} {\bibfnamefont {Kipp}\ \bibnamefont {Cannon}}, \bibinfo {author} {\bibfnamefont {Sarah}\ \bibnamefont {Caudill}}, \bibinfo {author} {\bibfnamefont {Chiwai}\ \bibnamefont {Chan}}, \bibinfo {author} {\bibfnamefont {Bryce}\ \bibnamefont {Cousins}}, \bibinfo {author} {\bibfnamefont {Jolien D.~E.}\ \bibnamefont {Creighton}}, \bibinfo {author} {\bibfnamefont {Becca}\ \bibnamefont {Ewing}}, \bibinfo {author} {\bibfnamefont {Heather}\ \bibnamefont {Fong}}, \bibinfo {author} {\bibfnamefont {Patrick}\ \bibnamefont {Godwin}}, \bibinfo {author} {\bibfnamefont {Chad}\ \bibnamefont {Hanna}}, \bibinfo {author} {\bibfnamefont {Shaun}\ \bibnamefont {Hooper}}, \bibinfo {author} {\bibfnamefont {Rachael}\ \bibnamefont {Huxford}}, \bibinfo {author} {\bibfnamefont {Ryan}\ \bibnamefont {Magee}}, \bibinfo {author} {\bibfnamefont {Duncan}\ \bibnamefont {Meacher}}, \bibinfo {author} {\bibfnamefont {Cody}\ \bibnamefont {Messick}}, \bibinfo {author} {\bibfnamefont {Soichiro}\ \bibnamefont
  {Morisaki}}, \bibinfo {author} {\bibfnamefont {Debnandini}\ \bibnamefont {Mukherjee}}, \bibinfo {author} {\bibfnamefont {Hiroaki}\ \bibnamefont {Ohta}}, \bibinfo {author} {\bibfnamefont {Alexander}\ \bibnamefont {Pace}}, \bibinfo {author} {\bibfnamefont {Stephen}\ \bibnamefont {Privitera}}, \bibinfo {author} {\bibfnamefont {Iris}\ \bibnamefont {de~Ruiter}}, \bibinfo {author} {\bibfnamefont {Surabhi}\ \bibnamefont {Sachdev}}, \bibinfo {author} {\bibfnamefont {Leo}\ \bibnamefont {Singer}}, \bibinfo {author} {\bibfnamefont {Divya}\ \bibnamefont {Singh}}, \bibinfo {author} {\bibfnamefont {Ron}\ \bibnamefont {Tapia}}, \bibinfo {author} {\bibfnamefont {Leo}\ \bibnamefont {Tsukada}}, \bibinfo {author} {\bibfnamefont {Daichi}\ \bibnamefont {Tsuna}}, \bibinfo {author} {\bibfnamefont {Takuya}\ \bibnamefont {Tsutsui}}, \bibinfo {author} {\bibfnamefont {Koh}\ \bibnamefont {Ueno}}, \bibinfo {author} {\bibfnamefont {Aaron}\ \bibnamefont {Viets}}, \bibinfo {author} {\bibfnamefont {Leslie}\ \bibnamefont {Wade}}, \ and\
  \bibinfo {author} {\bibfnamefont {Madeline}\ \bibnamefont {Wade}},\ }\href@noop {} {\enquote {\bibinfo {title} {Gstlal: A software framework for gravitational wave discovery},}\ } (\bibinfo {year} {2020}),\ \Eprint {http://arxiv.org/abs/2010.05082} {arXiv:2010.05082 [astro-ph.IM]} \BibitemShut {NoStop}%
\bibitem [{\citenamefont {Usman}\ \emph {et~al.}(2016)\citenamefont {Usman}, \citenamefont {Nitz}, \citenamefont {Harry}, \citenamefont {Biwer}, \citenamefont {Brown}, \citenamefont {Cabero}, \citenamefont {Capano}, \citenamefont {Canton}, \citenamefont {Dent}, \citenamefont {Fairhurst}, \citenamefont {Kehl}, \citenamefont {Keppel}, \citenamefont {Krishnan}, \citenamefont {Lenon}, \citenamefont {Lundgren}, \citenamefont {Nielsen}, \citenamefont {Pekowsky}, \citenamefont {Pfeiffer}, \citenamefont {Saulson}, \citenamefont {West},\ and\ \citenamefont {Willis}}]{Usman_2016}%
  \BibitemOpen
  \bibfield  {author} {\bibinfo {author} {\bibfnamefont {Samantha~A}\ \bibnamefont {Usman}}, \bibinfo {author} {\bibfnamefont {Alexander~H}\ \bibnamefont {Nitz}}, \bibinfo {author} {\bibfnamefont {Ian~W}\ \bibnamefont {Harry}}, \bibinfo {author} {\bibfnamefont {Christopher~M}\ \bibnamefont {Biwer}}, \bibinfo {author} {\bibfnamefont {Duncan~A}\ \bibnamefont {Brown}}, \bibinfo {author} {\bibfnamefont {Miriam}\ \bibnamefont {Cabero}}, \bibinfo {author} {\bibfnamefont {Collin~D}\ \bibnamefont {Capano}}, \bibinfo {author} {\bibfnamefont {Tito~Dal}\ \bibnamefont {Canton}}, \bibinfo {author} {\bibfnamefont {Thomas}\ \bibnamefont {Dent}}, \bibinfo {author} {\bibfnamefont {Stephen}\ \bibnamefont {Fairhurst}}, \bibinfo {author} {\bibfnamefont {Marcel~S}\ \bibnamefont {Kehl}}, \bibinfo {author} {\bibfnamefont {Drew}\ \bibnamefont {Keppel}}, \bibinfo {author} {\bibfnamefont {Badri}\ \bibnamefont {Krishnan}}, \bibinfo {author} {\bibfnamefont {Amber}\ \bibnamefont {Lenon}}, \bibinfo {author} {\bibfnamefont {Andrew}\
  \bibnamefont {Lundgren}}, \bibinfo {author} {\bibfnamefont {Alex~B}\ \bibnamefont {Nielsen}}, \bibinfo {author} {\bibfnamefont {Larne~P}\ \bibnamefont {Pekowsky}}, \bibinfo {author} {\bibfnamefont {Harald~P}\ \bibnamefont {Pfeiffer}}, \bibinfo {author} {\bibfnamefont {Peter~R}\ \bibnamefont {Saulson}}, \bibinfo {author} {\bibfnamefont {Matthew}\ \bibnamefont {West}}, \ and\ \bibinfo {author} {\bibfnamefont {Joshua~L}\ \bibnamefont {Willis}},\ }\bibfield  {title} {\enquote {\bibinfo {title} {The pycbc search for gravitational waves from compact binary coalescence},}\ }\href {\doibase 10.1088/0264-9381/33/21/215004} {\bibfield  {journal} {\bibinfo  {journal} {Classical and Quantum Gravity}\ }\textbf {\bibinfo {volume} {33}},\ \bibinfo {pages} {215004} (\bibinfo {year} {2016})}\BibitemShut {NoStop}%
\bibitem [{\citenamefont {Gabbard}\ \emph {et~al.}(2018)\citenamefont {Gabbard}, \citenamefont {Williams}, \citenamefont {Hayes},\ and\ \citenamefont {Messenger}}]{PhysRevLett.120.141103}%
  \BibitemOpen
  \bibfield  {author} {\bibinfo {author} {\bibfnamefont {Hunter}\ \bibnamefont {Gabbard}}, \bibinfo {author} {\bibfnamefont {Michael}\ \bibnamefont {Williams}}, \bibinfo {author} {\bibfnamefont {Fergus}\ \bibnamefont {Hayes}}, \ and\ \bibinfo {author} {\bibfnamefont {Chris}\ \bibnamefont {Messenger}},\ }\bibfield  {title} {\enquote {\bibinfo {title} {Matching matched filtering with deep networks for gravitational-wave astronomy},}\ }\href {\doibase 10.1103/PhysRevLett.120.141103} {\bibfield  {journal} {\bibinfo  {journal} {Phys. Rev. Lett.}\ }\textbf {\bibinfo {volume} {120}},\ \bibinfo {pages} {141103} (\bibinfo {year} {2018})}\BibitemShut {NoStop}%
\bibitem [{\citenamefont {Ma}\ \emph {et~al.}(2024)\citenamefont {Ma}, \citenamefont {Wang}, \citenamefont {Wang},\ and\ \citenamefont {Cao}}]{PhysRevD.109.043009}%
  \BibitemOpen
  \bibfield  {author} {\bibinfo {author} {\bibfnamefont {CunLiang}\ \bibnamefont {Ma}}, \bibinfo {author} {\bibfnamefont {Sen}\ \bibnamefont {Wang}}, \bibinfo {author} {\bibfnamefont {Wei}\ \bibnamefont {Wang}}, \ and\ \bibinfo {author} {\bibfnamefont {Zhoujian}\ \bibnamefont {Cao}},\ }\bibfield  {title} {\enquote {\bibinfo {title} {Using deep learning to predict matched signal-to-noise ratio of gravitational waves},}\ }\href {\doibase 10.1103/PhysRevD.109.043009} {\bibfield  {journal} {\bibinfo  {journal} {Phys. Rev. D}\ }\textbf {\bibinfo {volume} {109}},\ \bibinfo {pages} {043009} (\bibinfo {year} {2024})}\BibitemShut {NoStop}%
\bibitem [{\citenamefont {{Luo}}\ \emph {et~al.}(2020)\citenamefont {{Luo}}, \citenamefont {{Lin}}, \citenamefont {{Chen}},\ and\ \citenamefont {{Huang}}}]{2020FrPhy..1514601L}%
  \BibitemOpen
  \bibfield  {author} {\bibinfo {author} {\bibfnamefont {Hua-Mei}\ \bibnamefont {{Luo}}}, \bibinfo {author} {\bibfnamefont {Wenbin}\ \bibnamefont {{Lin}}}, \bibinfo {author} {\bibfnamefont {Zu-Cheng}\ \bibnamefont {{Chen}}}, \ and\ \bibinfo {author} {\bibfnamefont {Qing-Guo}\ \bibnamefont {{Huang}}},\ }\bibfield  {title} {\enquote {\bibinfo {title} {{Extraction of gravitational wave signals with optimized convolutional neural network}},}\ }\href {\doibase 10.1007/s11467-019-0936-x} {\bibfield  {journal} {\bibinfo  {journal} {Frontiers of Physics}\ }\textbf {\bibinfo {volume} {15}},\ \bibinfo {eid} {14601} (\bibinfo {year} {2020})}\BibitemShut {NoStop}%
\bibitem [{\citenamefont {Skliris}\ \emph {et~al.}(2024)\citenamefont {Skliris}, \citenamefont {Norman},\ and\ \citenamefont {Sutton}}]{skliris2024realtimedetectionunmodelledgravitationalwave}%
  \BibitemOpen
  \bibfield  {author} {\bibinfo {author} {\bibfnamefont {Vasileios}\ \bibnamefont {Skliris}}, \bibinfo {author} {\bibfnamefont {Michael R.~K.}\ \bibnamefont {Norman}}, \ and\ \bibinfo {author} {\bibfnamefont {Patrick~J.}\ \bibnamefont {Sutton}},\ }\href {https://arxiv.org/abs/2009.14611} {\enquote {\bibinfo {title} {Real-time detection of unmodelled gravitational-wave transients using convolutional neural networks},}\ } (\bibinfo {year} {2024}),\ \Eprint {http://arxiv.org/abs/2009.14611} {arXiv:2009.14611 [astro-ph.IM]} \BibitemShut {NoStop}%
\bibitem [{\citenamefont {et~al.}(2016)}]{WaveNet}%
  \BibitemOpen
  \bibfield  {author} {\bibinfo {author} {\bibfnamefont {A.~{van den Oord}}\ \bibnamefont {et~al.}},\ }\bibfield  {title} {\enquote {\bibinfo {title} {{WaveNet: A Generative Model for Raw Audio}},}\ }\href {\doibase 10.48550/arXiv.1609.03499} {\bibfield  {journal} {\bibinfo  {journal} {arXiv e-prints}\ ,\ \bibinfo {eid} {arXiv:1609.03499}} (\bibinfo {year} {2016})},\ \Eprint {http://arxiv.org/abs/1609.03499} {arXiv:1609.03499 [cs.SD]} \BibitemShut {NoStop}%
\bibitem [{\citenamefont {Graves}\ \emph {et~al.}(2013)\citenamefont {Graves}, \citenamefont {rahman Mohamed},\ and\ \citenamefont {Hinton}}]{graves2013speech}%
  \BibitemOpen
  \bibfield  {author} {\bibinfo {author} {\bibfnamefont {Alex}\ \bibnamefont {Graves}}, \bibinfo {author} {\bibfnamefont {Abdel}\ \bibnamefont {rahman Mohamed}}, \ and\ \bibinfo {author} {\bibfnamefont {Geoffrey}\ \bibnamefont {Hinton}},\ }\href@noop {} {\enquote {\bibinfo {title} {Speech recognition with deep recurrent neural networks},}\ } (\bibinfo {year} {2013}),\ \Eprint {http://arxiv.org/abs/1303.5778} {arXiv:1303.5778 [cs.NE]} \BibitemShut {NoStop}%
\bibitem [{\citenamefont {et~al.}(2019{\natexlab{b}})}]{PyTorch}%
  \BibitemOpen
  \bibfield  {author} {\bibinfo {author} {\bibfnamefont {A.~Paszke}\ \bibnamefont {et~al.}},\ }\bibfield  {title} {\enquote {\bibinfo {title} {Pytorch: An imperative style, high-performance deep learning library},}\ }in\ \href {http://papers.neurips.cc/paper/9015-pytorch-an-imperative-style-high-performance-deep-learning-library.pdf} {\emph {\bibinfo {booktitle} {Advances in Neural Information Processing Systems 32}}}\ (\bibinfo  {publisher} {Curran Associates, Inc.},\ \bibinfo {year} {2019})\ pp.\ \bibinfo {pages} {8024--8035}\BibitemShut {NoStop}%
\bibitem [{\citenamefont {Bengio}\ \emph {et~al.}(2009)\citenamefont {Bengio}, \citenamefont {Louradour}, \citenamefont {Collobert},\ and\ \citenamefont {Weston}}]{curriculum}%
  \BibitemOpen
  \bibfield  {author} {\bibinfo {author} {\bibfnamefont {Yoshua}\ \bibnamefont {Bengio}}, \bibinfo {author} {\bibfnamefont {J\'{e}r\^{o}me}\ \bibnamefont {Louradour}}, \bibinfo {author} {\bibfnamefont {Ronan}\ \bibnamefont {Collobert}}, \ and\ \bibinfo {author} {\bibfnamefont {Jason}\ \bibnamefont {Weston}},\ }\bibfield  {title} {\enquote {\bibinfo {title} {Curriculum learning},}\ }in\ \href {\doibase 10.1145/1553374.1553380} {\emph {\bibinfo {booktitle} {Proceedings of the 26th Annual International Conference on Machine Learning}}},\ \bibinfo {series and number} {ICML '09}\ (\bibinfo  {publisher} {Association for Computing Machinery},\ \bibinfo {address} {New York, NY, USA},\ \bibinfo {year} {2009})\ p.\ \bibinfo {pages} {41–48}\BibitemShut {NoStop}%
\bibitem [{\citenamefont {Pan}\ and\ \citenamefont {Yang}(2010)}]{transfer}%
  \BibitemOpen
  \bibfield  {author} {\bibinfo {author} {\bibfnamefont {Sinno~Jialin}\ \bibnamefont {Pan}}\ and\ \bibinfo {author} {\bibfnamefont {Qiang}\ \bibnamefont {Yang}},\ }\bibfield  {title} {\enquote {\bibinfo {title} {A survey on transfer learning},}\ }\href {\doibase 10.1109/TKDE.2009.191} {\bibfield  {journal} {\bibinfo  {journal} {IEEE Transactions on Knowledge and Data Engineering}\ }\textbf {\bibinfo {volume} {22}},\ \bibinfo {pages} {1345--1359} (\bibinfo {year} {2010})}\BibitemShut {NoStop}%
\bibitem [{\citenamefont {{LIGO Scientific Collaboration}}\ \emph {et~al.}(2018)\citenamefont {{LIGO Scientific Collaboration}}, \citenamefont {{Virgo Collaboration}},\ and\ \citenamefont {{KAGRA Collaboration}}}]{lalsuite}%
  \BibitemOpen
  \bibfield  {author} {\bibinfo {author} {\bibnamefont {{LIGO Scientific Collaboration}}}, \bibinfo {author} {\bibnamefont {{Virgo Collaboration}}}, \ and\ \bibinfo {author} {\bibnamefont {{KAGRA Collaboration}}},\ }\href {\doibase 10.7935/GT1W-FZ16} {\enquote {\bibinfo {title} {{LVK} {A}lgorithm {L}ibrary - {LALS}uite},}\ }\bibinfo {howpublished} {Free software (GPL)} (\bibinfo {year} {2018})\BibitemShut {NoStop}%
\bibitem [{\citenamefont {Lopez}\ and\ \citenamefont {Schmidt}(2022)}]{gengliweb}%
  \BibitemOpen
  \bibfield  {author} {\bibinfo {author} {\bibfnamefont {M.}~\bibnamefont {Lopez}}\ and\ \bibinfo {author} {\bibfnamefont {S.}~\bibnamefont {Schmidt}},\ }\href@noop {} {\enquote {\bibinfo {title} {{Documentation of the gengli Package}},}\ }\bibinfo {howpublished} {\url{https://melissa.lopez.docs.ligo.org/gengli/index.html}} (\bibinfo {year} {2022})\BibitemShut {NoStop}%
\bibitem [{\citenamefont {et~al.}(2023)}]{pyCBC}%
  \BibitemOpen
  \bibfield  {author} {\bibinfo {author} {\bibfnamefont {A.~Nitz}\ \bibnamefont {et~al.}},\ }\href {\doibase 10.5281/zenodo.7547919} {\enquote {\bibinfo {title} {gwastro/pycbc: v2.0.6 release of pycbc},}\ } (\bibinfo {year} {2023})\BibitemShut {NoStop}%
\bibitem [{O3A(2022)}]{O3ASD}%
  \BibitemOpen
  \href@noop {} {\enquote {\bibinfo {title} {Noise curves used for simulations in the update of the observing scenarios paper},}\ }\bibinfo {howpublished} {\url{https://dcc.ligo.org/LIGO-T2000012/public}} (\bibinfo {year} {2022})\BibitemShut {NoStop}%
\bibitem [{\citenamefont {Abbott}\ \emph {et~al.}(2020{\natexlab{b}})\citenamefont {Abbott}, \citenamefont {Abbott}, \citenamefont {Abbott}, \citenamefont {Abraham}, \citenamefont {Acernese}, \citenamefont {Ackley}, \citenamefont {Adams}, \citenamefont {Adya}, \citenamefont {Affeldt}, \citenamefont {Agathos}, \citenamefont {Agatsuma}, \citenamefont {Aggarwal}, \citenamefont {Aguiar}, \citenamefont {Aiello}, \citenamefont {Ain}, \citenamefont {Ajith}, \citenamefont {Akutsu},\ and\ \citenamefont {Allen}}]{Abbott_2020b}%
  \BibitemOpen
  \bibfield  {author} {\bibinfo {author} {\bibfnamefont {B.~P.}\ \bibnamefont {Abbott}}, \bibinfo {author} {\bibfnamefont {R.}~\bibnamefont {Abbott}}, \bibinfo {author} {\bibfnamefont {T.~D.}\ \bibnamefont {Abbott}}, \bibinfo {author} {\bibfnamefont {S.}~\bibnamefont {Abraham}}, \bibinfo {author} {\bibfnamefont {F.}~\bibnamefont {Acernese}}, \bibinfo {author} {\bibfnamefont {K.}~\bibnamefont {Ackley}}, \bibinfo {author} {\bibfnamefont {C.}~\bibnamefont {Adams}}, \bibinfo {author} {\bibfnamefont {V.~B.}\ \bibnamefont {Adya}}, \bibinfo {author} {\bibfnamefont {C.}~\bibnamefont {Affeldt}}, \bibinfo {author} {\bibfnamefont {M.}~\bibnamefont {Agathos}}, \bibinfo {author} {\bibfnamefont {K.}~\bibnamefont {Agatsuma}}, \bibinfo {author} {\bibfnamefont {N.}~\bibnamefont {Aggarwal}}, \bibinfo {author} {\bibfnamefont {O.~D.}\ \bibnamefont {Aguiar}}, \bibinfo {author} {\bibfnamefont {L.}~\bibnamefont {Aiello}}, \bibinfo {author} {\bibfnamefont {A.}~\bibnamefont {Ain}}, \bibinfo {author} {\bibfnamefont {P.}~\bibnamefont
  {Ajith}}, \bibinfo {author} {\bibfnamefont {T.}~\bibnamefont {Akutsu}}, \ and\ \bibinfo {author} {\bibfnamefont {G.}~\bibnamefont {Allen}},\ }\bibfield  {title} {\enquote {\bibinfo {title} {Prospects for observing and localizing gravitational-wave transients with advanced ligo, advanced virgo and kagra},}\ }\href {\doibase 10.1007/s41114-020-00026-9} {\bibfield  {journal} {\bibinfo  {journal} {Living Reviews in Relativity}\ }\textbf {\bibinfo {volume} {23}} (\bibinfo {year} {2020}{\natexlab{b}}),\ 10.1007/s41114-020-00026-9}\BibitemShut {NoStop}%
\bibitem [{\citenamefont {{Loshchilov}}\ and\ \citenamefont {{Hutter}}(2017)}]{AdamW}%
  \BibitemOpen
  \bibfield  {author} {\bibinfo {author} {\bibfnamefont {I.}~\bibnamefont {{Loshchilov}}}\ and\ \bibinfo {author} {\bibfnamefont {F.}~\bibnamefont {{Hutter}}},\ }\bibfield  {title} {\enquote {\bibinfo {title} {{Decoupled Weight Decay Regularization}},}\ }\href {\doibase 10.48550/arXiv.1711.05101} {\bibfield  {journal} {\bibinfo  {journal} {arXiv e-prints}\ ,\ \bibinfo {eid} {arXiv:1711.05101}} (\bibinfo {year} {2017})},\ \Eprint {http://arxiv.org/abs/1711.05101} {arXiv:1711.05101 [cs.LG]} \BibitemShut {NoStop}%
\bibitem [{\citenamefont {Hastie}\ \emph {et~al.}(2009)\citenamefont {Hastie}, \citenamefont {Tibshirani},\ and\ \citenamefont {Friedman}}]{hastie2009elements}%
  \BibitemOpen
  \bibfield  {author} {\bibinfo {author} {\bibfnamefont {T.}~\bibnamefont {Hastie}}, \bibinfo {author} {\bibfnamefont {R.}~\bibnamefont {Tibshirani}}, \ and\ \bibinfo {author} {\bibfnamefont {J.H.}\ \bibnamefont {Friedman}},\ }\href {https://books.google.nl/books?id=eBSgoAEACAAJ} {\emph {\bibinfo {title} {The Elements of Statistical Learning: Data Mining, Inference, and Prediction}}},\ Springer series in statistics\ (\bibinfo  {publisher} {Springer},\ \bibinfo {year} {2009})\BibitemShut {NoStop}%
\bibitem [{\citenamefont {Pedregosa}\ \emph {et~al.}(2011)\citenamefont {Pedregosa}, \citenamefont {Varoquaux}, \citenamefont {Gramfort}, \citenamefont {Michel}, \citenamefont {Thirion}, \citenamefont {Grisel}, \citenamefont {Blondel}, \citenamefont {Prettenhofer}, \citenamefont {Weiss}, \citenamefont {Dubourg} \emph {et~al.}}]{pedregosa2011scikit}%
  \BibitemOpen
  \bibfield  {author} {\bibinfo {author} {\bibfnamefont {Fabian}\ \bibnamefont {Pedregosa}}, \bibinfo {author} {\bibfnamefont {Ga{\"e}l}\ \bibnamefont {Varoquaux}}, \bibinfo {author} {\bibfnamefont {Alexandre}\ \bibnamefont {Gramfort}}, \bibinfo {author} {\bibfnamefont {Vincent}\ \bibnamefont {Michel}}, \bibinfo {author} {\bibfnamefont {Bertrand}\ \bibnamefont {Thirion}}, \bibinfo {author} {\bibfnamefont {Olivier}\ \bibnamefont {Grisel}}, \bibinfo {author} {\bibfnamefont {Mathieu}\ \bibnamefont {Blondel}}, \bibinfo {author} {\bibfnamefont {Peter}\ \bibnamefont {Prettenhofer}}, \bibinfo {author} {\bibfnamefont {Ron}\ \bibnamefont {Weiss}}, \bibinfo {author} {\bibfnamefont {Vincent}\ \bibnamefont {Dubourg}},  \emph {et~al.},\ }\bibfield  {title} {\enquote {\bibinfo {title} {Scikit-learn: Machine learning in python},}\ }\href@noop {} {\bibfield  {journal} {\bibinfo  {journal} {Journal of machine learning research}\ }\textbf {\bibinfo {volume} {12}},\ \bibinfo {pages} {2825--2830} (\bibinfo {year}
  {2011})}\BibitemShut {NoStop}%
\bibitem [{\citenamefont {van~der Maaten}\ and\ \citenamefont {Hinton}(2008)}]{vanDerMaaten2008}%
  \BibitemOpen
  \bibfield  {author} {\bibinfo {author} {\bibfnamefont {Laurens}\ \bibnamefont {van~der Maaten}}\ and\ \bibinfo {author} {\bibfnamefont {Geoffrey}\ \bibnamefont {Hinton}},\ }\bibfield  {title} {\enquote {\bibinfo {title} {Visualizing data using {t-SNE}},}\ }\href {http://www.jmlr.org/papers/v9/vandermaaten08a.html} {\bibfield  {journal} {\bibinfo  {journal} {Journal of Machine Learning Research}\ }\textbf {\bibinfo {volume} {9}},\ \bibinfo {pages} {2579--2605} (\bibinfo {year} {2008})}\BibitemShut {NoStop}%
\bibitem [{\citenamefont {Lee}\ and\ \citenamefont {Verleysen}(2007)}]{nldrbook}%
  \BibitemOpen
  \bibfield  {author} {\bibinfo {author} {\bibfnamefont {John~A.}\ \bibnamefont {Lee}}\ and\ \bibinfo {author} {\bibfnamefont {Michel}\ \bibnamefont {Verleysen}},\ }\href@noop {} {\emph {\bibinfo {title} {Nonlinear Dimensionality Reduction}}},\ \bibinfo {edition} {1st}\ ed.\ (\bibinfo  {publisher} {Springer Publishing Company, Incorporated},\ \bibinfo {year} {2007})\BibitemShut {NoStop}%
\bibitem [{\citenamefont {{Meil{\u{a}}}}\ and\ \citenamefont {{Zhang}}(2024)}]{2024AnRSA..1140522M}%
  \BibitemOpen
  \bibfield  {author} {\bibinfo {author} {\bibfnamefont {Marina}\ \bibnamefont {{Meil{\u{a}}}}}\ and\ \bibinfo {author} {\bibfnamefont {Hanyu}\ \bibnamefont {{Zhang}}},\ }\bibfield  {title} {\enquote {\bibinfo {title} {{Manifold Learning: What, How, and Why}},}\ }\href {\doibase 10.1146/annurev-statistics-040522-115238} {\bibfield  {journal} {\bibinfo  {journal} {Annual Review of Statistics and Its Application}\ }\textbf {\bibinfo {volume} {11}},\ \bibinfo {eid} {annurev} (\bibinfo {year} {2024})},\ \Eprint {http://arxiv.org/abs/2311.03757} {arXiv:2311.03757 [stat.ML]} \BibitemShut {NoStop}%
\bibitem [{\citenamefont {Kullback}\ and\ \citenamefont {Leibler}(1951)}]{Kullback51klDivergence}%
  \BibitemOpen
  \bibfield  {author} {\bibinfo {author} {\bibfnamefont {S.}~\bibnamefont {Kullback}}\ and\ \bibinfo {author} {\bibfnamefont {R.~A.}\ \bibnamefont {Leibler}},\ }\bibfield  {title} {\enquote {\bibinfo {title} {On information and sufficiency},}\ }\href@noop {} {\bibfield  {journal} {\bibinfo  {journal} {Ann. Math. Statist.}\ }\textbf {\bibinfo {volume} {22}},\ \bibinfo {pages} {79--86} (\bibinfo {year} {1951})}\BibitemShut {NoStop}%
\bibitem [{\citenamefont {Rodgers}\ and\ \citenamefont {Nicewander}(1988)}]{corrcoefficients}%
  \BibitemOpen
  \bibfield  {author} {\bibinfo {author} {\bibfnamefont {Joseph~Lee}\ \bibnamefont {Rodgers}}\ and\ \bibinfo {author} {\bibfnamefont {W.~Alan}\ \bibnamefont {Nicewander}},\ }\bibfield  {title} {\enquote {\bibinfo {title} {Thirteen ways to look at the correlation coefficient},}\ }\href {\doibase 10.1080/00031305.1988.10475524} {\bibfield  {journal} {\bibinfo  {journal} {The American Statistician}\ }\textbf {\bibinfo {volume} {42}},\ \bibinfo {pages} {59--66} (\bibinfo {year} {1988})},\ \Eprint {http://arxiv.org/abs/https://doi.org/10.1080/00031305.1988.10475524} {https://doi.org/10.1080/00031305.1988.10475524} \BibitemShut {NoStop}%
\bibitem [{\citenamefont {Schmidt}\ \emph {et~al.}(2015)\citenamefont {Schmidt}, \citenamefont {Ohme},\ and\ \citenamefont {Hannam}}]{Schmidt_2015}%
  \BibitemOpen
  \bibfield  {author} {\bibinfo {author} {\bibfnamefont {Patricia}\ \bibnamefont {Schmidt}}, \bibinfo {author} {\bibfnamefont {Frank}\ \bibnamefont {Ohme}}, \ and\ \bibinfo {author} {\bibfnamefont {Mark}\ \bibnamefont {Hannam}},\ }\bibfield  {title} {\enquote {\bibinfo {title} {Towards models of gravitational waveforms from generic binaries: Ii. modelling precession effects with a single effective precession parameter},}\ }\href {\doibase 10.1103/physrevd.91.024043} {\bibfield  {journal} {\bibinfo  {journal} {Physical Review D}\ }\textbf {\bibinfo {volume} {91}} (\bibinfo {year} {2015}),\ 10.1103/physrevd.91.024043}\BibitemShut {NoStop}%
\bibitem [{\citenamefont {Davis}\ \emph {et~al.}(2020)\citenamefont {Davis}, \citenamefont {White},\ and\ \citenamefont {Saulson}}]{Davis_2020}%
  \BibitemOpen
  \bibfield  {author} {\bibinfo {author} {\bibfnamefont {Derek}\ \bibnamefont {Davis}}, \bibinfo {author} {\bibfnamefont {Laurel~V}\ \bibnamefont {White}}, \ and\ \bibinfo {author} {\bibfnamefont {Peter~R}\ \bibnamefont {Saulson}},\ }\bibfield  {title} {\enquote {\bibinfo {title} {Utilizing aligo glitch classifications to validate gravitational-wave candidates},}\ }\href {\doibase 10.1088/1361-6382/ab91e6} {\bibfield  {journal} {\bibinfo  {journal} {Classical and Quantum Gravity}\ }\textbf {\bibinfo {volume} {37}},\ \bibinfo {pages} {145001} (\bibinfo {year} {2020})}\BibitemShut {NoStop}%
\bibitem [{\citenamefont {Yosinski}\ \emph {et~al.}(2014)\citenamefont {Yosinski}, \citenamefont {Clune}, \citenamefont {Bengio},\ and\ \citenamefont {Lipson}}]{10.5555/2969033.2969197}%
  \BibitemOpen
  \bibfield  {author} {\bibinfo {author} {\bibfnamefont {Jason}\ \bibnamefont {Yosinski}}, \bibinfo {author} {\bibfnamefont {Jeff}\ \bibnamefont {Clune}}, \bibinfo {author} {\bibfnamefont {Yoshua}\ \bibnamefont {Bengio}}, \ and\ \bibinfo {author} {\bibfnamefont {Hod}\ \bibnamefont {Lipson}},\ }\bibfield  {title} {\enquote {\bibinfo {title} {How transferable are features in deep neural networks?}}\ }in\ \href@noop {} {\emph {\bibinfo {booktitle} {Proceedings of the 27th International Conference on Neural Information Processing Systems - Volume 2}}},\ \bibinfo {series and number} {NIPS'14}\ (\bibinfo  {publisher} {MIT Press},\ \bibinfo {address} {Cambridge, MA, USA},\ \bibinfo {year} {2014})\ p.\ \bibinfo {pages} {3320–3328}\BibitemShut {NoStop}%
\bibitem [{\citenamefont {Li}\ \emph {et~al.}(2015)\citenamefont {Li}, \citenamefont {Yosinski}, \citenamefont {Clune}, \citenamefont {Lipson},\ and\ \citenamefont {Hopcroft}}]{pmlr-v44-li15convergent}%
  \BibitemOpen
  \bibfield  {author} {\bibinfo {author} {\bibfnamefont {Yixuan}\ \bibnamefont {Li}}, \bibinfo {author} {\bibfnamefont {Jason}\ \bibnamefont {Yosinski}}, \bibinfo {author} {\bibfnamefont {Jeff}\ \bibnamefont {Clune}}, \bibinfo {author} {\bibfnamefont {Hod}\ \bibnamefont {Lipson}}, \ and\ \bibinfo {author} {\bibfnamefont {John}\ \bibnamefont {Hopcroft}},\ }\bibfield  {title} {\enquote {\bibinfo {title} {Convergent learning: Do different neural networks learn the same representations?}}\ }in\ \href {https://proceedings.mlr.press/v44/li15convergent.html} {\emph {\bibinfo {booktitle} {Proceedings of the 1st International Workshop on Feature Extraction: Modern Questions and Challenges at NIPS 2015}}},\ \bibinfo {series} {Proceedings of Machine Learning Research}, Vol.~\bibinfo {volume} {44},\ \bibinfo {editor} {edited by\ \bibinfo {editor} {\bibfnamefont {Dmitry}\ \bibnamefont {Storcheus}}, \bibinfo {editor} {\bibfnamefont {Afshin}\ \bibnamefont {Rostamizadeh}}, \ and\ \bibinfo {editor} {\bibfnamefont {Sanjiv}\
  \bibnamefont {Kumar}}}\ (\bibinfo  {publisher} {PMLR},\ \bibinfo {address} {Montreal, Canada},\ \bibinfo {year} {2015})\ pp.\ \bibinfo {pages} {196--212}\BibitemShut {NoStop}%
\end{thebibliography}%
\onecolumngrid

\end{document}